\documentclass[twoolumn,longauth,traditabstract]{aa} % for a paper on 1 column  

\def\setsymbol#1#2{\expandafter\def\csname #1\endcsname{#2}}
\def\getsymbol#1{\csname #1\endcsname}

\def\Planck{\textit{Planck}}

\def\all2013resultspapers{\nocite{planck2013-p01, planck2013-p02, planck2013-p02a, planck2013-p02d, planck2013-p02b, planck2013-p03, planck2013-p03c, planck2013-p03f, planck2013-p03d, planck2013-p03e, planck2013-p01a, planck2013-p06, planck2013-p03a, planck2013-pip88, planck2013-p08, planck2013-p11, planck2013-p12, planck2013-p13, planck2013-p14, planck2013-p15, planck2013-p05b, planck2013-p17, planck2013-p09, planck2013-p09a, planck2013-p20, planck2013-p19, planck2013-pipaberration, planck2013-p05, planck2013-p05a, planck2013-pip56, planck2013-p06b, planck2013-p01a}}

\newbox\tablebox    \newdimen\tablewidth
\def\leaderfil{\leaders\hbox to 5pt{\hss.\hss}\hfil}
\def\endPlancktable{\tablewidth=\columnwidth 
    $$\hss\copy\tablebox\hss$$
    \vskip-\lastskip\vskip -2pt}
\def\endPlancktablewide{\tablewidth=\textwidth 
    $$\hss\copy\tablebox\hss$$
    \vskip-\lastskip\vskip -2pt}
\def\tablenote#1 #2\par{\begingroup \parindent=0.8em
    \abovedisplayshortskip=0pt\belowdisplayshortskip=0pt
    \noindent
    $$\hss\vbox{\hsize\tablewidth \hangindent=\parindent \hangafter=1 \noindent
    \hbox to \parindent{$^#1$\hss}\strut#2\strut\par}\hss$$
    \endgroup}
\def\doubleline{\vskip 3pt\hrule \vskip 1.5pt \hrule \vskip 5pt}

\def\L2{\ifmmode L_2\else $L_2$\fi}

\def\DeltaT{\ifmmode \Delta T\else $\Delta T$\fi}
\def\deltat{\ifmmode \Delta t\else $\Delta t$\fi}
\def\fknee{\ifmmode f_{\rm knee}\else $f_{\rm knee}$\fi}
\def\Fmax{\ifmmode F_{\rm max}\else $F_{\rm max}$\fi}
\def\solar{\ifmmode{\rm M}_{\mathord\odot}\else${\rm M}_{\mathord\odot}$\fi}
\def\Msolar{\ifmmode{\rm M}_{\mathord\odot}\else${\rm M}_{\mathord\odot}$\fi}
\def\Lsolar{\ifmmode{\rm L}_{\mathord\odot}\else${\rm L}_{\mathord\odot}$\fi}
\def\inv{\ifmmode^{-1}\else$^{-1}$\fi}
\def\mo{\ifmmode^{-1}\else$^{-1}$\fi}
\def\sup#1{\ifmmode ^{\rm #1}\else $^{\rm #1}$\fi}
\def\expo#1{\ifmmode \times 10^{#1}\else $\times 10^{#1}$\fi}
\def\,{\thinspace}
\def\lsim{\mathrel{\raise .4ex\hbox{\rlap{$<$}\lower 1.2ex\hbox{$\sim$}}}}
\def\gsim{\mathrel{\raise .4ex\hbox{\rlap{$>$}\lower 1.2ex\hbox{$\sim$}}}}

\def\simprop{\mathrel{\raise .4ex\hbox{\rlap{$\propto$}\lower 1.2ex\hbox{$\sim$}}}}
\def\deg{\ifmmode^\circ\else$^\circ$\fi}
\def\pdeg{\ifmmode $\setbox0=\hbox{$^{\circ}$}\rlap{\hskip.11\wd0 .}$^{\circ}
          \else \setbox0=\hbox{$^{\circ}$}\rlap{\hskip.11\wd0 .}$^{\circ}$\fi}
\def\arcs{\ifmmode {^{\scriptstyle\prime\prime}}
          \else $^{\scriptstyle\prime\prime}$\fi}
\def\arcm{\ifmmode {^{\scriptstyle\prime}}
          \else $^{\scriptstyle\prime}$\fi}
\newdimen\sa  \newdimen\sb
\def\parcs{\sa=.07em \sb=.03em
     \ifmmode \hbox{\rlap{.}}^{\scriptstyle\prime\kern -\sb\prime}\hbox{\kern -\sa}
     \else \rlap{.}$^{\scriptstyle\prime\kern -\sb\prime}$\kern -\sa\fi}
\def\parcm{\sa=.08em \sb=.03em
     \ifmmode \hbox{\rlap{.}\kern\sa}^{\scriptstyle\prime}\hbox{\kern-\sb}
     \else \rlap{.}\kern\sa$^{\scriptstyle\prime}$\kern-\sb\fi}
\def\ra[#1 #2 #3.#4]{#1\sup{h}#2\sup{m}#3\sup{s}\llap.#4}
\def\dec[#1 #2 #3.#4]{#1\deg#2\arcm#3\arcs\llap.#4}
\def\deco[#1 #2 #3]{#1\deg#2\arcm#3\arcs}
\def\rra[#1 #2]{#1\sup{h}#2\sup{m}}

\def\dots{\relax\ifmmode \ldots\else $\ldots$\fi}
\def\WHzsr{\ifmmode $W\,Hz\mo\,sr\mo$\else W\,Hz\mo\,sr\mo\fi}
\def\mHz{\ifmmode $\,mHz$\else \,mHz\fi}
\def\GHz{\ifmmode $\,GHz$\else \,GHz\fi}
\def\mKs{\ifmmode $\,mK\,s$^{1/2}\else \,mK\,s$^{1/2}$\fi}
\def\muKs{\ifmmode \,\mu$K\,s$^{1/2}\else \,$\mu$K\,s$^{1/2}$\fi}
\def\muKRJs{\ifmmode \,\mu$K$_{\rm RJ}$\,s$^{1/2}\else \,$\mu$K$_{\rm RJ}$\,s$^{1/2}$\fi}
\def\muKHz{\ifmmode \,\mu$K\,Hz$^{-1/2}\else \,$\mu$K\,Hz$^{-1/2}$\fi}
\def\MJysr{\ifmmode \,$MJy\,sr\mo$\else \,MJy\,sr\mo\fi}
\def\MJysrmK{\ifmmode \,$MJy\,sr\mo$\,mK$_{\rm CMB}\mo\else \,MJy\,sr\mo\,mK$_{\rm CMB}\mo$\fi}
\def\microns{\ifmmode \,\mu$m$\else \,$\mu$m\fi}
\def\micron{\microns}
\def\muK{\ifmmode \,\mu$K$\else \,$\mu$\hbox{K}\fi}
\def\microK{\ifmmode \,\mu$K$\else \,$\mu$\hbox{K}\fi}
\def\muW{\ifmmode \,\mu$W$\else \,$\mu$\hbox{W}\fi}
\def\kms{\ifmmode $\,km\,s$^{-1}\else \,km\,s$^{-1}$\fi}
\def\kmsMpc{\ifmmode $\,\kms\,Mpc\mo$\else \,\kms\,Mpc\mo\fi}
\providecommand{\sorthelp}[1]{}

\usepackage{graphicx,amsmath}
\usepackage{txfonts}
\usepackage{url}
\usepackage{natbib}
\bibpunct{(}{)}{;}{a}{}{,}  % to follow the A&A style
\usepackage{color}
\usepackage{colortbl} 
\usepackage[usenames,dvipsnames]{xcolor}
\usepackage{subfig} 
\usepackage[breaklinks, colorlinks, citecolor=blue]{hyperref}
\usepackage[switch]{lineno}
\usepackage{fixltx2e} %FB
\usepackage{multirow}
\def\Xpol{{\tt Xpol}}
\def\Xpure{{\tt Xpure}}
\def\Xspect{{\tt Xspect}}

\def\MJysr{${\rm MJy}\,{\rm sr}^{-1}$}
\def\kkms{${\rm K}\,{\rm km}\,{\rm s}^{-1}$}
\def\fsky{\ifmmode f_{\rm sky} \else $f_{\rm sky}$ \fi}
\def\fskyeff{\ifmmode f_{\rm sky}^{\rm eff} \else $f_{\rm sky}^{\rm eff}$ \fi}

\def\clte{$C_\ell^{TE}$}
\def\cltb{$C_\ell^{TB}$}
\def\clee{$C_\ell^{EE}$}
\def\clbb{$C_\ell^{BB}$}
\def\cleb{$C_\ell^{EB}$}
\def\dlee{${\cal D}_\ell^{EE}$}
\def\dlbb{${\cal D}_\ell^{BB}$}
\def\dlte{${\cal D}_\ell^{TE}$}
\def\dltb{${\cal D}_\ell^{TB}$}
\def\dleb{${\cal D}_\ell^{EB}$}
\def\nhi{$N_{\rm \ion{H}{i}}$}

\def\idust{$\left\langle I_{353}\right\rangle$}
\def\pte{{\rm PTE}}

\newcommand{\hi}{\ion{H}{i}}
\newcommand{\hatn}{\vec{\hat{n\,}}}
\newcommand{\hQ}[1]{}
\newcommand{\hpeter}[1]{}
\newcommand{\hfrancois}[1]{}
\newcommand{\hjonathan}[1]{}
\newcommand{\hdouglas}[1]{}

\def\window{window}

\def\window{region}
\def\lwdpref{LW}
\def\lwdpref{LR}
\def\patc{patch}
\def\patcs{patches}
\def\flds{patches}

\def\flds{fields}
\def\sten{field}
\def\stens{fields}
\def\mb2{$M_{\rm B2}$}

\def\betadisp{0.17} 

\def\alphaps{-2.42}
\def\alphapsuncer{0.02}
\def\alphaexponent{exponent}
\def\alphaexponents{exponents}
\def\bicep{BICEP2}

\def\pesb{1.59}
\def\pest{19.6}

\def\lmin{40}
\def\lminfit{60}

\begin{document}
%
%%%%%%%%%%%%%%%%%%%%%%%%%%%%%%%%%%%%%%%%%%%%%%%%%%%%%%%%%%%%%%%%%%%%%%%%%%%%%%%%%%%%%%%%%%%%%%%%%%%%%%%%%%%%%%%%%%

%put page numbers in the left and right margins
 % \setpagewiselinenumbers 
 % \modulolinenumbers[5] 
 % \linenumbers
 
%This author list corresponds to \title{Author list for SVN PIP\_97\_Aumont: The angular power spectrum of polarized dust emission at intermediate and high Galactic latitudes}
%Prepared by R. Leonardi (rleonardi@sciops.esa.int), ESAC/ESA
%This version is from Tue Nov 11 10:55:22 2014 CET
%\subtitle{There are 233 co-authors in this list}
\author{\small
Planck Collaboration:
R.~Adam\inst{80}
\and
P.~A.~R.~Ade\inst{91}
\and
N.~Aghanim\inst{64}
\and
M.~Arnaud\inst{78}
\and
J.~Aumont\inst{64}
\and
C.~Baccigalupi\inst{90}
\and
A.~J.~Banday\inst{99, 10}
\and
R.~B.~Barreiro\inst{70}
\and
J.~G.~Bartlett\inst{1, 72}
\and
N.~Bartolo\inst{33, 71}
% ...
}
\institute{\small
APC, AstroParticule et Cosmologie, Universit\'{e} Paris Diderot, CNRS/IN2P3, CEA/lrfu, Observatoire de Paris, Sorbonne Paris Cit\'{e}, 10, rue Alice Domon et L\'{e}onie Duquet, 75205 Paris Cedex 13, France\goodbreak
\and
African Institute for Mathematical Sciences, 6-8 Melrose Road, Muizenberg, Cape Town, South Africa\goodbreak
\and
Agenzia Spaziale Italiana Science Data Center, Via del Politecnico snc, 00133, Roma, Italy\goodbreak
\and
Agenzia Spaziale Italiana, Viale Liegi 26, Roma, Italy\goodbreak
\and
Aix Marseille Universit\'{e}, CNRS, LAM (Laboratoire d'Astrophysique de Marseille) UMR 7326, 13388, Marseille, France\goodbreak
\and
Astrophysics Group, Cavendish Laboratory, University of Cambridge, J J Thomson Avenue, Cambridge CB3 0HE, U.K.\goodbreak
\and
Astrophysics \& Cosmology Research Unit, School of Mathematics, Statistics \& Computer Science, University of KwaZulu-Natal, Westville Campus, Private Bag X54001, Durban 4000, South Africa\goodbreak
\and
Atacama Large Millimeter/submillimeter Array, ALMA Santiago Central Offices, Alonso de Cordova 3107, Vitacura, Casilla 763 0355, Santiago, Chile\goodbreak
\and
CITA, University of Toronto, 60 St. George St., Toronto, ON M5S 3H8, Canada\goodbreak
\and
CNRS, IRAP, 9 Av. colonel Roche, BP 44346, F-31028 Toulouse cedex 4, France\goodbreak
\and
CRANN, Trinity College, Dublin, Ireland\goodbreak
\and
California Institute of Technology, Pasadena, California, U.S.A.\goodbreak
\and
Centre for Theoretical Cosmology, DAMTP, University of Cambridge, Wilberforce Road, Cambridge CB3 0WA, U.K.\goodbreak
\and
Centro de Estudios de F\'{i}sica del Cosmos de Arag\'{o}n (CEFCA), Plaza San Juan, 1, planta 2, E-44001, Teruel, Spain\goodbreak
\and
Computational Cosmology Center, Lawrence Berkeley National Laboratory, Berkeley, California, U.S.A.\goodbreak
\and
Consejo Superior de Investigaciones Cient\'{\i}ficas (CSIC), Madrid, Spain\goodbreak
\and
DSM/Irfu/SPP, CEA-Saclay, F-91191 Gif-sur-Yvette Cedex, France\goodbreak
\and
DTU Space, National Space Institute, Technical University of Denmark, Elektrovej 327, DK-2800 Kgs. Lyngby, Denmark\goodbreak
\and
D\'{e}partement de Physique Th\'{e}orique, Universit\'{e} de Gen\`{e}ve, 24, Quai E. Ansermet,1211 Gen\`{e}ve 4, Switzerland\goodbreak
\and
Departamento de F\'{\i}sica, Universidad de Oviedo, Avda. Calvo Sotelo s/n, Oviedo, Spain\goodbreak
\and
Department of Astronomy and Astrophysics, University of Toronto, 50 Saint George Street, Toronto, Ontario, Canada\goodbreak
\and
Department of Astrophysics/IMAPP, Radboud University Nijmegen, P.O. Box 9010, 6500 GL Nijmegen, The Netherlands\goodbreak
\and
Department of Physics \& Astronomy, University of British Columbia, 6224 Agricultural Road, Vancouver, British Columbia, Canada\goodbreak
\and
Department of Physics and Astronomy, Dana and David Dornsife College of Letter, Arts and Sciences, University of Southern California, Los Angeles, CA 90089, U.S.A.\goodbreak
\and
Department of Physics and Astronomy, University College London, London WC1E 6BT, U.K.\goodbreak
\and
Department of Physics, Florida State University, Keen Physics Building, 77 Chieftan Way, Tallahassee, Florida, U.S.A.\goodbreak
\and
Department of Physics, Gustaf H\"{a}llstr\"{o}min katu 2a, University of Helsinki, Helsinki, Finland\goodbreak
\and
Department of Physics, Princeton University, Princeton, New Jersey, U.S.A.\goodbreak
\and
Department of Physics, University of California, Berkeley, California, U.S.A.\goodbreak
\and
Department of Physics, University of California, One Shields Avenue, Davis, California, U.S.A.\goodbreak
\and
Department of Physics, University of California, Santa Barbara, California, U.S.A.\goodbreak
\and
Department of Physics, University of Illinois at Urbana-Champaign, 1110 West Green Street, Urbana, Illinois, U.S.A.\goodbreak
\and
Dipartimento di Fisica e Astronomia G. Galilei, Universit\`{a} degli Studi di Padova, via Marzolo 8, 35131 Padova, Italy\goodbreak
\and
Dipartimento di Fisica e Scienze della Terra, Universit\`{a} di Ferrara, Via Saragat 1, 44122 Ferrara, Italy\goodbreak
\and
Dipartimento di Fisica, Universit\`{a} La Sapienza, P. le A. Moro 2, Roma, Italy\goodbreak
\and
Dipartimento di Fisica, Universit\`{a} degli Studi di Milano, Via Celoria, 16, Milano, Italy\goodbreak
\and
Dipartimento di Fisica, Universit\`{a} degli Studi di Trieste, via A. Valerio 2, Trieste, Italy\goodbreak
\and
Dipartimento di Fisica, Universit\`{a} di Roma Tor Vergata, Via della Ricerca Scientifica, 1, Roma, Italy\goodbreak
\and
Dipartimento di Matematica, Universit\`{a} di Roma Tor Vergata, Via della Ricerca Scientifica, 1, Roma, Italy\goodbreak
\and
Discovery Center, Niels Bohr Institute, Blegdamsvej 17, Copenhagen, Denmark\goodbreak
\and
Dpto. Astrof\'{i}sica, Universidad de La Laguna (ULL), E-38206 La Laguna, Tenerife, Spain\goodbreak
\and
European Southern Observatory, ESO Vitacura, Alonso de Cordova 3107, Vitacura, Casilla 19001, Santiago, Chile\goodbreak
\and
European Space Agency, ESAC, Planck Science Office, Camino bajo del Castillo, s/n, Urbanizaci\'{o}n Villafranca del Castillo, Villanueva de la Ca\~{n}ada, Madrid, Spain\goodbreak
\and
European Space Agency, ESTEC, Keplerlaan 1, 2201 AZ Noordwijk, The Netherlands\goodbreak
\and
Facolt\`{a} di Ingegneria, Universit\`{a} degli Studi e-Campus, Via Isimbardi 10, Novedrate (CO), 22060, Italy\goodbreak
\and
Gran Sasso Science Institute, INFN, viale F. Crispi 7, 67100 L'Aquila, Italy\goodbreak
\and
HGSFP and University of Heidelberg, Theoretical Physics Department, Philosophenweg 16, 69120, Heidelberg, Germany\goodbreak
\and
Haverford College Astronomy Department, 370 Lancaster Avenue, Haverford, Pennsylvania, U.S.A.\goodbreak
\and
Helsinki Institute of Physics, Gustaf H\"{a}llstr\"{o}min katu 2, University of Helsinki, Helsinki, Finland\goodbreak
\and
INAF - Osservatorio Astronomico di Padova, Vicolo dell'Osservatorio 5, Padova, Italy\goodbreak
\and
INAF - Osservatorio Astronomico di Roma, via di Frascati 33, Monte Porzio Catone, Italy\goodbreak
\and
INAF - Osservatorio Astronomico di Trieste, Via G.B. Tiepolo 11, Trieste, Italy\goodbreak
\and
INAF/IASF Bologna, Via Gobetti 101, Bologna, Italy\goodbreak
\and
INAF/IASF Milano, Via E. Bassini 15, Milano, Italy\goodbreak
\and
INFN, Sezione di Bologna, Via Irnerio 46, I-40126, Bologna, Italy\goodbreak
\and
INFN, Sezione di Roma 1, Universit\`{a} di Roma Sapienza, Piazzale Aldo Moro 2, 00185, Roma, Italy\goodbreak
\and
INFN, Sezione di Roma 2, Universit\`{a} di Roma Tor Vergata, Via della Ricerca Scientifica, 1, Roma, Italy\goodbreak
\and
INFN/National Institute for Nuclear Physics, Via Valerio 2, I-34127 Trieste, Italy\goodbreak
\and
IPAG: Institut de Plan\'{e}tologie et d'Astrophysique de Grenoble, Universit\'{e} Grenoble Alpes, IPAG, F-38000 Grenoble, France, CNRS, IPAG, F-38000 Grenoble, France\goodbreak
\and
IUCAA, Post Bag 4, Ganeshkhind, Pune University Campus, Pune 411 007, India\goodbreak
\and
Imperial College London, Astrophysics group, Blackett Laboratory, Prince Consort Road, London, SW7 2AZ, U.K.\goodbreak
\and
Infrared Processing and Analysis Center, California Institute of Technology, Pasadena, CA 91125, U.S.A.\goodbreak
\and
Institut Universitaire de France, 103, bd Saint-Michel, 75005, Paris, France\goodbreak
\and
Institut d'Astrophysique Spatiale, CNRS (UMR8617) Universit\'{e} Paris-Sud 11, B\^{a}timent 121, Orsay, France\goodbreak
\and
Institut d'Astrophysique de Paris, CNRS (UMR7095), 98 bis Boulevard Arago, F-75014, Paris, France\goodbreak
\and
Institute for Space Sciences, Bucharest-Magurale, Romania\goodbreak
\and
Institute of Astronomy, University of Cambridge, Madingley Road, Cambridge CB3 0HA, U.K.\goodbreak
\and
Institute of Theoretical Astrophysics, University of Oslo, Blindern, Oslo, Norway\goodbreak
\and
Instituto de Astrof\'{\i}sica de Canarias, C/V\'{\i}a L\'{a}ctea s/n, La Laguna, Tenerife, Spain\goodbreak
\and
Instituto de F\'{\i}sica de Cantabria (CSIC-Universidad de Cantabria), Avda. de los Castros s/n, Santander, Spain\goodbreak
\and
Istituto Nazionale di Fisica Nucleare, Sezione di Padova, via Marzolo 8, I-35131 Padova, Italy\goodbreak
\and
Jet Propulsion Laboratory, California Institute of Technology, 4800 Oak Grove Drive, Pasadena, California, U.S.A.\goodbreak
\and
Jodrell Bank Centre for Astrophysics, Alan Turing Building, School of Physics and Astronomy, The University of Manchester, Oxford Road, Manchester, M13 9PL, U.K.\goodbreak
\and
Kavli Institute for Cosmology Cambridge, Madingley Road, Cambridge, CB3 0HA, U.K.\goodbreak
\and
LAL, Universit\'{e} Paris-Sud, CNRS/IN2P3, Orsay, France\goodbreak
\and
LAPTh, Univ. de Savoie, CNRS, B.P.110, Annecy-le-Vieux F-74941, France\goodbreak
\and
LERMA, CNRS, Observatoire de Paris, 61 Avenue de l'Observatoire, Paris, France\goodbreak
\and
Laboratoire AIM, IRFU/Service d'Astrophysique - CEA/DSM - CNRS - Universit\'{e} Paris Diderot, B\^{a}t. 709, CEA-Saclay, F-91191 Gif-sur-Yvette Cedex, France\goodbreak
\and
Laboratoire Traitement et Communication de l'Information, CNRS (UMR 5141) and T\'{e}l\'{e}com ParisTech, 46 rue Barrault F-75634 Paris Cedex 13, France\goodbreak
\and
Laboratoire de Physique Subatomique et de Cosmologie, Universit\'{e} Joseph Fourier Grenoble I, CNRS/IN2P3, Institut National Polytechnique de Grenoble, 53 rue des Martyrs, 38026 Grenoble cedex, France\goodbreak
\and
Laboratoire de Physique Th\'{e}orique, Universit\'{e} Paris-Sud 11 \& CNRS, B\^{a}timent 210, 91405 Orsay, France\goodbreak
\and
Lawrence Berkeley National Laboratory, Berkeley, California, U.S.A.\goodbreak
\and
Max-Planck-Institut f\"{u}r Astrophysik, Karl-Schwarzschild-Str. 1, 85741 Garching, Germany\goodbreak
\and
McGill Physics, Ernest Rutherford Physics Building, McGill University, 3600 rue University, Montr\'{e}al, QC, H3A 2T8, Canada\goodbreak
\and
National University of Ireland, Department of Experimental Physics, Maynooth, Co. Kildare, Ireland\goodbreak
\and
Niels Bohr Institute, Blegdamsvej 17, Copenhagen, Denmark\goodbreak
\and
Observational Cosmology, Mail Stop 367-17, California Institute of Technology, Pasadena, CA, 91125, U.S.A.\goodbreak
\and
Optical Science Laboratory, University College London, Gower Street, London, U.K.\goodbreak
\and
SB-ITP-LPPC, EPFL, CH-1015, Lausanne, Switzerland\goodbreak
\and
SISSA, Astrophysics Sector, via Bonomea 265, 34136, Trieste, Italy\goodbreak
\and
School of Physics and Astronomy, Cardiff University, Queens Buildings, The Parade, Cardiff, CF24 3AA, U.K.\goodbreak
\and
School of Physics and Astronomy, University of Nottingham, Nottingham NG7 2RD, U.K.\goodbreak
\and
Space Research Institute (IKI), Russian Academy of Sciences, Profsoyuznaya Str, 84/32, Moscow, 117997, Russia\goodbreak
\and
Space Sciences Laboratory, University of California, Berkeley, California, U.S.A.\goodbreak
\and
Special Astrophysical Observatory, Russian Academy of Sciences, Nizhnij Arkhyz, Zelenchukskiy region, Karachai-Cherkessian Republic, 369167, Russia\goodbreak
\and
Sub-Department of Astrophysics, University of Oxford, Keble Road, Oxford OX1 3RH, U.K.\goodbreak
\and
Theory Division, PH-TH, CERN, CH-1211, Geneva 23, Switzerland\goodbreak
\and
UPMC Univ Paris 06, UMR7095, 98 bis Boulevard Arago, F-75014, Paris, France\goodbreak
\and
Universit\'{e} de Toulouse, UPS-OMP, IRAP, F-31028 Toulouse cedex 4, France\goodbreak
% ...
}

%%%%%%%%%%%%%%%%%%%%%%%%%%%%%%%
% PIP preliminary number removed in version for BICEP2 team
%%%%%%%%%%%%%%%%%%%%%%%%%%%%%%%
\title{\textit{Planck\/} intermediate results. XXX.\\
The angular power spectrum of polarized dust emission\\
at intermediate and high Galactic latitudes}

\offprints{Jonathan~Aumont ({\tt jonathan.aumont@ias.u-psud.fr})}

\authorrunning{Planck Collaboration}
\titlerunning{Dust polarization at high latitudes}
%
%%%%%%%%%%%%%%%%%%%%%%%%%%%%%%%%%%%%%%%%%%%%%%%%%%%%%%%%%%%%%%%%%%%%%%%%%%%%%%%%%%%%%%%%%%%%%%%%%%%%%%%%%%%%%%%%%%

\date{Received 19 September 2014; accepted 1 December 2014}
%\date{Preprint online version: xx November 2014}

\abstract{ The polarized thermal emission from diffuse Galactic dust
  is the main foreground present in measurements of the polarization
  of the cosmic microwave background (CMB) at frequencies above 100\,GHz.
  In this paper we exploit the uniqueness of the \Planck\ HFI
  polarization data from 100 to 353\,GHz to measure
  the polarized dust angular power spectra \clee\ and \clbb\ 
  over the multipole range $\lmin<\ell<600$ well away from the Galactic
  plane. These measurements will bring new insights into interstellar
  dust physics and allow a precise determination of the level of
  contamination for CMB polarization experiments. Despite the
  non-Gaussian and anisotropic nature of Galactic dust, we show that
  general statistical properties of the emission can be characterized
  accurately over large fractions of the sky using angular power spectra. The
  polarization power spectra of the dust are well described by power
  laws in multipole, $C_\ell\propto \ell^\alpha$, with \alphaexponents\
  $\alpha^{EE,BB}=\alphaps\pm\alphapsuncer$. The amplitudes of the polarization
  power spectra vary with the average brightness in a way similar to
  the intensity power spectra. The frequency dependence of the
  dust polarization spectra is consistent with modified blackbody
  emission with $\beta_{\rm d}=\pesb$ and $T_{\rm d}=\pest\,$K down to the
  lowest \Planck\ HFI frequencies. We find a systematic difference between the
  amplitudes of the Galactic $B$- and $E$-modes, $C_\ell^{BB} /C_\ell^{EE} =
  0.5$. We verify that these general properties are preserved towards
  high Galactic latitudes with low dust column densities. We show that
  even in the faintest dust-emitting regions there are no ``clean''
  windows in the sky where primordial CMB $B$-mode polarization
  measurements could be made without subtraction of foreground emission. 
  Finally, we investigate the level of dust polarization in
  the specific field recently targeted by the \bicep\ experiment.
  Extrapolation of the \Planck\ 353\,GHz data to 150\,GHz gives a dust power
  ${\cal D_\ell^{BB}}\equiv\ell(\ell+1)C_\ell^{BB}/(2\pi)$ of
  $1.32 \times 10^{-2}\,\mu$K$_{\rm CMB}^2$ over the multipole
  range of the primordial recombination bump ($40<\ell<120$);
  the statistical uncertainty is
  $\pm0.29 \times 10^{-2}\,\mu$K$_{\rm CMB}^2$ and there is an additional
  uncertainty $(+0.28,-0.24) \times 10^{-2}\,\mu$K$_{\rm CMB}^2$ from the
  extrapolation.
  This level is the same magnitude as reported by \bicep\ over this $\ell$
  range, which highlights the need for assessment of the polarized dust signal
  even in the cleanest windows of the sky.
  % The present uncertainties are large and will be reduced through an ongoing,
  % joint analysis of the \Planck\ and \bicep\ data sets.
}


\maketitle

\section{Introduction}

The sky at high Galactic latitude and frequencies above about 100\,GHz is
dominated by thermal emission from the Galactic interstellar medium,
specifically arising from dust grains of size about 0.1\micron.  Asymmetrical
dust grains align with the Galactic magnetic field to produce polarized
emission.  This polarized submillimetre emission has been measured from
ground-based and balloon-borne telescopes
\citep[e.g.,][]{Hildebrand,ArcheopsDust,ArcheopsDustPS,Vaillancourt,Matthews}.
The observed polarization relates to the nature, size and shape of dust grains
and the mechanisms of alignment,
discussed for example by \citet{Draine04} and \citet{Martin07}.
It also probes the structure of the Galactic magnetic field, which is an
essential component of models of Galactic dust
polarization \citep{Bacci,Fauvet1,Fauvet2,Odea,Jaffe,PSM}.

The polarized emission from dust is also of interest in the context of
foregrounds \citep{TucciB,Dunkley,WMAP7Galactic}
to the cosmic microwave background (CMB).  On angular scales between 10\arcmin\
and a few tens of degrees, cosmological $B$-mode polarization signals
may be present that were imprinted during the epoch of inflation.  The
discovery of a primordial \mbox{$B$-mode} polarization signature is a major
scientific goal of a large number of CMB experiments.  These include
ground-based experiments (ACTPol, \citealt{ACTpolConcept};
\bicep, \citealt{BICEP2Instru}; Keck-array, \citealt{KeckConcept};
POLARBEAR, \citealt{POLARBEARConcept}; QUBIC, \citealt{QUBICConcept};
QUIJOTE, \citealt{QUIJOTE}; and SPTpol, \citealt{SPTpolConcept}), stratospheric
balloon missions (EBEX, \citealt{EBEXConcept}; and
SPIDER, \citealt{SPIDERConcept}) and the ESA
\Planck\footnote{\Planck\ (\url{http://www.esa.int/Planck}) is a project of
the European Space Agency (ESA) with instruments provided by two scientific
consortia funded by ESA member states (in particular the lead countries France
and Italy), with contributions from NASA (USA) and telescope reflectors
provided by a collaboration between ESA and a scientific consortium led and
funded by Denmark.} satellite \citep{PlanckPreLaunch}.  Accurate assessment,
and if necessary subtraction of foreground contamination is critical
to the measurement of CMB $E$- and $B$-mode polarization because
the expected signals from inflation and late-time reionization are expected
to be small.

\Planck\ has measured the all-sky dust polarization
at 353\,GHz, where the dust emission dominates over other
polarized signals.  These data have been presented in a first set of
publications in which the focus was on the structure of the Galactic magnetic
field and the characterization of dust polarization properties
\citep{planck2014-XIX,planck2014-XX,planck2014-XXI,planck2014-XXII}.
Here, we use the \Planck\ polarized data to compute
the \clee\ and \clbb\ power spectra of dust polarization over the
multipole range $\lmin<\ell<600$, on large fractions of the sky
away from the Galactic plane.  We also investigate dust polarization
in sky \patcs\ at high Galactic
latitude with sizes comparable to those surveyed by ground-based CMB
experiments.  We derive statistical properties of dust polarization
from these spectra, characterizing the shape of the spectra and
their amplitude with respect to both the observing frequency and
the mean dust intensity of the sky \window\ over which they are
computed.  We verify that these properties hold in low-column-density
\patcs\ at high Galactic latitude and we explore statistically the
potential existence of ``clean'' \patcs\ on the sky that might be
suitable for cosmology.

Our analysis of dust polarization is relevant to the present
generation of CMB polarization observations, as well as the design of
future experiments.  It gives a statistical description of Galactic
dust polarization, providing input for the modelling of Galactic dust
as part of component separation
methods and for CMB polarization likelihood analysis parameterization.

The \bicep\ collaboration has recently reported a significant detection
of the $B$-mode power spectrum around the expected angular scale of the
recombination bump, namely a few degrees
\citep{BICEP2Instru,BICEP2_PRL}.  Their analysis was based on dust
polarization models that predicted 
subdominant contamination of their $B$-mode
signal by dust polarization.  We use information from our detailed analysis of 
\Planck\ polarization data at $353\,$GHz to assess the potential dust
contamination.

The paper is organized as follows.  
In Sect.~\ref{part:data}, we present the \Planck\ HFI polarization data
used in this work and describe the general properties of the
polarization maps in terms of emission components and systematic effects.
In Sect.~\ref{part:computation}, we describe our method for computing
the dust \clee\ and \clbb\ angular power spectra, including the selected
science {\window}s of interest on the sky.
We assess and compare the two methods we use to compute the power spectra in
Appendix~\ref{methods_performance}.
In Sect.~\ref{intlat_spectra}, we present power spectra of dust
polarization for multipoles $\ell>\lmin$, computed with high signal-to-noise
ratio (S/N) on large
fractions of the sky, and characterize their shape and amplitude.  
Complementary angular power spectra involving temperature and
polarization, \clte and \cltb, and cross-polarization, \cleb\, are given in
Appendix~\ref{te-tb-eb}.
We extend this analysis to smaller sky \patcs\ at high Galactic
latitude in Sect.~\ref{highlat_spectra}.  
In Appendix~\ref{appendix_highlat} we discuss some complementary aspects of
this analysis of \patcs.
These results are used specifically in Sect.~\ref{sec:optimization} to build a
map of the expected dust contamination of the \clbb\ power spectrum at
150\,GHz and $\ell = 80$.  
In Sect.~\ref{sect:bicep2} we present a study of the polarized dust emission
in the vicinity of the \bicep\ region.
Systematic effects relating to the \Planck\ angular power spectrum estimates
are assessed in Appendix~\ref{methods_bicep}
and we discuss the decorrelation
of the dust signal between frequencies in Appendix~\ref{decorrelation}.
Section~\ref{part:conclusions} summarizes the main conclusions and
discusses the implications of this work for future CMB
experiments.\footnote{While this paper was in preparation three
papers have used publicly available polarization information from \Planck\
to infer potentially high levels of dust contamination in the \bicep\
field \citep{Flauger2014,Mortonson2014,Colley2014}.}

%%%%%%%%%%%%%%%%%%%%%%%%%%%%%%%%%%%%%%%%%%%%%%%%%%%%%%%%%%%%%%%%%%%%%
\section{\Planck\ polarization maps}
\label{part:data}

\subsection{\Planck\ data}
%%%%%%%%%%%%%%%%%%%%%%%%%%%%%%%%%%%%%%%%%%
\label{data_description}

The \Planck\ collaboration recently released the \Planck\ satellite nominal
mission temperature data and published a set of papers describing these data
and their cosmological interpretation
\citep[e.g.,][]{planck2013-p01,planck2013-p11}.  These results are based on the
data from the two instruments on-board the satellite \citep[LFI, Low
Frequency Instrument,][]{planck2011-1.4} and \citep[HFI, High
Frequency Instrument,][]{planck2011-1.5}.  The data processing of the
nominal mission data (Surveys~1 and 2, 14 months) was summarized in
\cite{planck2013-p02} and \cite{planck2013-p03}.

\Planck\ HFI measures the linear polarization at 100, 143,
217, and 353\,GHz \citep{rosset2010}. The properties of the detectors
(sensitivity, spectral response, noise properties, beams, etc.) are
described in detail in \cite{lamarre2010} and their in-flight
performance is reported in \cite{planck2011-1.5},
\cite{planck2013-p03c}, \cite{planck2013-p03f}, \cite{planck2013-p03d},
and \cite{planck2013-p03e}, while
\cite{planck2013-p03} describes the general processing applied to the
data to measure polarization.  In this paper, we make use of
full-mission (Surveys~1 to 5, 30 months, \citealt{planck2013-p01}),
polarization maps of the \Planck\ HFI (internal data release ``DX11d''),
projected into the {\tt HEALPix} pixelization
scheme \citep{gorski2005}.
This is one of the first publications to use
these maps, which will be described in the \Planck\ cosmology 2014 release.

To compute polarization angular power spectra, we use $Q$ and $U$ maps at
100, 143, 217, and 353\,GHz.  Specifically, we calculate power spectra
using the so-called ``Detector-Set'' maps (hereafter ``DetSets''),
constructed using two subsets of polarization sensitive
bolometers (PSBs) at a given frequency
\citep[see table~3 of][]{planck2013-p03}. Each DetSet
polarization map is constructed using data from two pairs of PSBs,
with the angle between the two PSBs in a pair being $90^\circ$, and
the angle between pairs being $45^\circ$.  In this paper we concentrate on the
$Q$ and $U$ maps at 353\,GHz.  The Stokes $Q$ and $U$ maps at lower
frequencies (100, 143, and 217\,GHz) are only used to determine the spectral
energy distribution (SED) of the dust emission in polarization.

To quantify systematic effects, we additionally use maps made from other data
subsets \citep{planck2013-p03}.  We use the ring halves, (hereafter
``HalfRing''), where the approximately 60 circles performed for each \Planck\
telescope ring (also called a stable pointing period) are divided
into two independent subsets of 30 circles.
Additionally we use observational years
(hereafter ``Years''), consisting of Surveys~1 and 2 on the one hand and
Surveys 3 and 4 on the other, to build two further
maps with independent noise.

The \Planck\ maps we use are in thermodynamic units (K$_{\rm CMB}$). To
characterize the SED of the dust emission in polarization we express
the data as the specific intensity (such as $I_{\rm d}(\nu)$ for
Stokes $I$ dust emission)
at the \Planck\ reference frequencies, using the
conversion factors and colour corrections from
\citet{planck2013-p03d}.\footnote{The conversion factor from K$_{\rm CMB}$ to
MJy\,sr$^{-1}$ is computed for a specific intensity $I_\nu \propto \nu^{\,-1}$. 
The colour correction depends on the dust SED; 
it is the scaling factor used to transform from the specific intensity of the
dust emission, at the reference frequency, to the \Planck\ brightness
in MJy\,sr$^{-1}$ \citep[see equation~19 in][]{planck2014-XXII}.
The conversion factors and the colour corrections are computed via
equation~(32) in \citet{planck2013-p03d} using the \Planck\ HFI filters and
the \Planck\ {\tt UcCC} software
available through the \Planck\ Explanatory Supplement
(\url{http://www.sciops.esa.int/wikiSI/planckpla/index.php?title=Unit_conversion_and_Color_correction&instance=Planck_Public_PLA});
we use the band-average values.}
For the average dust SED at intermediate Galactic latitudes,   
the colour correction factor is 1.12 at 353\,GHz
\citep[see Table 3 in][]{planck2014-XXII}.

As well as these basic products,
a \Planck\ CO map from \citet{planck2013-p03a}, the so-called ``Type~3'' map,
and the \Planck\ 857\,GHz map, are also used in the selection of the
large intermediate latitude analysis {\window}s (see Sect.~\ref{intlat_masks}).

%%%%%%%%%%%%%%%%%%%%%%%%%%%%%%%%%%%%%%%%%%
\subsection{Emission contributions to the \Planck\ HFI polarization maps}
\label{components}

\subsubsection{Polarized thermal dust emission} 
\label{dust}

Thermal dust emission is partially linearly polarized
\citep[e.g.,][]{Hildebrand,ArcheopsDust,ArcheopsDustPS,Vaillancourt}.
It is the dominant polarized foreground signal in the high frequency \Planck\
bands \citep{TucciB,WMAP5Likl,Fraisse,Fauvet1,planck2014-XXII}.

Dust polarization arises from alignment of non-spherical grains with the
interstellar magnetic field \citep[e.g.,][]{Hildebrand88,Draine04,Martin07}.
The structure of the dust polarization sky has already been described using
maps of the polarization fraction ($p$) and angle ($\psi$) derived from the
\Planck\ HFI 353\,GHz data  \citep{planck2014-XIX,planck2014-XX}.
The map of $p$ shows structure on all scales, with polarization fractions
ranging from low (less than 1\,\%) to high values (greater than 18\,\%).
\citet{planck2014-XIX} and \citet{planck2014-XX} report an
anti-correlation between $p$ and the local dispersion of $\psi$,
which indicates that variations in $p$ arise mainly
from depolarization associated with changes in the magnetic field orientation 
within the beam, rather than from changes in the efficiency of grain alignment.  

\citet{planck2014-XXII} showed that the SED of polarized dust emission
over the four \Planck\ HFI frequencies from 100 to 353\,GHz is consistent
with a modified blackbody emission law of the type
$I_{\rm d}(\nu)\propto\nu^{\,\beta_{\rm d}}B_\nu(T_{\rm d})$,
with spectral index $\beta_{\rm d}=\pesb$ for
$T_{\rm d}=\pest\,$K,\footnote{This spectral index was called
$\beta^{\,\rm p}_{\rm d,mm}$ in that paper, but we adopt a more compact
notation here.} and where $B_\nu$ is the Planck function.
About 39\,\% of the sky at intermediate Galactic latitudes was
analysed.\footnote{More specifically, for the latitude range
$10^\circ < |b| < 60^\circ$, with
patches contained within the region in Fig.~\ref{fig:masks} (below)
defined by $\fsky=0.8$ minus that with $\fsky=0.4$.}
Among 400 circular patches with $10^\circ$ radius (equivalent to a sky fraction
$\fskyeff = 0.0076$) the
$1\,\sigma$ dispersion of $\beta_{\rm d}$ was \betadisp\ 
for constant $T_{\rm d}=\pest\,$K.
We scale this uncertainty on $\beta_{\rm d}$
to larger sky areas by using the factor $(0.0076/\fskyeff)^{0.5}$.
This is a conservative choice because this uncertainty 
includes the effects of noise in the data and so is an upper limit to
the true regional variations of $\beta_{\rm d}$ on this scale.
This polarization spectral index
can be compared to variations in the spectral index
$\beta^{\rm I}_{\rm d,mm}$ for the intensity SED.
For that quantity the S/N of the data is higher than for polarization and
\citet{planck2014-XXII} report a dispersion of  
0.07 ($1\,\sigma$)  over the same sized circular patches.
\citet{planck2013-XVII} extend this analysis for intensity to high Galactic
latitudes in the southern Galactic cap, 
using the dust-\hi\ correlation to separate the faint emission of dust from 
the anisotropies of the cosmic infrared background, and find
a dispersion of about 0.10 in $\beta^{\rm I}_{\rm d,mm}$.
We expect spectral variations to be correlated in polarization and intensity,
unless the dust emission has a significant component that is unpolarized.

%============================================
\subsubsection{CMB} 
\label{CMB}

The CMB temperature anisotropies have been measured with unprecedented
accuracy by the \Planck\ collaboration
\citep{planck2013-p01,planck2013-p08}, and preliminary \Planck\
polarization results have been demonstrated to be in very 
good agreement with the cosmology inferred from temperature measurements
\citep{planck2013-p01,planck2013-p11}. 

For \clee, the $\Lambda$CDM
concordance model has been shown to be a very good fit to all the
available data (including preliminary \Planck\ results at
$\ell\gtrsim50$; see \citealt{BICEP1} for a recent compendium).
For 353\,GHz data at small
angular scales ($\ell\,{\gtrsim}\,400$), the $E$-mode CMB polarization is
comparable to the power of dust polarization at high Galactic latitudes. 

The CMB $B$-mode power, even for the highest 
primordial tensor perturbation models, is negligible with respect to the dust 
polarization at 353\,GHz.  Since no reliable published CMB polarization maps
are available, we have chosen not to
remove the CMB polarization from the \Planck\ HFI $Q$ and $U$ maps.
Nevertheless, when studying the \Planck\ HFI bands, because the
CMB $E$-mode polarization is significant with respect to the dust at 353 GHz
at high multipoles (and even at lower multipoles for the lower
frequencies), we subtract from the dust power spectra the \Planck\
best-fit $\Lambda$CDM \clee\ model (column 2 of
table 2 in \citealt{planck2013-p12}),
paying the price of an increased error due to sample variance. No CMB is
removed in this work when computing the dust \clbb\ spectra.

%============================================
\subsubsection{Synchrotron emission}
\label{sync}

Synchrotron emission is known to be significantly polarized \citep[up
to 75\,\% for typical relativistic electron spectra,][]{Rybicki}.
Since its specific intensity scaling with frequency follows
a power law with a spectral index close to $-3$
\citep{WMAP7Galactic,Macellari,Fuskeland}, synchrotron polarized
emission is expected to be subdominant in the \Planck\ HFI channels 
in general and negligible at 353\,GHz
\citep{TucciB,Dunkley,WMAP7Galactic,Fauvet1,Fuskeland,planck2014-XXII}.
Hence, we neither subtract nor mask any synchrotron contribution
before estimating the angular power spectra of dust polarization.  The
justification of this assumption will be demonstrated below by
studying the frequency dependence of the polarized dust power between
100 and 353\,GHz
(but also see Appendix~\ref{sec:bicep2_synchrotron}).

%============================================
\subsubsection{Polarized point sources} 
\label{polPS}

Radio sources have been shown to have a fractional polarization of a
few percent \citep[e.g.,][]{Battye,Massardi}.  Their contribution
to the polarization angular power spectra in the \Planck\ HFI bands is
expected to be negligible at low and intermediate multipoles
\citep{Battye,TucciPS}.  Upper limits have been set on the
polarization of infrared galaxies, and their contribution to the
polarization power spectra is also expected to be negligible
\citep[e.g.,][]{Seiffert}.  However, the brightest of the polarized
point sources can be responsible for ringing in the angular power
spectra estimation, and therefore need to be masked
(see Sect.~\ref{intlat_masks}).

%============================================
\subsubsection{CO emission} 
\label{CO}

The first three carbon monoxide (CO) Galactic emission lines at
115\,GHz ($J\,{=}\,1\,{\rightarrow}\,0$), 230\,GHz
($J\,{=}\,2\,{\rightarrow}\,1$), and 345\,GHz
($J\,{=}\,3\,{\rightarrow}\,2$), contribute significantly to the power
in the \Planck\ HFI bands at 100, 217, and 353\,GHz, respectively
\citep{planck2013-p03d}. The \Planck\ data were used to produce the
first all-sky maps of Galactic CO emission \citep{planck2013-p03a}.
It is known that CO emission can be intrinsically polarized
\citep{Goldreich82,PolarizedCO}.
Furthermore, CO emission can induce spurious polarization, due to the
differences in spectral transmission at the CO frequencies between
\Planck\ HFI detectors \citep{planck2013-p03d}.  For these reasons, we
mask CO-emitting regions (Sect.~\ref{intlat_masks}).  Outside
this mask, as has been shown in \cite{planck2013-p03a} by comparing
the \Planck\ CO maps to high Galactic latitude ground-based CO observations
\citep{Hartmann,Magnani}, the CO emission is negligible in the
\Planck\ channels (lower than one fourth of each channel noise rms
at 95\,\% CL).  We will check that our polarization analysis is not
contaminated by CO by examining the frequency dependence of the
polarized angular power spectra of the dust (see Sect.~\ref{SED}).

%%%%%%%%%%%%%%%%%%%%%%%%%%%%%%%%%%%%%%%%%%
\subsection{Systematics of the \Planck\ HFI polarization maps}
\label{systematics}

The first CMB polarization results from \Planck\ were presented in
\cite{planck2013-p01} and \cite{planck2013-p11}.  The $EE$ power
spectrum at $\ell > 200$ was found to be consistent with the
cosmological model derived from temperature anisotropies.  Systematic effects
in the data have so far limited the use of \Planck\ HFI polarization
data on large angular scales.  The polarization systematics in the
2013 data were discussed and estimated in \citet{planck2013-p03} (see
the power spectra shown in their figure~27).  The same data were used
in the \Planck\ Galactic polarization papers
\citep{planck2014-XIX,planck2014-XX,planck2014-XXI,planck2014-XXII}
and there the low brightness regions of the 353\,GHz sky were masked.

In this paper, we use a new set of \Planck\ polarization maps for
which the systematic effects have been significantly reduced. 
Corrections will be fully described and applied in the 
\Planck\ 2014 cosmology release, as well as the
remaining systematic effects that we describe briefly here.

Two main effects have been corrected in the time-ordered data prior to
mapmaking.  The correction for the nonlinearity of the
analogue-to-digital converters was improved, and we have also
corrected the data for very long time constants that were not
previously identified \citep[see][]{planck2013-p03,planck2013-p03e}.
After these corrections, then over the
multipole range $\lmin<\ell<600$ relevant to this analysis, the main
systematic effects result from leakage of intensity into polarization
maps.  Effects arising from polarization angle and polarization
efficiency uncertainties have been shown to be second order
\citep{planck2013-p03}.  The leakage effect can be expressed as
\begin{equation}
\label{eq:leakage}
 \Delta \{Q,U\}_\nu(\hatn) = \sum_s\sum_b \gamma^{s,b}_\nu\,
 \Gamma_\nu^{b,I\rightarrow \{Q,U\}}(\hatn)\,I_\nu^s(\hatn),
\end{equation}
where $\nu$ is the frequency, and $\Gamma_\nu^{b,I\rightarrow
\{Q,U\}}$, the leakage pattern for bolometer $b$, is multiplied by the
different $s$ leakage source maps $I_\nu^s$ and their associated scaling
coefficients $\gamma^{s,b}_\nu$.
The leakage patterns are fully determined by the scanning strategy.
They represent the cumulative result of all the systematic effects that lead
to a leakage of intensity to polarization.  In the \Planck\ HFI bands,
there are three main sources of intensity to polarization leakage:
(i) monopole differences between detectors not corrected by data destriping
(the intensity source term is constant over the sky);
(ii) bolometer inter-calibration mismatch (the source term is the full
intensity map, including the CMB dipole and the Galactic emission);
and (iii) a dust spectral mismatch term. The spectral bandpass varies from one
bolometer to another for a given band. As the bolometer gain is calibrated on
the CMB dipole, the differential gains on the dust emission produce the
bandpass mismatch term (for which the source term is the dust intensity map).

Results from a global fit of the $Q$ and $U$ maps with these three
leakage terms $\Gamma_\nu^{b,I\rightarrow \{Q,U\}}I_\nu^s$ are used to
quantify the leakage.  This fit yields estimates of the scaling
coefficients at each frequency $\nu$ in Eq.~(\ref{eq:leakage}), which
allow us to compute angular power spectra of the leakage terms in
Sect.~\ref{intlat_spectra}.  We point out that
the fit captures any emission in the maps that has a pattern on the
sky similar to one of the leakage patterns. These global fit maps are used
to assess the level of systematics, but are not removed from the data.

An independent and complementary estimate of
systematic effects in the data is provided by the null tests that we can build 
from the DetSets, HalfRings, and Years data subsets (see
Sect.~\ref{data_description}). These null tests are a good way of determining
the level of any systematics other than intensity to polarization leakage.
These are pursued in Sect.~\ref{spectra_description} and
Appendices~\ref{appendix_highlat_spectra} and \ref{bicep2_syste}.

%%%%%%%%%%%%%%%%%%%%%%%%%%%%%%%%%%%%%%%%%%%%%%%%%%%%%%%%%%%%%%%%%%%%%
\section{Computation of angular power spectra of polarized dust emission}
\label{part:computation}

%%%%%%%%%%%%%%%%%%%%%%%%%%%%%%%%%%%%%%%%%%
\subsection{Methods}
\label{method}

We can use the \Planck\ data to
compute the polarization angular power spectra (\clee\ and \clbb) of the
polarized dust emission within selected sky {\window}s. Even if the
statistical properties of the dust emission on the sky might not be
entirely captured by a 2-point function estimator, because the scope of this
paper is to assess the level of dust in the framework of CMB data analysis,
we follow the approximation that is generally made when processing such
data, i.e., that a large fraction of the information is contained in
power spectra.

On an incomplete
sky, a polarization field can be divided into three classes of modes:
pure $E$-modes; pure $B$-modes; and ``ambiguous'' modes, which are a
mixture of the true $E$- and $B$-modes \citep{Bunn}.

The ambiguous modes represent a cross-talk between $E$ and $B$, which is
often referred to as ``$E$-to-$B$ leakage'' for the CMB (because for the CMB
$C_\ell^{EE}\gg C_\ell^{BB}$).  Methods used to estimate the CMB angular
power spectrum for polarization account and correct analytically for
the incomplete sky coverage.  However, the presence of the ambiguous modes
yields a biased estimate of the variance of the spectra, unless
so-called ``pure'' power spectrum estimators are used \citep{Pure}.  For dust
polarization, the power in $E$- and $B$-modes are comparable, and we
do not expect a significant variance bias.

The two specific approaches we use are \Xpol, our main method, which
we describe as a ``classical'' pseudo-$C_\ell$
estimator\footnote{Pseudo-$C_\ell$ estimators compute an estimate of
the angular power spectra directly from the data (this being denoted the
``pseudo-power spectrum'') and then correct for sky coverage, beam smoothing,
data filtering, etc.}
and, for comparison, \Xpure, a pure pseudo-$C_\ell$ estimator.
Since (as we demonstrate below) they give similar results for the present
study, the former is chosen to be our main method because it is less
computationally expensive.

%%%%%%%%%%%%%%%%%%%%%%%%%%%%%%%%%%%%%%%%%%
\subsubsection{\Xpol}
\label{xpol}

\Xpol\ is an extension to polarization of the \Xspect\ method
\citep{XSPECT}.  \Xspect\ computes the pseudo power spectra and
corrects them for incomplete sky coverage, filtering effects, and pixel
and beam window functions.  Correction for incomplete sky coverage is
performed using a {\tt Master}-like algorithm \citep{MASTER},
consisting of the inversion of the mode-mode coupling matrix
$M_{\ell\ell'}$ that describes the effect of the partial sky coverage;
$M_{\ell\ell'}$ is computed directly from the power spectrum
of the mask that selects the data in the analysis region of
interest (Sect.~\ref{masks_definitions}).  The {\tt HEALPix} pixel
window functions \citep{gorski2005} are used to correct for pixelization
effects.  For the beams we use the \Planck\ HFI individual detector
beam transfer functions described in \cite{planck2013-p03c}.

\Xpol\ estimates the error bars analytically, without requiring Monte
Carlo simulations.  Using simulated data comprising
inhomogeneous white noise and a Gaussian map with a dust power spectrum,
we have checked that this analytical estimate is not biased for multipoles
$\ell>40$.  The analytical error
bars combine the contributions from instrumental noise and sample variance.  
The Gaussian approximation of the sample variance is
\begin{equation}
{\rm var}\left(C^{XX}_{\ell_{\rm bin}}\right)=\frac{2}{(2\ell_{\rm bin}+1)
 f_{\rm sky}\Delta\ell_{\rm bin}}\left(C^{XX}_{\ell_{\rm bin}}\right)^2,
\end{equation}
where $X=\{E,B\}$,  $f_{\rm sky}$ is the retained sky fraction (which can
be $f^{\rm eff}_{\rm sky}$ if the sky \sten\ is apodized), and
$\Delta\ell_{\rm bin}$ is the size of the multipole bin $\ell_{\rm bin}$.
To estimate the error relevant to the polarized dust signal measured within a
given region, we subtract quadratically this estimate of the contribution from
sample variance.
The performance of \Xpol\ on Gaussian simulations that have $EE$ and $BB$
dust-like angular power spectra is presented in
Appendix~\ref{methods_performance}.

%%%%%%%%%%%%%%%%%%%%%%%%%%%%%%%%%%%%%%%%%%
\subsubsection{\Xpure}
\label{xpure}

\Xpure\ is a numerical implementation of the pure
pseudo-spectral approach described and validated in \cite{Xpure}.  The
method is optimized for computing CMB $B$-mode power spectra over
small sky \patcs.  It uses a suitably chosen sky apodization that
vanishes (along with its first derivative) at the edges of the \patc, in
order to minimize the effects of $E$-to-$B$ leakage.  For the
estimation of the angular power spectra of the \Planck\ data we
compute the cross-correlation of two different DetSets.
Uncertainties are obtained by performing Monte Carlo inhomogeneous
white noise simulations, considering the diagonal terms of the
pixel-pixel covariance for the two data sets.

We have applied this algorithm to \Planck\ 353\,GHz maps and
simulations to estimate the $B$-mode power spectrum of the
Galactic thermal dust emission in regions ranging from 1\,\% to
30\,\% of the sky.  We have used \Xpure\ here as a cross-check for the
robustness of the \Xpol\ method presented in the previous section.
Validation and comparison of the performance of the two algorithms on
simulations is presented in Appendix~\ref{methods_performance} and application
to the data in Appendix~\ref{bicep2_xpol-xpure}.

%%%%%%%%%%% table properties of the selected large regions %%%%%%%%%%
\begin{table*}[htbp]
\begingroup
\newdimen\tblskip \tblskip=5pt
\caption{Properties of the large retained (\lwdpref) science {\window}s
described in Sect.~\ref{intlat_masks}.  For each \window, 
\fsky is the initial sky fraction,
$f_{\rm sky}^{\rm eff}$ its value after point source masking and apodization, 
\idust\ the mean specific intensity at 353\,GHz within the \window, in
${\rm MJy}\,{\rm sr}^{-1}$, 
and \nhi\ the mean \ion{H}{i} column density, in units of
$10^{20}\,{\rm cm}^{-2}$ \citep{LABHI}.  
For the power-law fits in multipole $\ell$, we also list 
the \alphaexponents\ $\alpha_{EE}$ and $\alpha_{BB}$ (Sect.~\ref{powerlawfit}), 
the $\chi^2$ of the fits with fixed \alphaexponents\
$\alpha_{EE}=\alpha_{BB}=\alphaps$, the value $A^{EE}$ of the fitted \dlee\
amplitude at $\ell=80$ (in $\mu$K$_{\rm CMB}^2$ at 353\,GHz,
Sect.~\ref{nhi-dependence}),
and the mean of the amplitude ratio $\big\langle A^{BB}/A^{EE}\big\rangle$
(see Sect.~\ref{EonB}).
}
\label{table_masks}
%\vskip -3mm
\footnotesize
\setbox\tablebox=\vbox{
 \newdimen\digitwidth
 \setbox0=\hbox{\rm 0}
 \digitwidth=\wd0
 \catcode`*=\active
 \def*{\kern\digitwidth}
 \newdimen\signwidth
 \setbox0=\hbox{+}
 \signwidth=\wd0
 \catcode`!=\active
 \def!{\kern\signwidth}
 \halign{\tabskip=0pt\hbox to 1.75in{#\leaderfil}\tabskip=1em&
% \halign{\tabskip=0pt\hbox to 1.75in{#\hfil}\tabskip=1em&
 \hfil#\hfil\tabskip=1em&
 \hfil#\hfil\tabskip=1em&
 \hfil#\hfil\tabskip=1em&
 \hfil#\hfil\tabskip=1em&
 \hfil#\hfil\tabskip=1em&
 \hfil#\hfil\tabskip=0pt\cr
\noalign{\doubleline}
 \omit& {\lwdpref}24& {\lwdpref}33& {\lwdpref}42& {\lwdpref}53& {\lwdpref}63&
 {\lwdpref}72\cr
\noalign{\vskip 4pt\hrule\vskip 6pt}
\fsky& 0.3& 0.4& 0.5& 0.6& 0.7& 0.8\cr
\fskyeff& 0.24& 0.33& 0.42& 0.53& 0.63& 0.72\cr
$\langle I_{353}\rangle/{\rm MJy}\,{\rm sr}^{-1}$& 0.068& 0.085& 0.106&
 0.133& 0.167& 0.227\cr
$N_{\rm \ion{H}{i}}/10^{20}\,{\rm cm}^{-2}$&   1.65& 2.12& 2.69& 3.45&
 4.41& 6.05\cr
\noalign{\vskip 4pt\hrule\vskip 6pt}
$\alpha_{EE}$& $-2.40\pm0.09$& $-2.38\pm0.07$& $-2.34\pm0.04$&
 $-2.36\pm0.03$& $-2.42\pm0.02$& $-2.43\pm0.02$\cr
$\alpha_{BB}$& $-2.29\pm0.15$& $-2.37\pm0.12$& $-2.46\pm0.07$&
 $-2.43\pm0.05$& $-2.44\pm0.03$& $-2.46\pm0.02$\cr
\noalign{\vskip 4pt\hrule\vskip 6pt}
$\chi_{EE}^2$ ($\alpha_{EE}=\alphaps$, $N_{\rm dof}=21$)& $26.3$& $28.1$&
 $31.8$& $38.3$& $32.7$& $44.8$\cr
\noalign{\vskip 2pt}
$\chi_{BB}^2$\, ($\alpha_{BB}=\alphaps,$ $N_{\rm dof}=21$)& $18.9$& $14.0$&
 $21.1$& $22.1$& $15.4$& $21.9$\cr
\noalign{\vskip 4pt\hrule\vskip 6pt}
$A^{EE}\ (\ell = 80)$& $37.5\pm1.6$&  $51.0\pm1.6$& $78.6\pm1.7$&
 $124.2\pm1.9$& $197.1\pm2.3$& $328.0\pm2.8$\cr
\noalign{\vskip 4pt\hrule\vskip 6pt}
$\big\langle A^{BB}/A^{EE}\big\rangle$& $0.49\pm0.04$& $0.48\pm0.03$&
 $0.53\pm0.02$& $0.54\pm0.02$& $0.53\pm0.01$& $0.53\pm0.01$\cr
\noalign{\vskip 4pt\hrule\vskip 6pt}
}}
\endPlancktablewide
\endgroup
\end{table*}

%%%%%%%%%%%%%%%%%%%%%%%%%%%%%%%%%%%%%%%%%%
\subsection{Computing cross-spectra}
\label{method_data_sets}

To avoid a bias arising from the noise, we compute all the \Planck\
power spectra from cross-correlations of DetSets maps (see
Sect.~\ref{part:data}).  The noise independence of the two DetSet maps at
a given frequency was quantified in \cite{planck2013-p08} and the
resulting level of noise bias in the cross-power spectra between them
has been shown to be negligible.  The cross-power spectrum at a given
frequency $\nu$ is
\begin{equation}
C_\ell(\nu\times\nu) \equiv C_\ell(D^1_\nu\times D^2_\nu),
\end{equation}
where $D^1_\nu$ and $D^2_\nu$ are the two independent DetSet~1 and
DetSet~2 maps at the frequency $\nu$.  The \Planck\ cross-band
spectrum between the frequencies $\nu$ and $\nu'$ is
\begin{eqnarray}
\qquad\quad C_\ell(\nu\times \nu') &\equiv&
 \frac{1}{4}\left[C_\ell(D^1_\nu\times D^1_{\nu'})
 + C_\ell(D^2_\nu\times D^2_{\nu'})\right. \nonumber\\
& &\quad \left. + C_\ell(D^1_\nu\times D^2_{\nu'})
 + C_\ell(D^2_\nu\times D^1_{\nu'})\right],
\end{eqnarray}
or equivalently the cross-spectrum between $\nu$ and $\nu'$ of the averaged
frequency maps $(D^1_\nu\ + D^2_{\nu})/2$.

We recall that from each computed \clee\ spectrum we subtract the \Planck\
best-fit $\Lambda$CDM \clee\ model \citep{planck2013-p11}, i.e., the
theoretical \clee\ model obtained from the temperature data fit,
while \clbb\ spectra are kept unaltered.

%%%%%%%%%%% figure of the selected large regions

\begin{figure}
\centering
\includegraphics[height=0.3\textwidth]{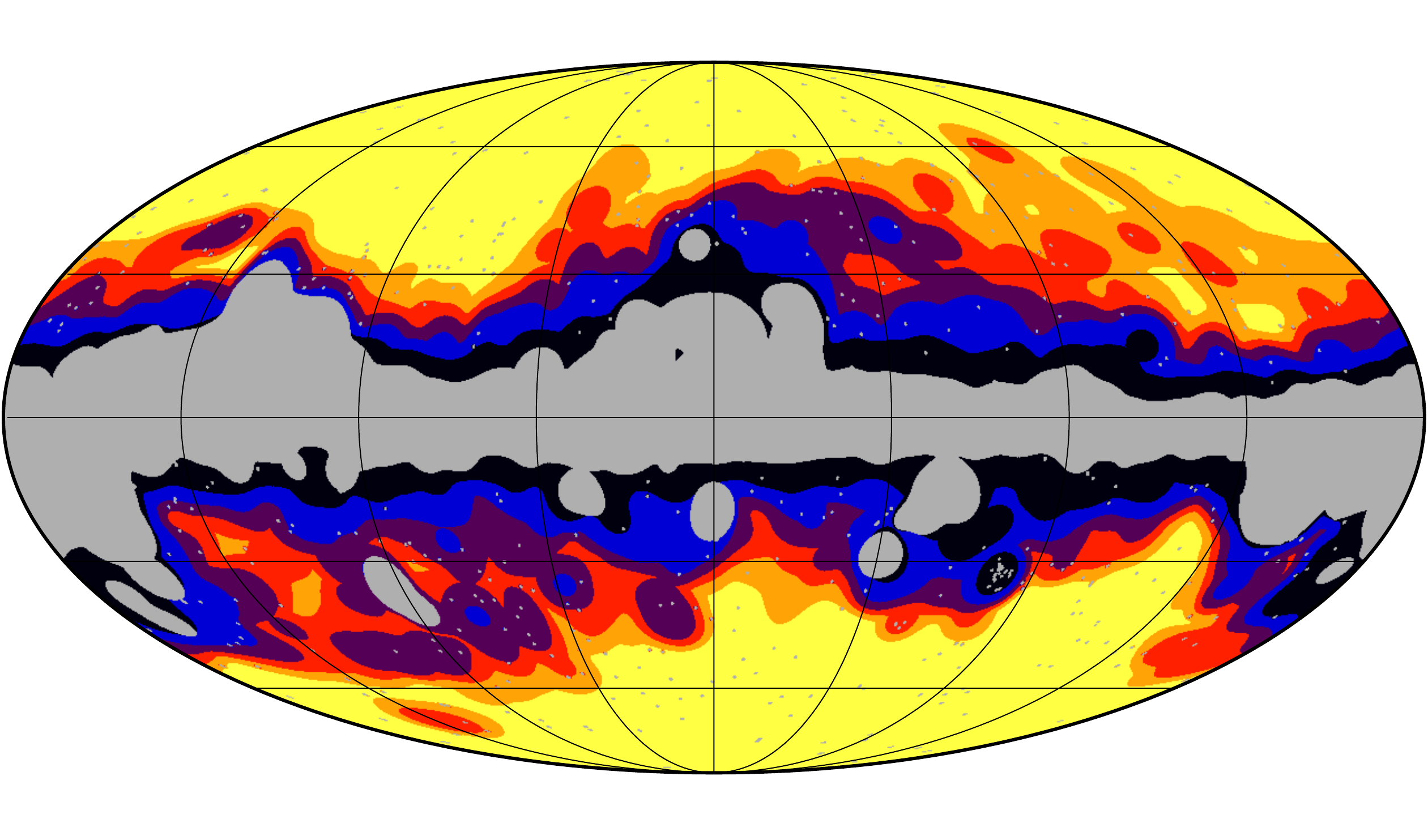}
\caption{
Masks and complementary selected large {\window}s that
retain fractional coverage of the sky \fsky from 0.8 to 0.3
(see details in Sect.~\ref{intlat_masks}).
The gray is the CO mask, whose complement is a selected \window\ with
$\fsky = 0.8$.
In increments of $f_{\rm sky}=0.1$, the retained {\window}s can be identified 
by the colours yellow (0.3) to black (0.8), inclusively.
Also shown is the (unapodized) point source mask used.
\label{fig:masks}
}
\end{figure}

%%%%%%%%%%%%%%%%%%%%%%%%%%%%%%%%%%%%%%%%%%
\subsection{Selection of {\window}s}
\label{masks_definitions}

To measure the dust polarization power spectra with high S/N,
we select six large {\window}s,
the analysis regions of interest at intermediate Galactic latitude, which
have effective coverage of the sky
from 24 to 72\,\% (see Sect.~\ref{intlat_masks}).\footnote{Although the
selection process is similar to that in \cite{planck2013-p08},
there are differences in detail.}
For statistical studies at high Galactic latitude, we compute spectra on a
complete set of smaller {\window}s or \patcs\ (Sect.~\ref{highlat_masks}),
similar in size to the \patcs\ observed in typical CMB experiments.

%=======================================
\subsubsection{Large {\window}s} 
\label{intlat_masks}

For selection of all of the large {\window}s, we used the \Planck\
CO map from \cite{planck2013-p03a}, smoothed to a $5^\circ$ resolution, to
mask the sky wherever the CO line brightness
$I_{\rm CO}\geq0.4\,$\kkms.\footnote{We use the CO
($J\,{=}\,1\,{\rightarrow}\,0$) ``Type~3'' map, which has the
highest signal-to-noise ratio. At this resolution and for this map,
the cut we apply corresponds to $S/N>8$.}
This mask is shown in Fig.~\ref{fig:masks}.
The complement to this mask by itself defines a preliminary \window\ that
retains a sky fraction $\fsky=0.8$. 

We then mask the sky above successively lower thresholds of $I_{857}$ in the
\Planck\ 857\,GHz intensity map, smoothed to a $5^\circ$ resolution, chosen
such that together with the CO mask we select five more preliminary {\window}s
that retain \fsky from 0.7 to 0.3 in steps of 0.1.
These six {\window}s are displayed in Fig.~\ref{fig:masks}. 

To avoid power leakage, these six masks are then apodized by convolving with
a $5^\circ$
FWHM Gaussian which alters the window function by gradually reducing the signal
towards the edges of the retained {\window}s and thus lowers the
effective retained sky coverage.  The \fskyeff value is simply defined as
the mean sky coverage of the window function map.

Finally, we mask data within a radius $2\,\sigma_{\rm beam}$ of point
sources selected from the \Planck\ Catalogue of Compact Sources
\citep[PCCS,][]{planck2013-p05} at 353\,GHz.  
\hpeter{New footnote.}
The selected sources have
${\rm S/N}>7$ and a flux density above 400\,mJy.\footnote{This included
the brightest point sources in the Large and Small Magellanic Clouds (LMC and
SMC).  We tested that masking the entirety of the LMC and SMC has no
significant effect on the spectra or on the conclusions that we derive from
them.}
Selection of spurious
infra-red sources in bright dust-emitting regions is avoided by using
contamination indicators of infrared cirrus listed in the PCCS description
\citep{planck2013-p05}. This point-source masking is done in order to prevent
the brightest polarized 
%radio 
sources from producing ringing in the power
spectrum estimation, while avoiding the removal of dust emitting regions and
their
statistical contribution to the angular power spectra. The details of this
source selection will be presented in the \Planck\ 2014 release papers.
The edges of the masks around point sources were apodized with a 30\arcmin\
FWHM Gaussian, further reducing the retained net effective sky coverage.

In combination these masking and apodization procedures result in six large
retained (\lwdpref) {\window}s, which we distinguish hereafter using the
percentage of the sky retained (the net effective fractional sky coverages,
\fskyeff, are listed in Table~\ref{table_masks}), e.g., 
{\lwdpref}72 for the largest \window\ and {\lwdpref}24 for the smallest.

Table~\ref{table_masks} also lists other properties of the {\window}s,
including \idust, the mean specific intensity at 353\,GHz within the \window\
in ${\rm MJy}\,{\rm sr}^{-1}$, and \nhi, the mean \ion{H}{i} column density
in units of $10^{20}\,{\rm cm}^{-2}$
computed on the LAB \ion{H}{i} survey data cube \citep{LABHI}.

%=======================================
\subsubsection{Small \patcs\ at high Galactic latitude}
\label{highlat_masks}

To examine the statistics of the angular power spectra at 
high Galactic latitude, we also analyse the \Planck\ polarization maps
within \patcs\ with a size similar to those of typical ground-based and
balloon-borne CMB experiments
\citep[e.g.,][]{QUIET,SPTPol,BICEP2_PRL,POLARBEAR_PRL,ACTPol}.
Specifically, we consider 400\,deg$^2$ circular areas (radius $11\pdeg3$)
centred on the central pixel positions of the {\tt HEALPix} $N_{\rm side}=8$
grid that have Galactic latitude $|b|>35^\circ$.  This results in 352 such
\patcs.  These are apodized with a $2^\circ$ FWHM Gaussian, which reduces
the retained sky fraction to $\fskyeff=0.0080$ for each \patc.
For these small \patcs, we do not mask point sources as was done in selecting
the large {\window}s, since we want to preserve the same \fskyeff for each
mask,
nor is the CO mask needed, since we use these masks only on the 353\,GHz high
Galactic latitude data.
Note that for $N_{\rm side}=8$, pixels have an area of about 54\,deg$^2$ and
characteristic centre-to-centre spacing of about $7\pdeg4$.
Therefore, this {\tt HEALPix} grid oversamples the sky relative to this \patc\
size, so that the \patcs\ overlap and are not independent.

%%%%%%%%%%%%%%%%%%%%%%%%%%%%%%%%%%%%%%%%%%%%%%%%%%%%%%%%%%%%
%%%%%%%%%%%%%%%%%%%%%%%%%%%%%%%%%%%%%%%%%%%%%%%%%%%%%%%%%%%%
%%%%%%%%%%%%%%%%%%%%%%%%%%%%%%%%%%%%%%%%%%%%%%%%%%%%%%%%%%%%

\section{Dust polarized angular power spectra at intermediate Galactic latitude}
\label{intlat_spectra}

In this section, we quantify the 353\,GHz dust polarization in the power
spectrum domain, achieving a high S/N through the use of \lwdpref\ {\window}s.
For convenience we present results for ${\cal D}_\ell^{EE}$ and
${\cal D}_\ell^{BB}$, where ${\cal D_\ell}\equiv\ell(\ell+1)C_\ell/(2\pi)$.

%%%%%%%%%%%%%%%%%%%%%%%%%%%%%%%%%%%%%%%%%%%%%%
\subsection{Description of the spectra}
\label{spectra_description}

Using \Xpol\ we have computed \dlee\ and \dlbb\ from the two DetSets at
353\,GHz, as a function of
the multipole $\ell$ in the range $\lmin$--600.\footnote{The spectra are
in units of $\mu$K$_{\rm CMB}^2$ at 353\,GHz.}
These represent the
first measurements of the thermal dust \dlee\ and \dlbb\ power spectra
on large fractions of the sky for $\ell>\lmin$
\citep[see][for earlier related studies]{ArcheopsDustPS,WMAP7Galactic}.

\begin{figure*}%[!h]
\centering
\includegraphics[height=0.6\textwidth]{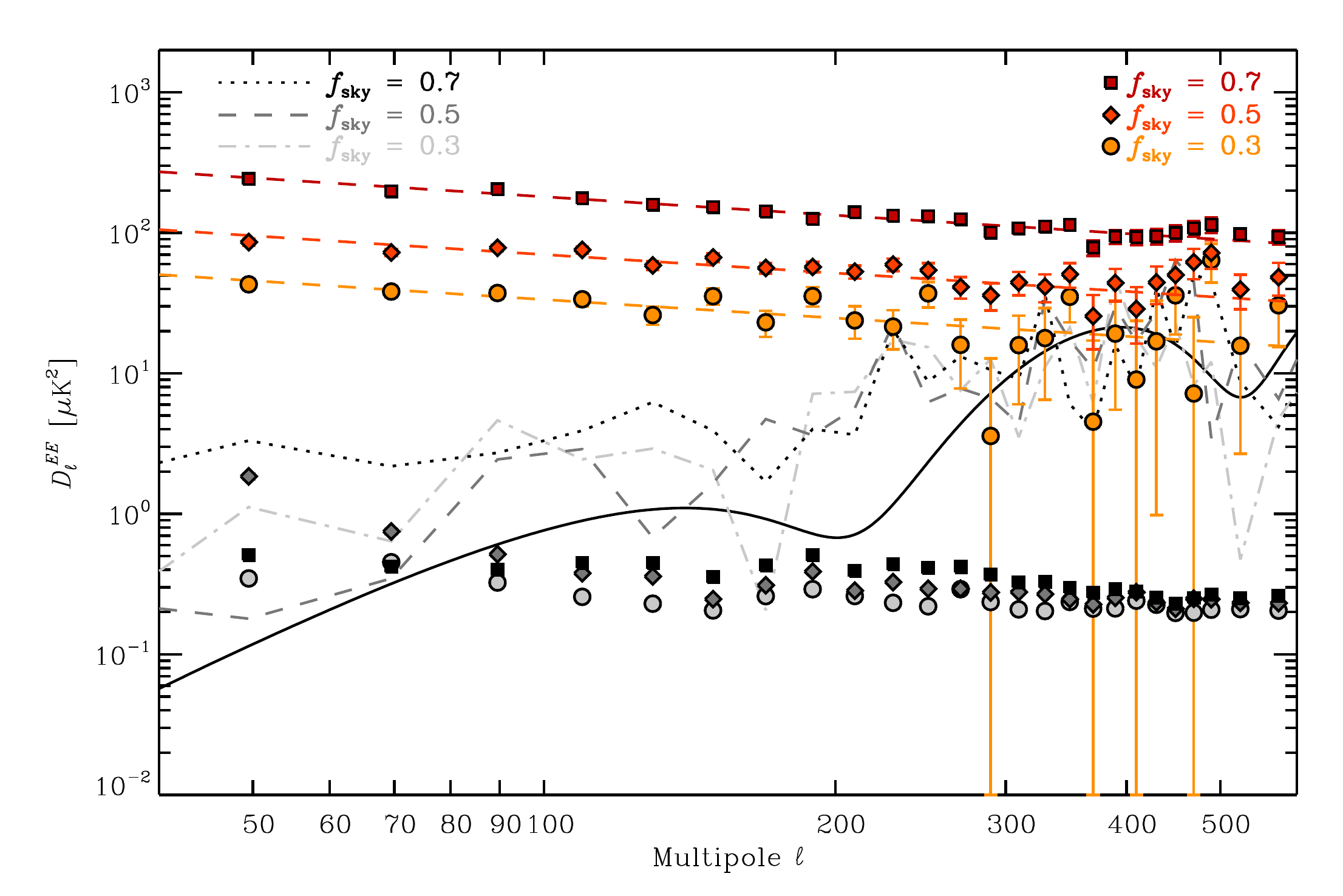}
\includegraphics[height=0.6\textwidth]{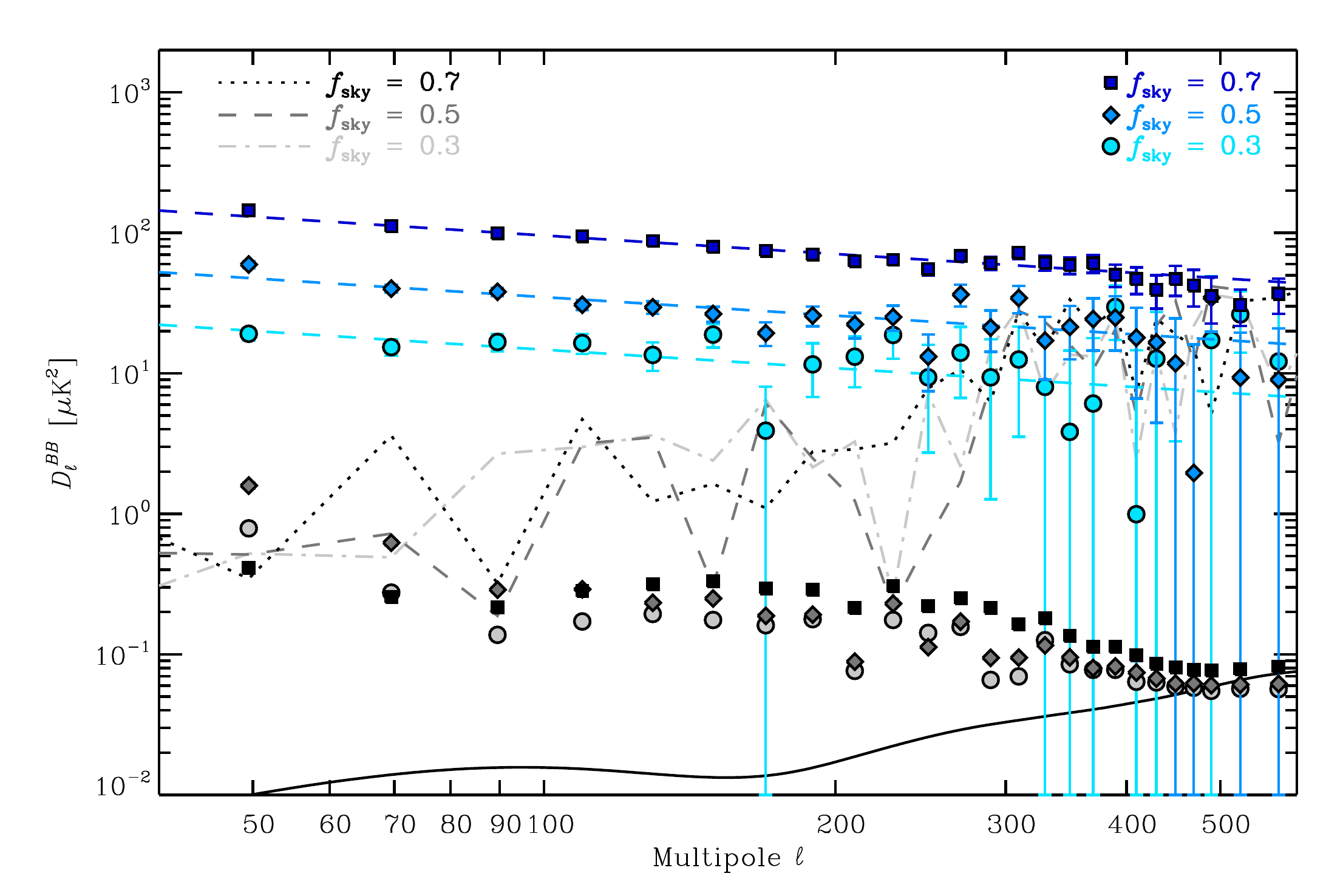}
\caption{
\footnotesize \Planck\ HFI 353\,GHz \dlee\
(red, top) and \dlbb\ (blue, bottom) power spectra (in $\mu$K$_{\rm CMB}^2$)
computed on three of the
selected \lwdpref\ analysis regions that have $\fsky=0.3$ (circles, lightest),
$\fsky=0.5$ (diamonds, medium) and $\fsky=0.7$ (squares, darkest).
The uncertainties shown are $\pm 1\,\sigma$.
The best-fit power laws in $\ell$ are displayed for each spectrum as a
dashed line of the corresponding colour. The \Planck\ 2013 best-fit
$\Lambda$CDM \dlee\ expectation \citep{planck2013-p11} and the
corresponding $r=0.2$ \dlbb\ CMB model are displayed as solid black lines; 
the rise for $\ell >200$ is from the lensing contribution.
In the lower parts of each panel, the global estimates of the power spectra of
the systematic effects responsible for intensity-to-polarization
leakage (Sect.~\ref{systematics}) are displayed
in different shades of grey, with the same symbols to identify the three
{\window}s. 
Finally, absolute values of
the null-test spectra anticipated in Sect.~\ref{systematics},
computed here from the cross-spectra of the HalfRing/DetSet differences
(see text), are represented as dashed-dotted, dashed, and dotted grey lines
for the three \lwdpref\ {\window}s.
\label{fig:eebb_spectra}
}
\end{figure*}

In Fig.~\ref{fig:eebb_spectra} we present the results for
$\fsky=\{0.3,0.5,0.7\}$.  The amplitudes of the spectra increase
with increasing \fsky because the polarized emission is brighter
on average when more sky is retained.  We point out that the
\lwdpref\ {\window}s presented in Sect.~\ref{intlat_masks} overlap,
since they form a nested set.  However, the power spectra derived from them
are almost independent of each other because each spectrum is dominated
by the brightest areas in the corresponding \window, namely the parts
closest to the Galactic plane and hence the areas in which each \lwdpref\
\window\ differs from those nested inside it.

The \dlee\ and \dlbb\ spectra are characterized by a power-law dependence on
multipole $\ell$ over the range $\ell=\lmin$ to $600$; furthermore, the slope
is similar for different {\window}s that retain from 24 to 63\,\% of the sky. 

Fig.~\ref{fig:eebb_spectra} also shows the \dlee\ power spectrum computed
from the \Planck\ 2013 best-fit
$\Lambda$CDM model of the CMB temperature data \citep{planck2013-p11},
and the much lower expectation for \dlbb\ from the CMB model with primordial
gravitational waves with amplitude
$r=0.2$.  At 353\,GHz, the \dlee\ angular power spectra
of the dust are about 3--4 orders of magnitude larger than the CMB model at
$\ell=30$, 1--2 orders of magnitude larger at $\ell=100$,
and about the same order of magnitude as the CMB  at $\ell>300$. 
At $353\,$GHz, the \dlbb\ angular power spectra for dust are much greater than
the CMB model power spectrum for all $\ell$ values in
Fig.~\ref{fig:eebb_spectra}.
The dust power spectra are larger than the $r=0.2$
CMB spectrum by 4--5 orders of magnitude at $\ell=30$, and by 3--4 orders
of magnitude at $\ell=100$.  At $\ell=500$, where the lensing of CMB
anisotropies is the dominant contribution to the CMB model spectrum,
the dust is still 2--3 orders of magnitude higher.

%
%%%%%%%%%%%%% EE, BB power spectra %%%%%%%%%%%%%%%%%%%%

As discussed in Appendix~\ref{methods_performance}, 
we do not expect any significant bias, or $E$-to-$B$ leakage, from
the computation of the dust angular power spectra using \Xpol.
Fig.~\ref{fig:eebb_spectra} also includes 
the \dlee\ and \dlbb\ spectra at 353\,GHz,
computed from our estimate of the leakage terms from intensity to polarization
(discussed in Sect.~\ref{systematics}),
for the same three \lwdpref\ {\window}s.
The dust spectra are much higher than the corresponding spectra for the
leakage, which represent the main systematic effects over the $\ell$ range of
interest for this work.  The largest contamination of the dust
signal by leakage is about 3.5\,\% at $\ell=50$,
for both the  \dlee\ and \dlbb\ spectra.  
Because we consider that our estimate of the leakage maps is conservative, 
we conclude that contamination of the dust \dlee\ and \dlbb\ power spectra
by systematic effects amounts to a maximum of 4\,\% at
$\ell=50$, and is less at higher multipoles.
Therefore, we have not corrected the \Planck\ data for
intensity-to-polarization leakage in this work.

Finally, in Fig.~\ref{fig:eebb_spectra} we 
present for the three \lwdpref\ {\window}s the absolute value of the 
null-test spectra anticipated in Sect.~\ref{systematics}, here
computed from the cross-spectra of the HalfRing/DetSet differences,
i.e., $(353_{\rm DS1,HR1}-353_{\rm DS1,HR2})/2
 \times(353_{\rm DS2,HR1}-353_{\rm DS2,HR2})/2$.
These \dlee\ and \dlbb\ spectra show a behaviour that is
close to what is expected from a white-noise dominated (thus $\ell^2$)
spectrum. The amplitudes of these error
estimates are consistent with the noise expectations; there is no evidence
for any effects of systematics.

For completeness, in Appendix~\ref{te-tb-eb} we present a further
quantification of the power spectrum of thermal dust emission at 353\,GHz via
the spectra involving temperature and polarization, \dlte\ and \dltb, and the
polarization cross-spectrum, \dleb.

\begin{figure}[t]
\centering
\includegraphics[height=0.33\textwidth]{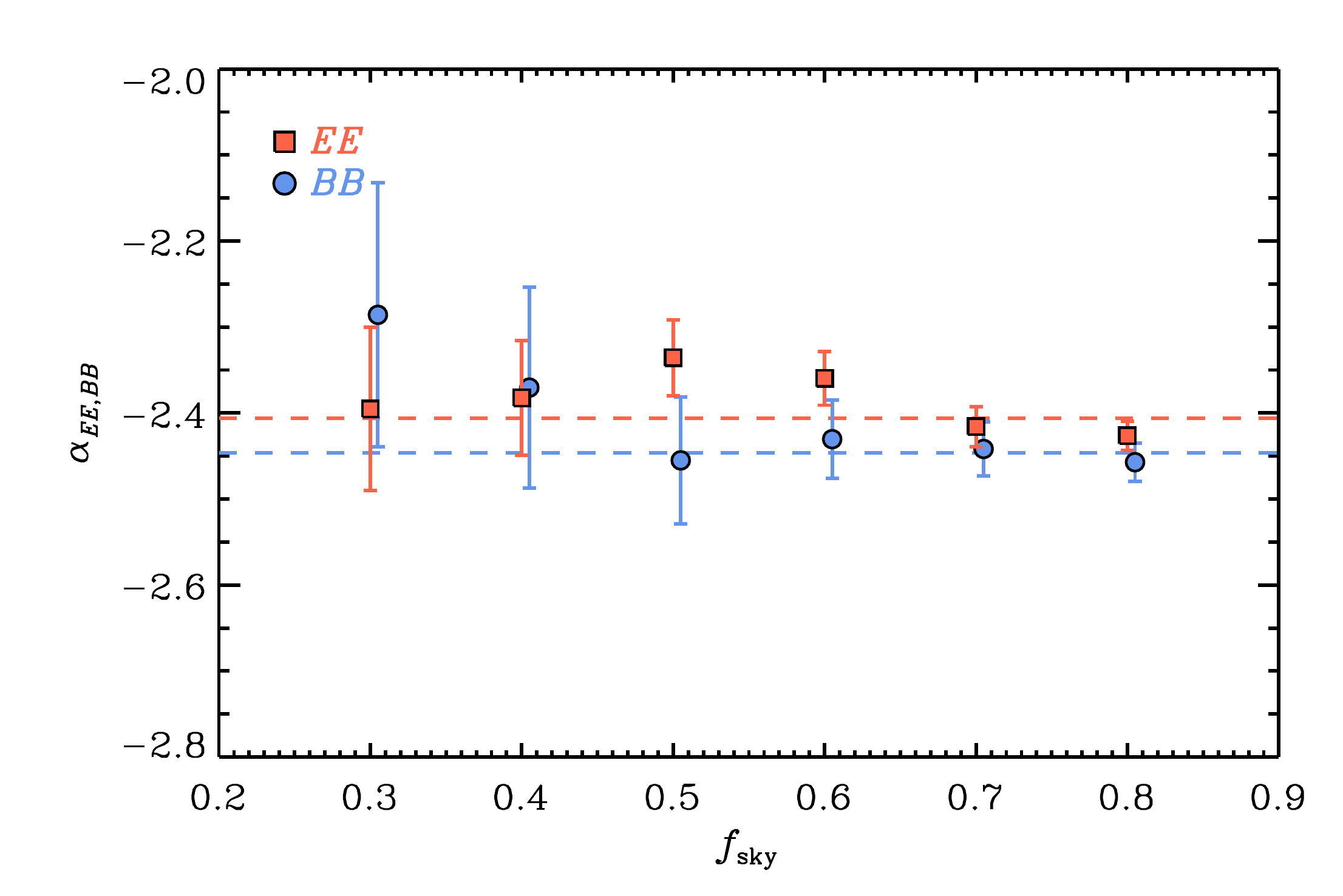}
\caption{
Best-fit power-law \alphaexponents\ $\alpha_{EE}$
(red squares) and $\alpha_{BB}$ (blue circles) fitted to the
353\,GHz dust \dlee and \dlbb\ power spectra for the different
\lwdpref\ {\window}s defined in
Sect.~\ref{intlat_masks}, distinguished here with $f_{\rm sky}$. 
Although the values in the {\window}s are not quite independent, simple means
have been calculated and are represented as red and blue dashed lines.
\label{fig:eebb_indices}
}
\end{figure}

%%%%%%%%%%%%%%%%%%%%%%%%%%%%%%%%%%%%%%%%%%%%%%
\subsection{Power-law fit}
\label{powerlawfit}

To assess the apparent power-law dependence
$C_\ell^{XX} \propto \ell^{\,\alpha_{XX}}$ quantitatively, we made a
$\chi^2$ fit to the spectra at 353\,GHz using the form
${\cal D}_\ell^{XX}=A^{XX}(\ell/80)^{\,\alpha_{XX}+2}$, where $\ X\in\{E,B\}$. 
For \dlee\ and \dlbb\ we fit the 22 band-powers in the range
$\lminfit<\ell<500$.
For both power spectra we restricted the fit to $\ell > \lminfit$ to avoid a
possible bias from systematic effects in the data,
and also because the angular power spectra of the dust polarization exhibit
more spatial variation, particularly on large angular scales, than
is expected for a Gaussian random field.

The \alphaexponents\ of the power-law fits, $\alpha_{XX}$,
are plotted in
Fig.~\ref{fig:eebb_indices} for each of the six \lwdpref\ {\window}s
identified with $\fsky$.  All \alphaexponents\ are
consistent with constant values of $\alpha_{EE}=-2.41\pm0.02$ and
of $\alpha_{BB}=-2.45\pm0.03$.
While there is a slight indication of a steeper slope for \dlbb\ than for
\dlee, hereafter we adopt the mean \alphaexponent\ $\alphaps\pm\alphapsuncer$.
This \alphaexponent\ is
consistent with the value $\alpha_{TT}$ fitted to the \Planck\ 353\,GHz dust
intensity power spectra in this range of $\ell$
\citep{planck2013-p08,planck2014-XXII} on similar-sized {\window}s at
intermediate Galactic latitude, but slightly flatter than the $\alpha_{TT}$
fitted at higher $\ell$ and higher frequency 
\citep{MAMD07,MAMD10,planck2013-p08}.

For the fits with fixed \alphaexponent\ $\alpha_{EE,BB} = \alphaps$, the
values of the $\chi^2$
(with number of degrees of freedom, $N_{\rm dof}=21$) are displayed in
Table~\ref{table_masks} for the six \lwdpref\ {\window}s.
For the \dlee\ spectra, the $\chi^2$ values range from 26.3
(probability to exceed, $\pte=0.2$) to 44.8 ($\pte=0.002$), with a trend for a
quality of fit that degrades with increasing $\fsky$. For the \dlbb\ spectra,
the $\chi^2$ values range from 14.0 ($\pte=0.87$)
to 22.1 ($\pte=0.39$), with a trend related to $\fsky$.

Possible explanations for this difference between \dlee\ and \dlbb\ spectrum
shape descriptions, are: the chance correlation between dust and CMB
polarization, which is not taken into account in the subtraction of the CMB
\dlee\ spectrum;
and the increasing S/N degrading the overall quality of the
fit when going from \dlbb\ to \dlee\ and for \dlee\ from $\fsky=0.3$ to
$\fsky=0.8$, as the amplitude of the dust polarized signal increases.
In any case, we stress that while the power law in $\ell$ is a good general
description of the shape of the dust polarized power spectra, a full
characterization would have to consider the detailed features that can be
discerned in Fig.~\ref{fig:eebb_spectra}.

\begin{figure}[t]
\centering
\includegraphics[height=0.33\textwidth]{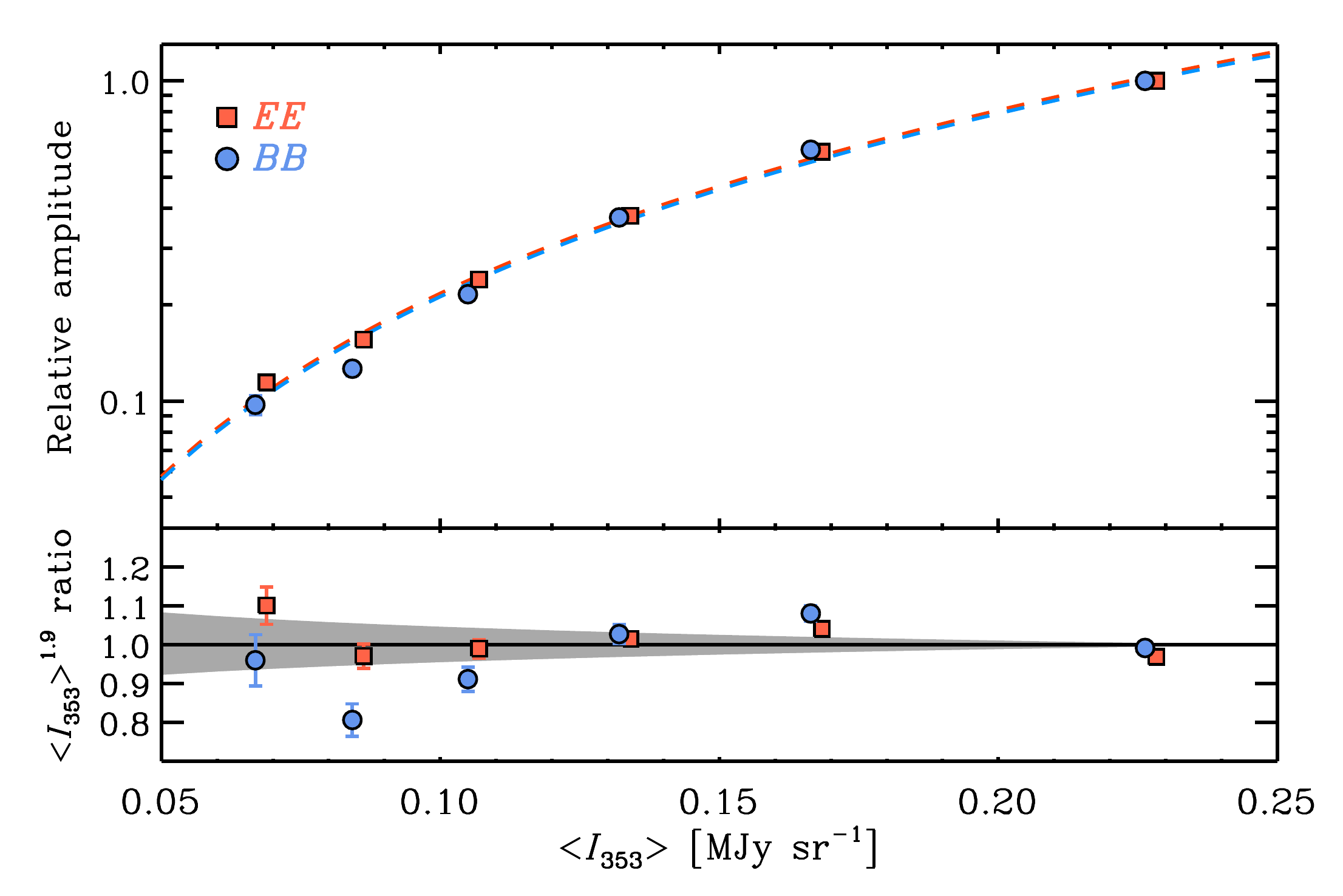}
\caption{
Amplitude of the dust $A^{EE}$ (red squares) and $A^{BB}$ (blue circles)
power spectra, normalized with respect to the largest amplitude for each mode.
These are plotted versus the mean dust intensity \idust\ for the six
\lwdpref\ {\window}s (top panel).  A power-law fit of the form 
$A^{XX}($\idust$)=K_{XX}$\idust$^{1.9}$, $X\in\{E,B\}$,
is overplotted as a dashed line of the corresponding colour (these almost
overlap). The bottom panel presents the ratio of the data and
the fitted \idust$^{1.9}$ power law; the range associated with
the $\pm1\,\sigma$ uncertainty in the power-law exponent of 1.9
is displayed in grey.  For details see Sect.~\ref{nhi-dependence}.
\label{fig:eebb_nhi}
}
\end{figure}

%%%%%%%%%%%%%%%%%%%%%%%%%%%%%%%%%%%%%%%%%%%%%%
\subsection{Amplitude dependence on \idust}
\label{nhi-dependence}

Similar to what has been done for intensity power spectra in e.g.,
\cite{Gautier} and 
\cite{MAMD07}, we investigate how the amplitude of the dust polarization
power spectrum scales with the dust intensity.
To quantify the dust emission at $353\,$GHz, we use the model map derived from
a modified blackbody fit to the \Planck\ data at $\nu \ge 353\,$GHz and
{\it IRAS\/} at $\lambda = 100\,\mu$m, presented in \citet{planck2013-p06b}. 
This map is corrected for zodiacal emission 
and the brightest extragalactic point sources are subtracted.  
The Galactic reference offsets of the underlying {\it IRAS\/} and \Planck\
data were obtained through a method based on correlation with 21-cm data from
the LAB \hi\ survey \citep{LABHI}
integrated in velocity, effectively removing the CIB monopole in the model map.
The mean dust intensity, \idust, listed for each
\lwdpref\ \window\ in Table~\ref{table_masks}, ranges from $0.068$ to
$0.227\,{\rm MJy}\,{\rm sr}^{-1}$ for increasing $\fskyeff$. The mean
column density calculated from the LAB survey data
is also listed in Table~\ref{table_masks}, with values ranging from
$1.65\times10^{20}\,{\rm cm}^{-2}$ to
$6.05\times10^{20}\,{\rm cm}^{-2}$.

For all of the \lwdpref\ {\window}s, we fit the \dlee\ and \dlbb\ spectra with 
a power law in $\ell$, using the fixed
\alphaexponent\ $\alpha_{EE,BB}=\alphaps$, over the $\ell-$ranges
defined in Sect.~\ref{powerlawfit}.  The amplitudes $A^{EE}$ derived from
these fits are listed in Table~\ref{table_masks} (the $A^{BB}$ amplitudes can
be retrieved from the $A^{BB}/A^{EE}$ ratio), and plotted as a function of
\idust\ in Fig.~\ref{fig:eebb_nhi},
after normalization by the maximum value found for the largest \window\
({\lwdpref}72).  We fit the empirical dependence of these amplitudes on
\idust\ as a power law of the form
$A^{XX}($\idust$)=K_{XX}$\idust$^{\epsilon_{XX}}$ where $X\in\{E,B\}$.
The two fitted exponents are quite
similar, $\epsilon_{EE}=1.88\pm0.02$ and $\epsilon_{BB}=1.90\pm0.02$.
The exponent that we find for polarization is close to the one observed in
the diffuse interstellar medium for the dust intensity, consistent with
$A^{TT}_\nu \propto\left\langle I_\nu\right\rangle^2$,
where $\left\langle I_\nu\right\rangle$
is the mean value of the dust specific intensity \citep{MAMD07}.
Values close to 2 are expected, because we compute angular power spectra, which
deal with squared quantities. 

Although the data points roughly follow this \idust$^{1.9}$ dependence, the
empirical law fails to fully describe individual dust amplitudes (e.g., the
estimate is off by about 20\,\% for \dlbb\ on {\lwdpref}33).  The scaling can
help to asses the order of magnitude of the dust polarization level on a
specific \window, but is not a substitute for actually characterizing the
polarized angular power spectra.

\begin{figure}[t]
\centering
\includegraphics[height=0.33\textwidth]{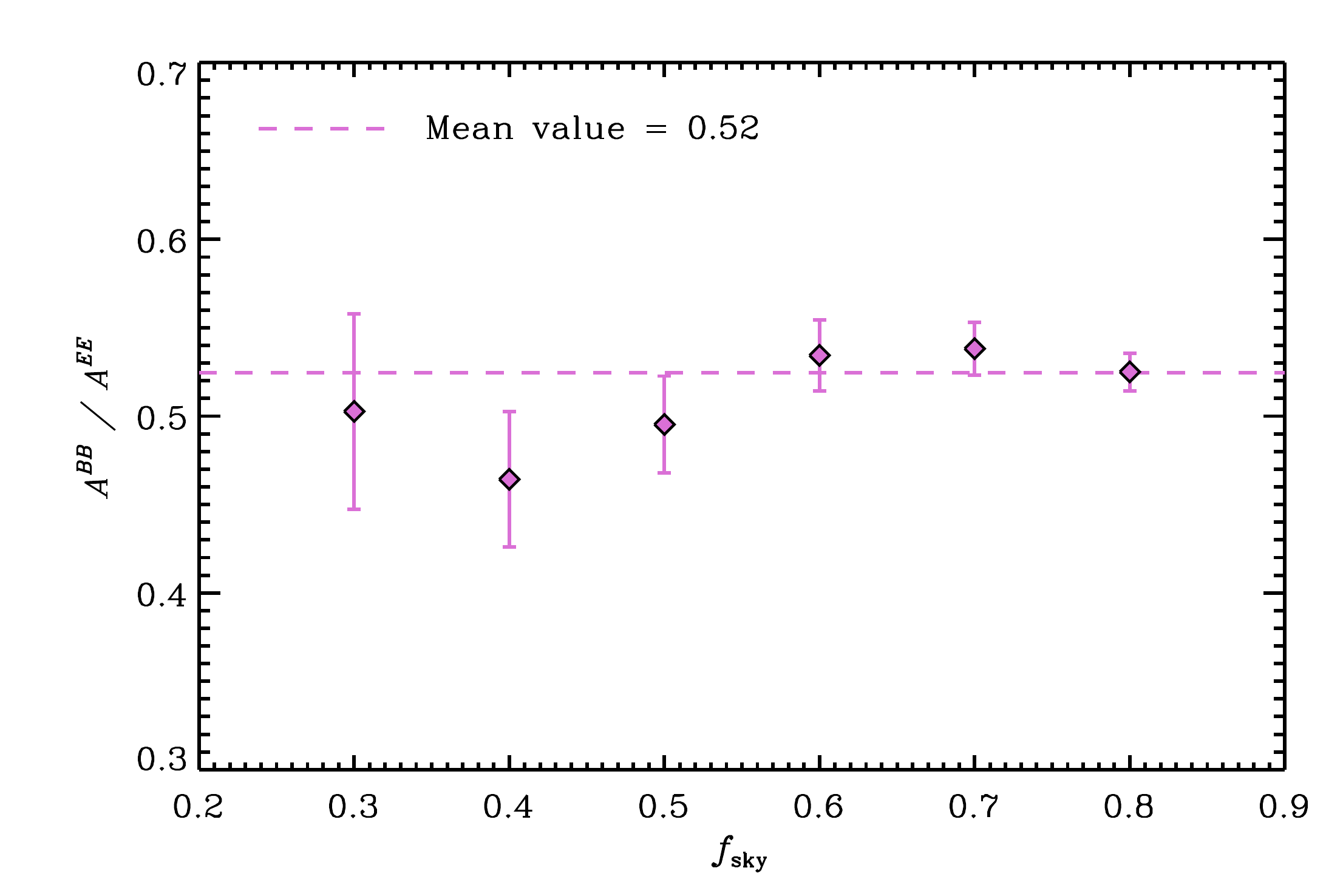}
\caption{
Ratio of the amplitudes of the \dlbb\ and \dlee\ dust
power spectra at 353\,GHz for the different \lwdpref\ {\window}s defined in
Sect.~\ref{intlat_masks}, distinguished here with $f_{\rm sky}$. 
The mean value $\big\langle A^{BB}/A^{EE}\big\rangle=0.52$
is plotted as a dashed line.
\label{fig:eebb_ratio}
}
\end{figure}

%%%%%%%%%%%%%%%%%%%%%%%%%%%%%%%%%%%%%%%%%%%%%%%
\subsection{Amplitude of \dlbb\ relative to \dlee}
\label{EonB}

We examine the ratio of the amplitudes of the fitted power laws found
in Sect.~\ref{nhi-dependence}.
The $A^{BB}/A^{EE}$ ratios are listed in Table~\ref{table_masks}, 
and plotted for different values of \fsky in Fig.~\ref{fig:eebb_ratio}. 
For all of the \lwdpref\ {\window}s, we observe more power in the \dlee\
dust spectrum than in \dlbb. 
All ratios are consistent with a value of $A^{BB}/A^{EE}=0.52\pm0.03$,
significantly different from unity, over various large fractions of the
intermediate latitude sky.

This result is not taken into account in existing models
of polarized microwave dust emission that have been developed
to test component separation methods. . 
For example, we have computed the \dlee\ and \dlbb\ spectra over the \lwdpref\ 
{\window}s for the \Planck\ Sky Model \citep{PSM}
and the model of \citet{Odea}; for both models and all \lwdpref\ {\window}s
we find a ratio $A^{BB}/A^{EE}$ close to 1. However, these two models
are based on a very simplified picture of the Galactic magnetic field geometry
and assumptions on how the polarized emission depends on it.
Further insight into the structure of the dust polarization sky is required
to account for the observed ratio.

%%%%%%%%%%%%%%%%%%%%%%%%%%%%%%%%%%%%%%%%%%%%%%%
\subsection{Amplitude dependence on frequency}
\label{SED}

Finally, we explore the frequency dependence of the amplitude of the
angular power spectra.  We compute the \dlee\ and
\dlbb\ angular power spectra from the $Q$ and $U$ DetSet maps at
100, 143, 217, and 353\,GHz (see Sect.~\ref{method_data_sets}).
From these four sets of polarization maps, we compute ten power spectra:
$100\times100$;
$100\times143$; $100\times217$; $100\times353$; $143\times143$;
$143\times217$; $143\times353$; $217\times217$; $217\times353$; and
$353\times353$.

The ten angular cross-power spectra are consistent with a power law in
$\ell$, with the \alphaexponent\ $\alpha_{EE,BB}=\alphaps$ measured at
$353\,$GHz (Sect.~\ref{powerlawfit}).  
Therefore, to each of these spectra we fit the amplitudes of a power-law
function that has a fixed \alphaexponent\ $\alpha_{EE,BB}=\alphaps$,
in the range $\lmin<\ell<500$, for \dlee\ and \dlbb. As an
illustration of the quality of the fit, for the smallest \window\
({\lwdpref}24, $\fsky=0.3$) the averages and dispersions
of the $\chi^2$ (21 degrees of freedom) of the fits 
are $\chi_{EE}^2=13.4\pm8.2$ and $\chi_{BB}^2=12.8\pm6.9$
for the ten cross-frequency spectra.

\begin{figure}[t]
\centering
\includegraphics[height=0.33\textwidth]{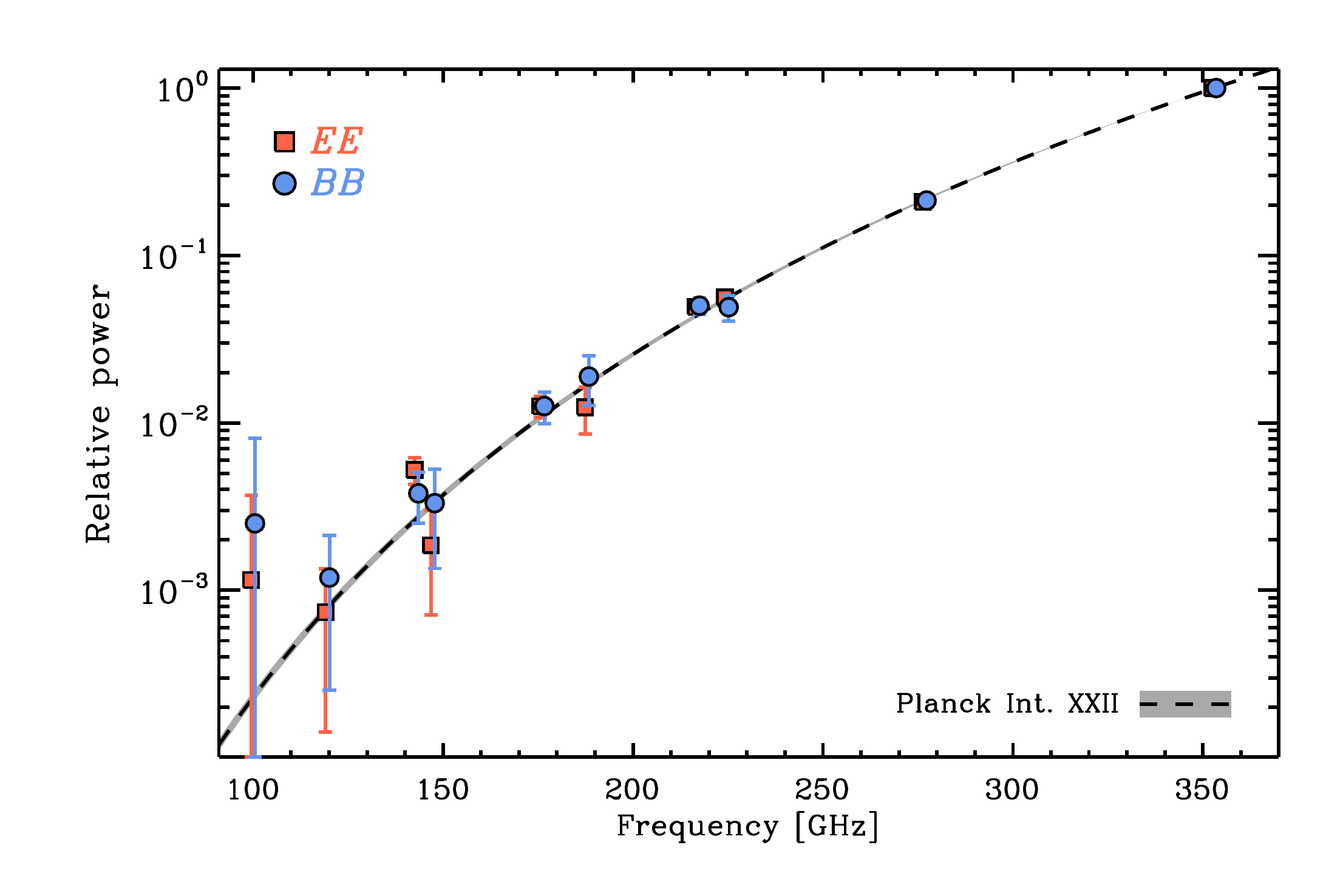}
\caption{Frequency dependence of the amplitudes $A^{EE,BB}$ of the
angular power spectra, relative to 353\,GHz (see details in Sect.~\ref{SED}).
Results for \dlee\ (red squares) and \dlbb\
(blue circles) for the smallest \window,
{\lwdpref}24. These include evaluations from
cross-spectra involving polarization data at two frequencies,
plotted at the geometric mean frequency. The square of the adopted relative
SED for dust polarization, which is a
modified blackbody spectrum with $\beta_{\rm d}=\pesb$ and
$T_{\rm d}=\pest\,$K, is displayed as a black dashed line.
The $\pm1\,\sigma$ uncertainty area from the expected dispersion of
$\beta_{\rm d}$, 0.03 for the size of {\lwdpref}24 as inferred from
\cite{planck2014-XXII} (see Sect.~\ref{dust}), is displayed in grey.
\label{fig:eebb_sed}
}
\end{figure}

To compare the frequency dependence of the results of the fits to
that expected from the SED for dust polarization from \cite{planck2014-XXII},
we converted the fitted amplitudes
$A^{EE,BB}$ from $\mu$K$_{\rm CMB}^2$ to units of
$({\rm MJy}\,{\rm sr}^{-1})^2$, taking into account
the \Planck\ colour corrections.\footnote{Conversion factors were computed
as described in Sect.~\ref{data_description},
here using colour corrections corresponding to a dust modified blackbody
spectrum with $\beta_{\rm d}=\pesb$ and $T_{\rm d}=\pest$\,K.}
For all {\window}s, we examined the frequency dependence 
by plotting the amplitudes normalized to unity at $353\,$GHz, versus the
effective frequency.\footnote{For a cross-spectrum between data at
frequency $\nu_1$ and frequency $\nu_2$
the effective frequency is taken for convenience as the geometric mean,
$\nu_{\rm eff}\equiv\sqrt{\nu_1\nu_2}$.}
A representative example is shown in Fig.~\ref{fig:eebb_sed} 
for the smallest \window, {\lwdpref}24 ($\fsky=0.3$).

For all of the \lwdpref\ {\window}s the frequency dependence found is
in good agreement with the square of the adopted dust SED, which is a
modified blackbody
spectrum having $\beta_{\rm d}=\pesb$ and $T_{\rm d}=\pest\,$K
\citep{planck2014-XXII}.  Note that no fit was performed at this stage and
that the square of the adopted dust SED goes through the 353\,GHz data point. 
However, if we fit the amplitude of the dust frequency dependence, with
fixed $\beta_{\rm d}$ and $T_{\rm d}$, the $\chi^2$ ($N_{\rm dof}=9$)
is 13.1 for $EE$ ($\pte=0.16$) and 3.2 for $BB$ ($\pte=0.96$).
This good agreement supports the assumption made in the present work
about the faintness of the synchrotron and CO emission at high
Galactic latitude (see Sect.~\ref{components}).
The residuals to the fit do not show any evidence for an excess power at
$100\,$GHz, which might arise from polarized synchrotron emission.  This
result is consistent with the study of
synchrotron polarization at high Galactic latitudes by \cite{Fuskeland}, and
confirmed in Appendix~\ref{sec:bicep2_synchrotron}.
Furthermore, we do not see any excess at either 100 or $217\,$GHz, such as
could arise from leakage, and/or polarization, associated with CO line
emission in these bands (Sect.~\ref{CO}).

\begin{figure*}
\centering
\includegraphics[height=0.6\textwidth]{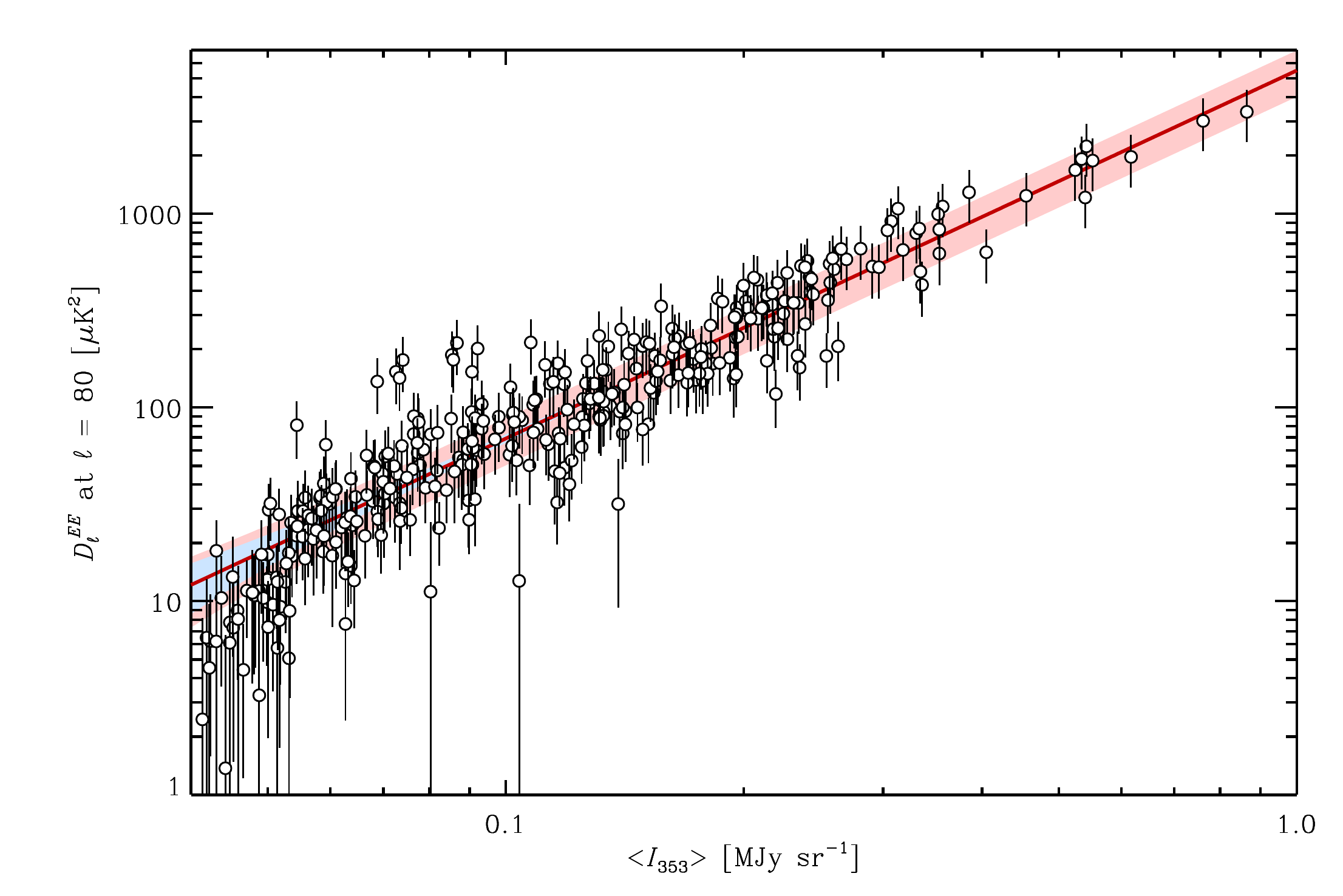}
\includegraphics[height=0.6\textwidth]{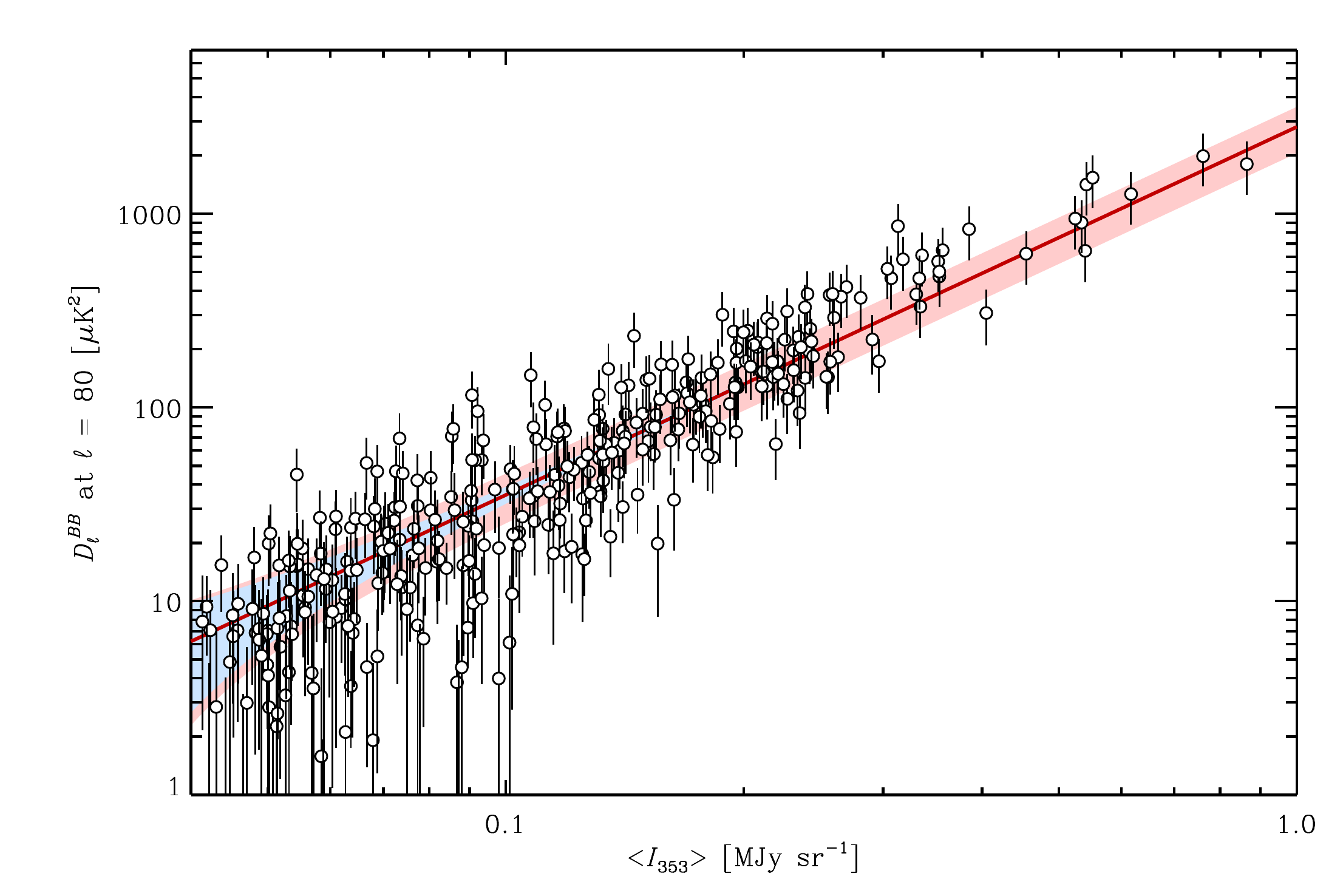}
\caption{
Fitted dust \dlee\ (top panel) and \dlbb\ (bottom panel) amplitudes
($A^{EE}$ and $A^{BB}$) at $\ell=80$, in $\mu$K$^2$ 
for the 400\,deg$^2$ \patcs\ as a function of their mean \idust.
The empirical scaling law, $A^{EE,BB}\propto$ \idust$^{1.9}$,
adjusted in amplitude to the data points, is over-plotted as a red line.
The $\pm3\,\sigma$ statistical error on this relation from Monte Carlo
simulations of \Planck\ inhomogeneous noise
(see Appendix~\ref{appendix_highlat_spectra}) is represented as a light-blue
shaded area and the total $\pm3\,\sigma$ error, including statistical noise
plus Gaussian sample variance, is represented as a light-red shaded area.
Points are computed for all 352 \patcs, but note that, as described in
Sect.~\ref{highlat_masks}, the \patcs\ overlap and so their properties are not
independent.
\label{fig:highlat_bb}
}
\end{figure*}

%%%%%%%%%%%%%%%%%%%%%%%%%%%%%%%%%%%%%%%%%%%%%%
\section{Statistical study of the dust power spectra at high
Galactic latitudes}
\label{highlat_spectra}

In Sect.~\ref{intlat_spectra}, statistical properties of dust polarization 
were derived from angular power spectra computed with 
the \Planck\ $353\,$GHz data 
on large fractions of the sky (\fskyeff from 24\,\% to 72\,\%).
Most of the CMB experiments target 
\flds\ at high Galactic latitude with a smaller size than this.  So now we
evaluate the dust $B$-mode power in such \patcs.

We perform a statistical analysis computing the \dlee\ and \dlbb\ power spectra
of the $353\,$GHz \Planck\ polarization data for high Galactic latitude
circular \patcs\ of size 400\,deg$^2$, similar
to the size of the \flds\ observed in ground-based and balloon-borne CMB
polarization experiments (see Sect.~\ref{highlat_masks}).
We verify that the empirical scaling law of the power spectra amplitudes as
a function of the mean intensity, derived in Sect.~\ref{nhi-dependence},
holds on small independent fractions of the sky that are not nested.

%%%%%%%%%%%%%%%%%%%%%%%%%%%%%%%%%%%%%%%%%%%%%%%
\subsection{Data processing}
\label{highlat_spectra_processing}

We compute the dust \dlee\ and \dlbb\ angular power spectra at
353\,GHz from the cross-spectra of the two DetSets with independent noise
in this channel, using \Xpol\ 
on all of the 352 high Galactic latitude ($|b|>35^\circ$) \patcs\ defined in
Sect.~\ref{highlat_masks}. As shown in
Appendix~\ref{methods_performance}, we do not expect any significant cut-sky
leakage from $E$ to $B$ polarization for the dust in such \patcs.
The \dlee\ and \dlbb\ spectra are computed in the range $40<\ell<370$, using
top-hat binning in the intervals defined between multipoles
40, 70, 110, 160, 220, 290, and 370.

To each of the 352 \dlee\ and \dlbb\ spectra, we fit the power law in $\ell$
presented in Sect.~\ref{powerlawfit} over the range 
$40<\ell<370$, with a slope fixed to $\alpha_{EE,BB}=\alphaps$
in order to compute the amplitudes $A^{EE,BB}$ and their
associated errors.  Since we are interested in quantifying the
possible contamination by dust polarization for ground-based CMB experiments,
here we express the amplitude of the fitted power laws at the position of the
recombination bump at $\ell=80$
($A^{EE,BB}_{\ell=80}=A^{EE,BB}\times80^{-0.42}$).
In Appendix~\ref{appendix_highlat_spectra}, using a null test, we show that
the potential systematic effects in the data do not affect the computation
of the \dlee\ and \dlbb\ spectrum amplitudes.

%%%%%%%%%%%%%%%%%%%%%%%%%%%%%%%%%%%%%%%%%%%%%%%
\subsection{Results}
\label{highlat_spectra_results}

For \patcs\ covering 1\,\% or more of the sky, we do not observe any
significant departure in the shapes of the \dlee\ and \dlbb\ spectra
from the results presented in Sect.~\ref{powerlawfit}, even if a large
dispersion of the power-law \alphaexponent\ is observed due to the low S/N
for some \patcs. The \dlbb/\dlee\ ratios are on average consistent with what
has been derived on the larger
\lwdpref\ {\window}s in Sects.~\ref{EonB}. These ratios are presented in
Appendix~\ref{appendix_highlat_BonE}.

The fitted dust \dlee\ and \dlbb\ amplitudes ($A^{EE}$ and $A^{BB}$) at
$\ell=80$ are presented in Fig.~\ref{fig:highlat_bb} in units of
$\mu$K$_{\rm CMB}^2$ at 353\,GHz and as a function of the mean dust intensity
\idust\ of each \patc. 

We see a clear correlation of the
dust \dlee\ and \dlbb\ amplitudes with \idust. The
data points are consistent with the same \idust\ dependence of the scaling
law found in
Sect.~\ref{nhi-dependence}, $A^{EE,BB}\propto$ \idust$^{1.9}$, indicating that
this empirical law also applies in the faintest regions of the sky
and holds reasonably well for the description of the amplitudes for any
approximately 1\,\% \patc\ of the sky at latitudes above 35$^\circ$.

The amplitudes of the two empirical scaling laws are adjusted to the data
points by computing the median over all the \patcs\ of
$A^{EE,BB}/$\idust$^{1.9}$.  The amplitudes that
we derive from Fig.~\ref{fig:highlat_bb} are in agreement with the amplitudes
reported for the \lwdpref\ \window{s} is Sect.~\ref{nhi-dependence} and
Table~\ref{table_masks}.

The dust brightness \idust\ in the cleanest 400\,deg$^2$ \patcs\
(\idust\  down to 0.038\,\MJysr) is about a factor of 2 lower than the value 
listed in Table~\ref{table_masks} for the {\lwdpref}24 \window\
(0.068\,\MJysr).  Applying the empirical scaling
derived in Sect.~\ref{nhi-dependence}, the expected level of the dust
polarization angular power spectrum is a factor of
(0.068/0.038)$^{1.9}\,{=}\,3$ higher on 24\,\% of the sky than on the faintest
\patcs.

A dispersion around this empirical power law for the \patcs\ is observed.
The shaded regions in Fig.~\ref{fig:highlat_bb} show the expected scatter
in this relation coming from the
noise, together with Gaussian sample variance.  The former is computed from
\Planck\ inhomogeneous noise Monte Carlo simulations presented in
Appendix~\ref{appendix_highlat_spectra} and the latter is computed given
the \fskyeff and the binning used to determine the spectra
(see Sect.~\ref{highlat_spectra_processing}).  Thus the general trend of
the dust \dlee\ and \dlbb\ amplitudes among these \patcs\ follows the
empirical scaling in \idust, but the inhomogeneous nature of the
polarized dust emission, including variations in magnetic field
orientation and grain alignment, is responsible for
a large dispersion around it, larger than expected for a stationary Gaussian
process plus instrumental noise. 

Even at low \idust, the statistical error on the \dlbb\ amplitude at
$\ell=80$ for such small \patcs\ is at least $7.5$\,$\mu$K$_{\rm CMB}^2$
(3$\,\sigma$).

\begin{figure*}[!ht]
\centering
\includegraphics[height=.5\textwidth]{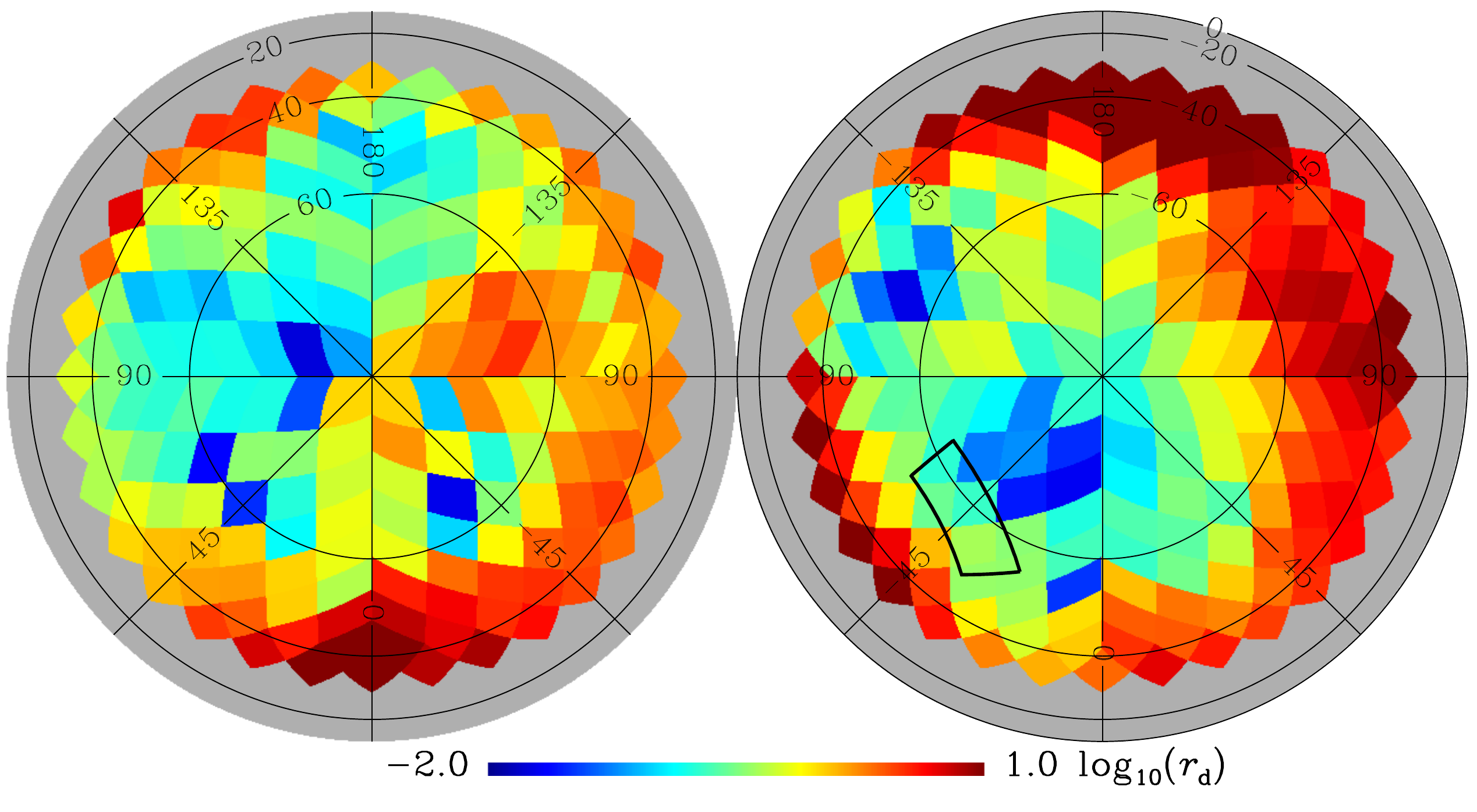}
\includegraphics[height=.5\textwidth]{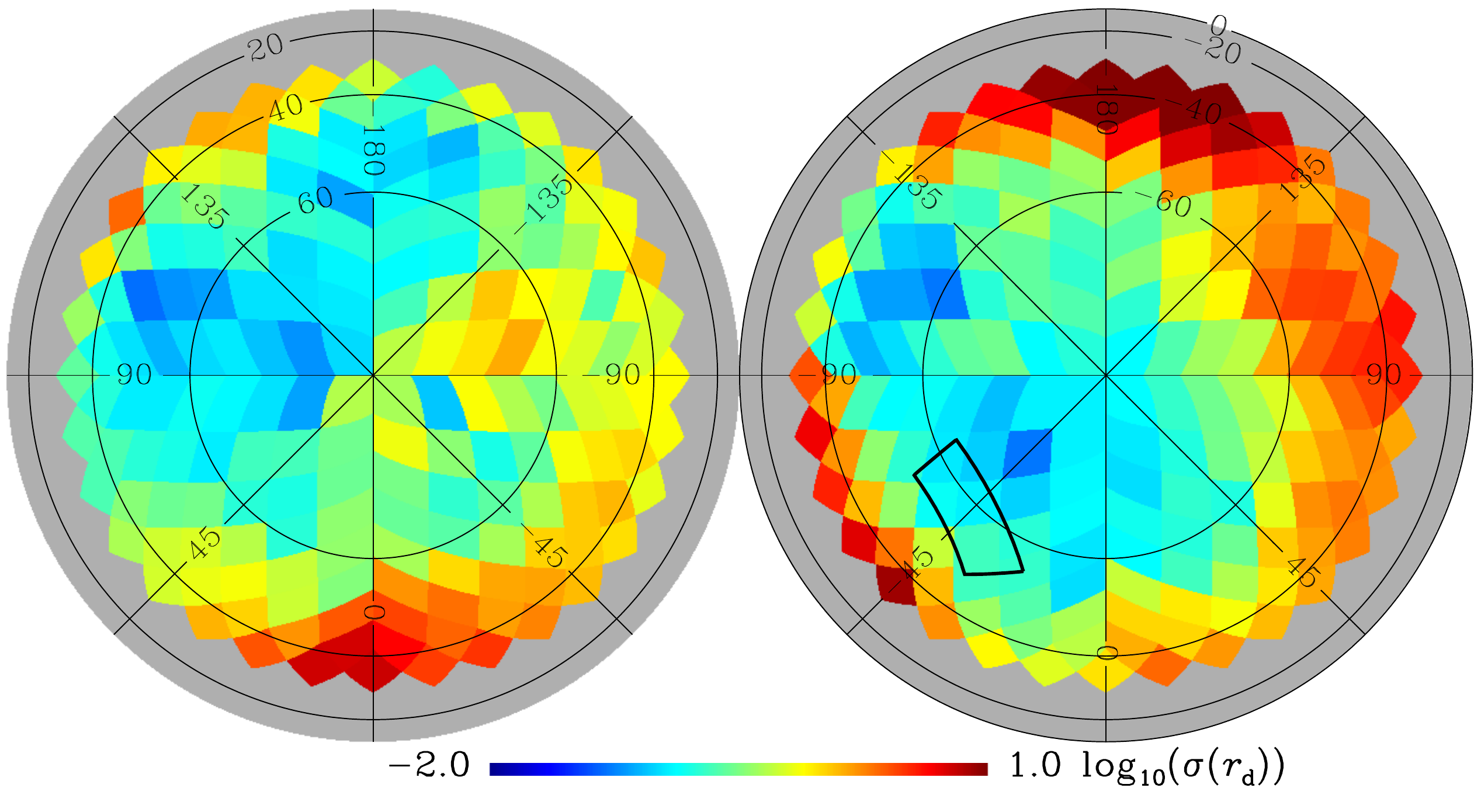}
\caption{{\it Top}: map in orthographic projection of the 150\,GHz
\dlbb\ amplitudes at $\ell=80$, computed from the \Planck\ 353\,GHz data,
extrapolated to 150\,GHz, and normalized by the CMB expectation for
tensor-to-scalar ratio $r=1$.  The colours represent the estimated
contamination from dust in $r_{\rm d}$ units
(see details in Sect.~\ref{sec:optimization}).
The logarithm of the absolute value of $r_{\rm d}$ for a 400\,deg$^2$ \patc\
is presented in the pixel on which the \patc\ is centred.  
As described in Sect.~\ref{highlat_masks}, the \patcs\ overlap and so their
properties are not independent.
The northern (southern) Galactic hemisphere is on the left (right).
The thick black contour outlines the approximate \bicep\ deep-field region
(see Sect.~\ref{sect:bicep2}).  
{\it Bottom}: associated uncertainty, $\sigma(r_{\rm d})$.
\label{fig:bb_map}}
\end{figure*}

%%%%%%%%%%%%%%%%%%%%%%%%%%%%%%%%%%%%%%%%%%%%%%%%%%%

\subsection{Optimizing the search for a primordial $B$-mode signal at
the recombination peak }
\label{sec:optimization}

In Sect.~\ref{highlat_spectra_processing} the 353\,GHz $A^{BB}$ amplitudes
were obtained for the 352 patches of 400\,deg$^2$ at Galactic latitudes
$|b|>35^\circ$.  These amplitudes are expressed at $\ell=80$, the approximate
position of the maximum of the CMB tensor $B$-mode recombination peak.
In order to extrapolate these amplitudes to 150\,GHz, an observing frequency
typical of CMB experiments searching for $B$-modes,
we use the dust SED of \cite{planck2014-XXII} presented in Sect.~\ref{dust},
which is a modified blackbody spectrum with 
$\beta_{\rm d}=\pesb\pm\betadisp$  and $T_{\rm d}=\pest$\,K. 
The conversion factor from the \Planck\ 353\,GHz band to the 150\,GHz central
frequency for this extrapolation, taking into account the \Planck\ colour
corrections and the statistical uncertainty of $\beta_{\rm d}$, is
$0.0395^{+0.0045}_{-0.0035}$ in the SED, and this squared in \dlbb.

These extrapolated estimates are divided by the value of the $r=1$ primordial
tensor CMB \dlbb\ spectrum at $\ell=80$,
$6.71\times10^{-2}$\,$\mu$K$_{\rm CMB}^2$,
to express the estimated power in units that we denote $r_{\rm d}$.
Because the CMB primordial tensor $B$-mode power
scales linearly with $r$,\footnote{This spectrum does not include the
CMB lensing $B$-mode signal, which would
become dominant even at $\ell=80$ for a very low $r$.}
a value $r_{\rm d}=0.1$ would mean that the expected
contamination from dust at $\ell=80$ is equal to the amplitude of the
primordial tensor CMB
\dlbb\ for $r=0.1$.  For each of these estimates we also compute
$\sigma(r_{\rm d})$, the quadratic sum of the fit errors on $A^{BB}$ and the
above uncertainty from the extrapolation to 150\,GHz.  Note that the
fitted amplitudes $A^{BB}$ for five of these \patcs\ are
negative,\footnote{Negative values can arise in cross-spectra, as computed
here.}
but are consistent with $r_{\rm d}=0$ at 1$\,\sigma$.
The $r_{\rm d}$ values vary from $-0.17$ to more than
10 and $\sigma(r_{\rm d})$ ranges from as small as $5.6\times10^{-2}$ to
larger than 10.  Taking the smallest value of $\sigma(r_{\rm d})$ we see
that \Planck\ measurements of the dust \dlbb\ spectra for such small \patcs\
have, at best, a statistical uncertainty of 
0.17 (3\,$\sigma$) in $r_{\rm d}$ units.

To reveal the spatial dependence over the high latitude sky, a map of (the
absolute value of) $r_{\rm d}$ is shown in Fig.~\ref{fig:bb_map}.
Each computed value of $r_{\rm d}$ is given in a pixel at the position of the
centre of each of the 352 \patcs\ defined in Sect.~\ref{highlat_masks} on an
$N_{\rm side}=8$ {\tt HEALPix} map.  The accompanying map of
$\sigma(r_{\rm d})$ is also presented.

We can see in Fig.~\ref{fig:bb_map} that there is a high latitude region in
the southern Galactic hemisphere for which $r_{\rm d}$ is
quite low.  This region
is also associated with a small estimated uncertainty.  For example, the six
{\tt HEALPix} $N_{\rm side}=8$ pixels numbered 741, 742, 754, 755, 762, and 763
(in the {\tt HEALPix} ``ring'' ordering scheme)
have $r_{\rm d} = 0.053 \pm 0.096$, $0.027 \pm 0.098$, $-0.062 \pm 0.052$,
$-0.020 \pm 0.127$, $0.057 \pm 0.122$, and $ -0.031 \pm 0.121$, respectively.
These pixels are located around Galactic coordinates $l=-30^\circ$,
$b=-70^\circ$.

We stress that the expectation for this low level of dust contamination is
valid only for these particular \patcs, including their positions, sizes,
shapes, and apodizations. 
In addition, because we found the amplitudes of the dust \dlbb\ spectra
associated with these \patcs\ based on a power-law fit, our estimate does
not take into
account possible features in the power spectra that might alter the precise
value of dust contamination. 
Nevertheless, there are clearly some \patcs\ that appear to be optimal, i.e.\
cleaner than the others.
But it needs to be emphasized that finding the cleanest areas of the polarized
sky for primordial $B$-mode searches cannot be accomplished accurately using
the \Planck\ total intensity maps alone.

%%%%%%%%%%%%%%%%%%%%%%%%%%%%%%%%%%%%%%%%%%%%%%%%%%%%%%%%%%%

\section{\textit{BB} angular power spectrum of dust in the \bicep\ \sten}
\label{sect:bicep2}

In this section, we use the \Planck\ data and the results presented above to
assess the dust polarization in the 
\sten\ observed by the \bicep\ experiment \citep{BICEP2Instru,BICEP2_PRL}.
As above, finding \dlbb\ at 150\,GHz involves two steps, measuring the power
spectrum at 353\,GHz and then extrapolating the amplitude to 150\,GHz.

%%%%%%%%%%%%%%%%%%%%%%%%%%%%%%%%%%%%%%%%%%%%%%
\subsection{An approximation to the observed \bicep\ \sten}
\label{bicep2_mask}

To define a \sten\ similar to and representative of the actual
\bicep\ \sten,
we carried out the following steps (at the {\tt HEALPix} $N_{\rm side}=2048$
pixelization): 
(i) we constructed a mask $M$ by filling the inside of the \bicep\ deep-field
outline.\footnote{\url{http://bicepkeck.org}
This outline encloses the effective 373\,deg$^2$
of deep integration (87 nK-degrees, \citealt{BICEP2_PRL}).
}
with 1 and the outside with 0; 
(ii) we took the complement of this mask, $M'=1-M$; 
(iii) inside $M'$ we computed the distance to the border 
using the {\tt HEALPix} ``iprocess\_mask'' procedure;
(iv) we smoothed this ``distance map'' with a $7\pdeg5$ FWHM Gaussian;
(v) we apodized, computing the $7\pdeg5$ FWHM Gaussian weight
from the distance map to obtain $M''$; 
and (vi) we took the complement of $M''$ to be our definition of the \bicep\
\sten\ (i.e., $M_{\rm B2}=1-M''$). 
The resulting \sten, which in the rest of this paper we will refer to as
\mb2, has $\fskyeff=0.017$ (689\,deg$^2$) and \idust$=0.060$\,\MJysr.
Its \fskyeff is larger than the \bicep\ $373\,{\rm deg}^2$ deep-field region,
but \mb2\ is similar to the \bicep\ inverse noise variance map presented in
\cite{BICEP2Instru}, even if it extends further in declination but less in
right ascension. 
In Appendix~\ref{bicep2_mask_dependence} we show
that the main results in this section do not depend significantly on the
definition of this \sten.
However, note that the various applications of filtering in the \bicep\
pipeline mean that their observed spatial modes are a subset of those present
within their nominal observed region.

%%%%%%%%%%%%%%%%%%%% fig 9 %%%%%%%%%%%%%%%%%%%%%%%%%%%%%%%%

\begin{figure*}[t!]
\centering
\includegraphics[height=0.55\textwidth]{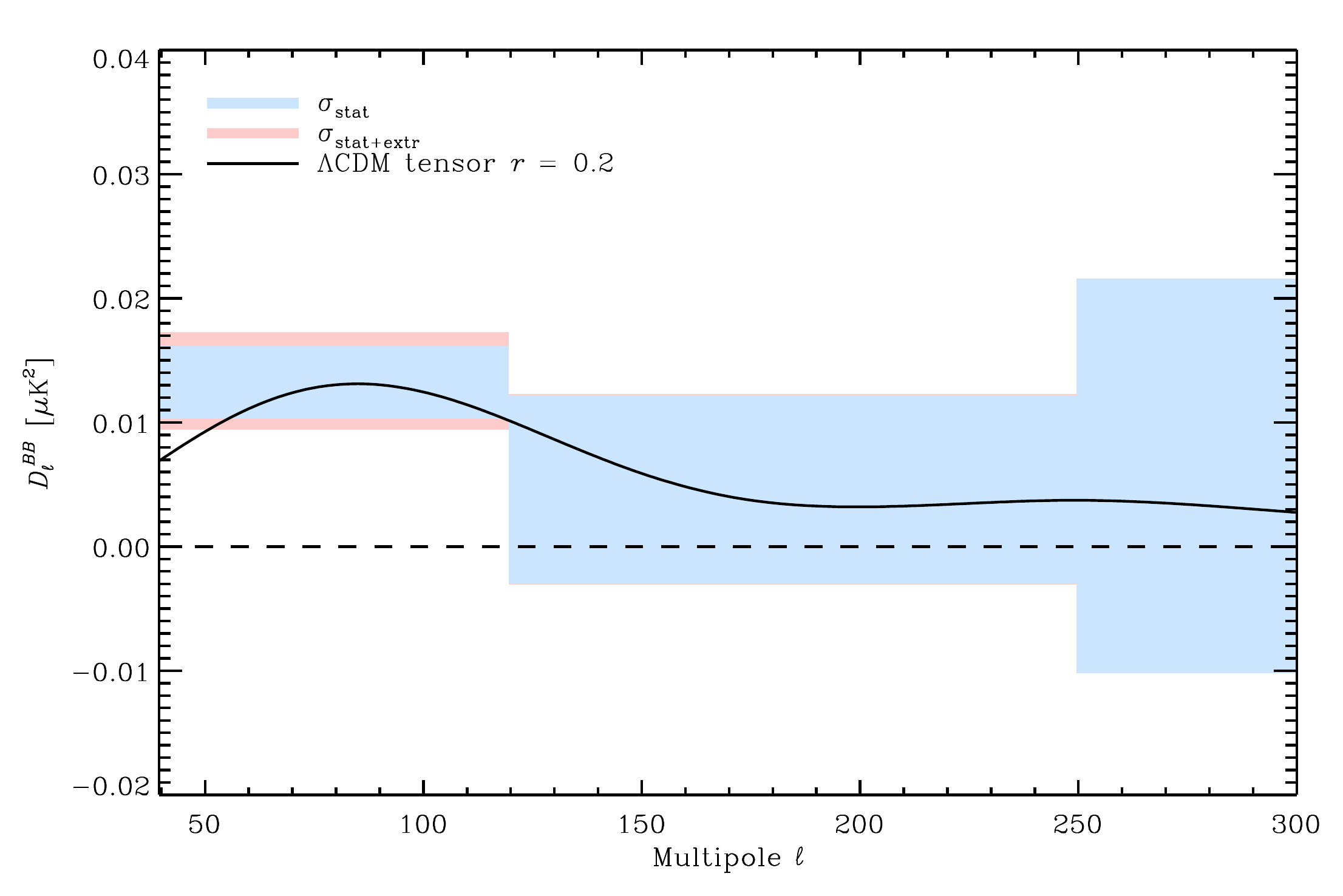}
\caption{
\Planck\ 353\,GHz \dlbb\ angular power spectrum computed
on \mb2\ defined in Sect.~\ref{bicep2_mask} and
extrapolated to 150\,GHz (box centres).  
The shaded boxes represent the $\pm1\,\sigma$ uncertainties: blue for the
statistical uncertainties from noise; and red adding in quadrature the
uncertainty from the extrapolation to 150\,GHz.
The \Planck\ 2013 best-fit $\Lambda$CDM \dlbb\
CMB model based on temperature anisotropies, with a tensor amplitude fixed
at $r=0.2$, is overplotted as a black line.
\label{fig:bicep2_bb}
}
\end{figure*}

%%%%%%%%%%%%%%%%%%%%%%%%%%%%%%%%%%%%%%%%%%%%%%
\subsection{Statistical estimate of the dust $B$-mode level in the \bicep\
\sten}
\label{bicep2_extrapolation}

We have seen in Sect.~\ref{highlat_spectra_results} that the dust \dlbb\
spectra amplitudes statistically follow an empirical scaling law of the mean
dust intensity \idust\ for the \patc\ on which they are computed.  We can use
this scaling law to assess the most probable dust \dlbb\ value for \mb2.
For this \sten, which has a mean dust intensity of \idust=0.060\,\MJysr,
the expected value at 353\,GHz is $(13.4\pm0.26)$\,$\mu$K$_{\rm CMB}^2$
at $\ell=80$, taking into account the uncertainty on the
fitted \idust$^{1.9}$ amplitude from Sect.~\ref{nhi-dependence}.

In order to extrapolate this value to the \bicep\ observing frequency,
150\,GHz, we proceed as in Sect.~\ref{sec:optimization}, using the same
typical dust SED from \cite{planck2014-XXII}.  
For \patcs\ the size of \mb2\ the expected dispersion of $\beta_{\rm d}$
(Sect.~\ref{dust}) is 0.11, which introduces an uncertainty in the
extrapolation.
The extrapolation, unlike in that section, is made to 150\,GHz
taking into account the \bicep\ bandpass.\footnote{\url{http://bicepkeck.org}}
The conversion factor from the \Planck\ 353\,GHz band to the 150\,GHz \bicep\
band, taking into account both \Planck\ and \bicep\ colour
corrections, as well as the statistical error on $\beta_{\rm d}$, is thus
$0.0408^{+0.0046}_{-0.0036}$ in the SED, and therefore this value
squared in \dlbb.

The resulting expected dust \dlbb\ amplitude at 150\,GHz is
$2.23^{+0.55}_{-0.45}\times10^{-2}$\,$\mu$K$_{\rm CMB}^2$ at $\ell=80$.  The 
uncertainty quoted here applies to the most probable value for patches with a
similar \idust, but it ignores the intrinsic dispersion around this value
(see Sect.~\ref{highlat_spectra_results}). Therefore, a
direct measurement of the dust \dlbb\ spectrum in \mb2\ is required to
complete the assessment of its polarized dust level.

In Sect.~\ref{sec:optimization}, we have localized the \dlbb\ amplitudes from
the analysis of \patcs\ on a map.  The pixels falling inside the approximate
\bicep\ deep-field region displayed in Fig.~\ref{fig:bb_map} give a mean value
of $r_{\rm d}=0.207$, i.e., an expected dust power of
\dlbb$=1.39\times10^{-2}$\,$\mu$K$_{\rm CMB}^2$ 
at 150\,GHz and $\ell=80$ (note that in this region the \Planck\ estimate of
the dust contamination is significantly higher,
by a factor of about 2, than for some more optimal
patches). Nevertheless, a more accurate estimate of the dust \dlbb\
polarization amplitude in the \bicep\ observed \sten\ requires the direct
computation of the angular power spectrum on a similar \window.

\begin{figure*}[!ht]
\centering
\includegraphics[height=0.55\textwidth]{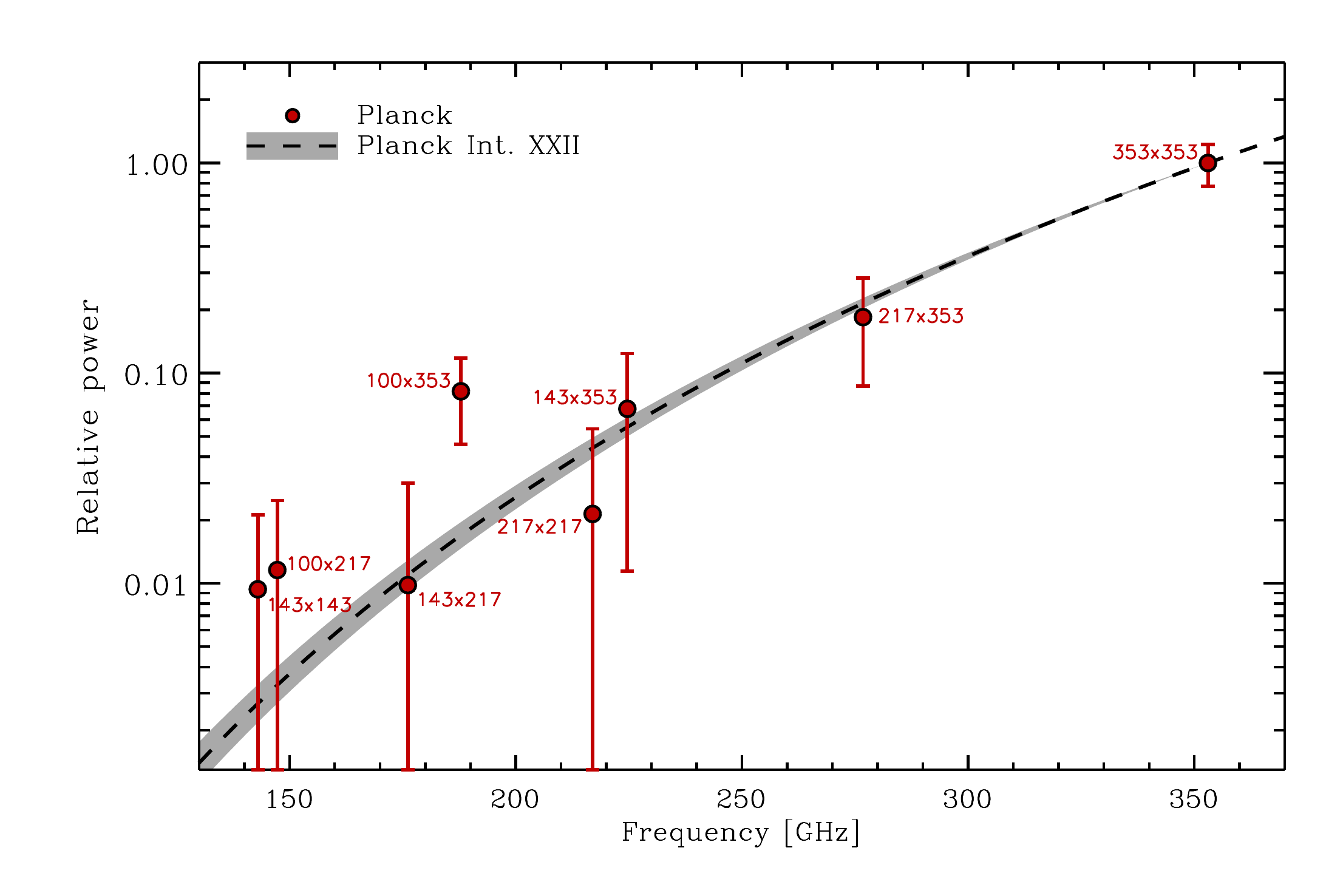}
\caption{
Frequency dependence of the amplitude $A^{BB}$ of the angular power spectrum
\dlbb\ computed on \mb2\ defined in
Sect.~\ref{bicep2_mask}, normalized to the 353\,GHz
amplitude (red points); amplitudes for cross-power spectra are plotted at the
geometric mean frequency.
The square of the adopted dust SED, a modified blackbody
spectrum with $\beta_{\rm d}=\pesb$ and $T_{\rm d}=\pest\,$K,
is over-plotted as a black
dashed-line, again normalized to the 353\,GHz point. The $\pm1\,\sigma$ error
area arising from the expected dispersion of $\beta_{\rm d}$,
0.11 for the \mb2\ \patc\ size (Sect.~\ref{dust}),
is displayed in light grey.
\label{fig:bb_sed}
}
\end{figure*}

%%%%%%%%%%%%%%%%%%%%%%%%%%%%%%%%%%%%%%%%%%%%%%
\subsection{$BB$ angular power spectrum of dust in the \bicep\ \sten}
\label{bicep2_ps353}

We now compute the \dlbb\ spectrum of the dust at 150\,GHz on
\mb2\ presented in Sect.~\ref{bicep2_mask}, using the \Planck\ 353\,GHz data.
For this purpose, we compute the 353\,GHz \dlbb\ spectrum by cross-correlating
the two DetSets that have independent noise at this frequency in the three
$\ell$-bins defined by the intervals between multipoles
40, 120, 250, and 400. 
The $\ell$-bin sizes were increased with respect to
the previous sections in order to increase the S/N, especially
around the range of the CMB recombination bump.
Appendix~\ref{bicep2_xpol-xpure} confirms that the result does not depend on
the method of computing the power spectrum.

This power spectrum is extrapolated to 150\,GHz as in
Sect.~\ref{bicep2_extrapolation}, with an extrapolation uncertainty estimated
from the inferred dispersion of $\beta_{\rm d}$.
Our final estimate of the \dlbb\ spectrum is presented in
Fig.~\ref{fig:bicep2_bb}, together with its $1\,\sigma$ error budget.  For the
first bin, $\ell\,{=}\,40$--120, the expected level of dust polarized \dlbb,
as extrapolated to 150\,GHz, is $1.32 \times 10^{-2}$\,$\mu$K$_{\rm CMB}^2$
(Fig.~\ref{fig:bicep2_bb}). The statistical error, estimated from Monte Carlo
simulations of inhomogeneous \Planck\
noise (presented in Appendix~\ref{methods_performance} for this particular
binning), is $\pm\,0.29 \times 10^{-2}$\,$\mu$K$_{\rm CMB}^2$, so that the dust
\dlbb\ spectrum is statistically detected at 4.5$\,\sigma$ in this broad
$\ell$ bin. 

In order to assess the potential contribution from systematics, we have
computed the dust \dlbb\ spectrum on \mb2\ on different subsets of the data
and performed null tests, which are presented in Appendix~\ref{bicep2_syste}.
In this lowest bin of $\ell$, we do not observe any departure from what is
allowed by noise.
Nevertheless, we stress that below the noise level our cross-spectra could be
subject to a positive or negative bias due to systematic effects.  For example,
if instead of taking the DetSets cross-spectra (as we have done throughout
this paper) we take the mean value computed from the DetSets, HalfRings, and
Years cross-spectra (presented in Appendix~\ref{bicep2_syste}), the
statistical significance of our measurement is decreased from 4.5$\,\sigma$
to 3.6$\,\sigma$.

The uncertainty coming from the \mb2\ definition 
(presented in Appendix~\ref{bicep2_mask_dependence}) is
$0.04 \times 10^{-2}$\,$\mu$K$_{\rm CMB}^2$ for this bin, thus 
much less than the statistical error. For this reason, it is not added to
the error budget.
However, the spectral extrapolation to 150\,GHz adds an additional uncertainty
$(+0.28, -0.24) \times 10^{-2}$\,$\mu$K$_{\rm CMB}^2$ to the estimated power
in \mb2, added in quadrature in Fig.~\ref{fig:bicep2_bb}.

The expected value in this lowest-$\ell$ bin from direct computation of
the \dlbb\ power spectrum on \mb2, as shown in Fig.~\ref{fig:bicep2_bb}, 
is lower than (but consistent with) the
statistical expectation from the analysis of the 352 high Galactic latitude
\patcs\ presented in Sects.~\ref{highlat_spectra_results} and
\ref{bicep2_extrapolation}.  This indicates that \mb2\ is not one of the
outliers of Fig.~\ref{fig:highlat_bb} and therefore its dust $B$-mode power
is well represented by its mean dust intensity through the empirical scaling
law ${\cal D}\propto\,$\idust$^{1.9}$. 

These values of the \dlbb\ amplitude in the $\ell$ range of the primordial
recombination bump are of the same magnitude as those reported by
\citet{BICEP2_PRL}.
Our results emphasize the need for a dedicated joint analysis of the $B$-mode
polarization in this region incorporating all pertinent observational details
of the \Planck\ and \bicep\ data sets, which is in progress.

%%%%%%%%%%%%%%%%%%%%%%%%%%%%%%%%%%%%%%%%%%%%%%
\subsection{Frequency dependence}
\label{sec:bice2p_sed}

We complement the power spectrum analysis of the $353\,$GHz map with \Planck\
data at lower frequencies.  As in the analysis in Sect.~\ref{SED}, we compute
the frequency dependence of the $BB$ power measured by
\Planck\ at HFI frequencies in the \bicep\ \sten, using 
the \patc\ \mb2\ as defined in Sect.~\ref{bicep2_mask}.

We compute on \mb2\ the \Planck\ \dlbb\ auto- and 
cross-power spectra from the three \Planck\ HFI bands at 100, 143, 217, 
and 353\,GHz, using the two DetSets with independent noise at each frequency,
resulting in ten angular power spectra 
($100\times100$, $100\times143$, $100\times217$, $100\times353$,
 $143\times143$, $143\times217$, $143\times353$, $217\times217$,
 $217\times353$, and $353\times353$), constructed by combining the
cross-spectra as presented in Sect.~\ref{method_data_sets}.
We use the same multipole binning as in Sect.~\ref{bicep2_ps353}.
To each of these \dlbb\ spectra, we fit the amplitude of a
power law in $\ell$ with a fixed \alphaexponent\ $\alpha_{\rm BB}=-0.42$
(see Sect.~\ref{powerlawfit}).  In Fig.~\ref{fig:bb_sed} we plot
these amplitudes as a function of the effective
frequency from 143 to 353\,GHz, in units of sky brightness squared,
like in Sect.~\ref{SED}.
Data points at effective frequencies below 143\,GHz are not presented,
because the dust polarization is not detected at these frequencies.
An upper limit on the synchrotron contribution at 150\,GHz from the
 \Planck\ LFI data is given in Appendix~\ref{sec:bicep2_synchrotron}.

We can see that the frequency dependence of the amplitudes of the \Planck\
HFI \dlbb\ spectra is in very good agreement with a squared dust
modified blackbody spectrum having $\beta_{\rm d}=\pesb$ and
$T_{\rm d}=\pest\,$K \citep{planck2014-XXII}. 
We note that this emission model was normalized only to the
353\,GHz point and that no global fit has been performed. 
Nevertheless, the $\chi^2$ value from the amplitudes relative to this model
is 4.56 ($N_{\rm dof}=7$).
This shows that dust dominates in the specific \mb2\ region defined where these
cross-spectra have been computed. 
This result emphasizes the need for a dedicated joint \Planck{--}\bicep\
analysis. 

%%%%%%%%%%%%%%%%%%%%%%%%%%%%%%%%%%%%%%%%%%%%%%%%%%%%%%%%%%%%%%%%%%%%%

\hpeter{The clearpage here is just to avoid a latex problem.}
%\clearpage

\section{Conclusions}
\label{part:conclusions}

We have presented the first nearly all-sky statistical analysis of the
polarized emission from interstellar dust, focussing mostly on the
characterization of this emission as a foreground contaminant 
at frequencies above 100\,GHz.  Our quantitative analysis of the angular
dependence of the dust polarization
relies on measurements at 353\,GHz of the \clee\ and \clbb\
(alternatively \dlee\ and \dlbb) angular power spectra
for multipoles $40<\ell<500$. At this
frequency only two polarized components are present: dust emission;
and the CMB, which is subdominant in this multipole range.
We have found that the
statistical, spatial, and spectral distribution properties can be 
represented accurately by a simple model over most of the sky, and for all
frequencies at which \Planck\ HFI measures polarization.

\begin{itemize}

\item The angular power spectra \clee\ and \clbb\ at 353\,GHz are well fit
  by power laws in $\ell$ with
  \alphaexponents\ consistent with $\alpha_{EE,BB}=\alphaps\pm\alphapsuncer$, 
  for sky fractions ranging from 24\,\% to 72\,\% for the \lwdpref\
  {\window}s used.

\item The amplitudes of \dlee\ and \dlbb\ in the \lwdpref\ {\window}s vary with
  mean dust intensity at 353 GHz, \idust, roughly as \idust$^{1.9}$.

\item The frequency dependence of the dust \dlee\ and \dlbb\ from 353\,GHz
  down to 100\,GHz, obtained after removal of the \dlee\ prediction from
  the \Planck\ best-fit CMB model \citep{planck2013-p11}, is
  accurately described by the modified blackbody dust emission law
  derived in \cite{planck2014-XXII}, with $\beta_{\rm d}=\pesb$ and
  $T_{\rm d}=\pest\,$K.

\item The ratio between the amplitudes of the two polarization power
  spectra is $C_\ell^{BB}/C_\ell^{EE}=0.53$, which is not consistent with
  the simplest theoretical models.

\item Dust \dlee\ and \dlbb\ spectra computed for 352
  high Galactic latitude 400\,deg$^2$ \patcs\ satisfy the above
  general properties at 353\,GHz and have the same frequency dependence.

\end{itemize}

We have shown that \Planck's determination of the 353\,GHz dust
polarization properties is unaffected by systematic errors for $\ell>40$.
This enables us to draw the following conclusions relevant for CMB
polarization experiments aimed at detection of primordial CMB tensor $B$-modes.

\begin{itemize}

\item Extrapolating the \Planck\ 353\,GHz \dlbb\ spectra computed
  on the 400\,deg$^2$ circular \patcs\ at high Galactic
  latitude to 150\,GHz shows that we expect significant contamination
  by dust over most of the high Galactic latitude sky in the $\ell$ range
  of interest for detecting a primordial \dlbb\ spectrum.

\item Even for the cleanest of these regions, the \Planck\ statistical error
  on the estimate of \dlbb\ amplitude at $\ell = 80$ for such small regions
  is at best $0.17$ (3$\,\sigma$) in units of $r_{\rm d}$.
  
\item  Our results show that subtraction of polarized
  dust emission will be essential for detecting primordial $B$-modes at a level
  of $r=0.1$ or below.

\item There is a significant dispersion of the polarization \dlbb\ amplitude
  for a given dust total intensity. Choices of the cleanest areas of the
  polarized sky cannot be made accurately using the \Planck\ total intensity
  maps alone.

\item Component separation, or template cleaning, 
  can best be done at present with the
  \Planck\ HFI 353\,GHz data, but the accuracy of such cleaning is limited
  by \Planck\ noise in small fields.  Ground-based or
  balloon-borne experiments should include dust channels at high frequency.
  Alternatively, if they intend to rely on the \Planck\
  data to remove the dust emission, they should optimize the
  integration time and area so as to have a similar
  signal-to-noise level for the CMB and dust power spectra.

\end{itemize}

Turning specifically to the part of the sky mapped by the \bicep\ experiment,
our analysis of the \mb2\ region indicates the following results.

\begin{itemize}

\item Over the multipole range $40<\ell<120$, the \Planck\ 353\,GHz \dlbb\
 power spectrum extrapolated to 150\,GHz yields a value $1.32 \times
 10^{-2}$\,$\mu$K$_{\rm CMB}^2$, with statistical error $\pm0.29 \times
 10^{-2}$\,$\mu$K$_{\rm CMB}^2$
 and a further uncertainty $(+0.28, -0.24) \times
 10^{-2}$\,$\mu$K$_{\rm CMB}^2$
 from the extrapolation. 
This value is comparable in magnitude to the \bicep\ measurements at these
multipoles that correspond to the recombination bump. 

\item The frequency dependence of \dlbb\ across the
  \Planck\ bands is consistent
  with the typical SED of dust polarization \citep{planck2014-XXII}. 

\item  Assessing the dust contribution to the $B$-mode power measured by the
 \bicep\ experiment requires a dedicated joint analysis with \Planck,
 incorporating all pertinent observational details of the two data sets, such
 as masking, filtering, and colour corrections.

\item We have identified regions in which the dust polarization \dlbb\
 amplitude may be significantly lower, 
 by about a factor of 2, than in the \bicep\ observing region.

\end{itemize}

%TEMPORARY FILLER:  LATEX PROBLEM RELATED TO THE LONG URL BELOW.

%%%%%%%%%%%%%%%%%%%%%%%%%%%%%%%%%%%%%%%%%%%%%%%%%%%%%%%%%%%%%%%%%%%%

\begin{acknowledgements}
The development of \Planck\ has been supported by: ESA; CNES and
CNRS/INSU-IN2P3-INP (France); ASI, CNR, and INAF (Italy); NASA and DoE
(USA); STFC and UKSA (UK); CSIC, MICINN, JA, and RES (Spain); Tekes,
AoF, and CSC (Finland); DLR and MPG (Germany); CSA (Canada); DTU Space
(Denmark); SER/SSO (Switzerland); RCN (Norway); SFI (Ireland);
FCT/MCTES (Portugal); and PRACE (EU).
A description of the Planck Collaboration and a list of its members,
including the technical or scientific activities in which they have
been involved, can be found at
\url{http://www.rssd.esa.int/index.php?project=PLANCK&page=Planck_Collaboration}.
Some of the results in this paper have been derived using the {\tt HEALPix}
package. 
The research leading to these results has received funding from the European
Research Council under the European Union's Seventh Framework Programme
(FP7/2007-2013) / ERC grant agreement No.\ 267934.
%We thank the BICEP2 team for helpful comments and discussions on this paper.
\end{acknowledgements}

\bibliography{Dust_Polarized_Cl.bib,Planck_bib.bib}

%%%%%%%%%%%%%%%%%%%%%%%%%%%%%%%%%%%%%%%%%%%%%%%%%%%%%%%%%%%%%%%%%%%%%
\appendix

%%%%%%%%%%%%%%%%%%%%%%%%%%%%%%%%%%%%%%%%%%%%%%%%%%%%%%%%%%%%%%%%%%%%%
\section{Power spectrum estimator performance}
\label{methods_performance}

Using \Xpol\ (see Sect.~\ref{xpol}) we have estimated the dust
polarization \dlee\ and \dlbb\ angular power spectrum, not on the full
sky, but on particular cuts of the sky, such as the {\lwdpref}24
\window\ defined in Sect.~\ref{intlat_masks} and the much smaller
\mb2\ \patc\ defined in Sect.~\ref{bicep2_mask}.  To validate the performance
of \Xpol\ on such cuts we compare the results of this algorithm to those from
\Xpure\ (see Sect.~\ref{xpure}), using in both cases simulated data.

\hQ{ORIGINAL. In order to validate the performance of \Xpol\ (see Sect.~\ref{xpol})
for estimation of the dust polarization \dlee\ and \dlbb\ angular
power spectrum on a cut sky, we compare the
results of this algorithm to those from \Xpure\ (see
Sect.~\ref{xpure}) using simulated data. }

\begin{figure}
\centering
\includegraphics[height=0.33\textwidth]{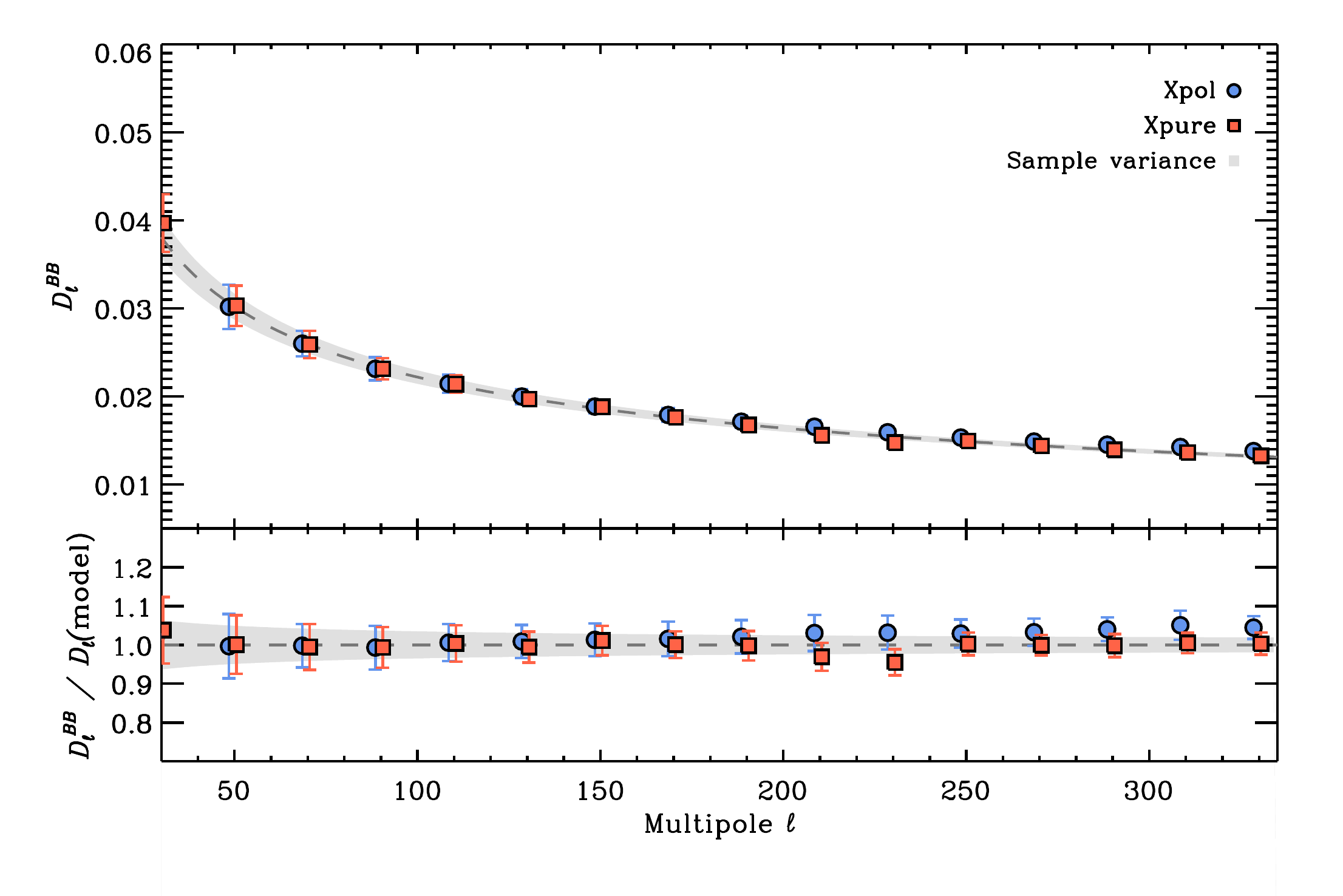}
\includegraphics[height=0.33\textwidth]{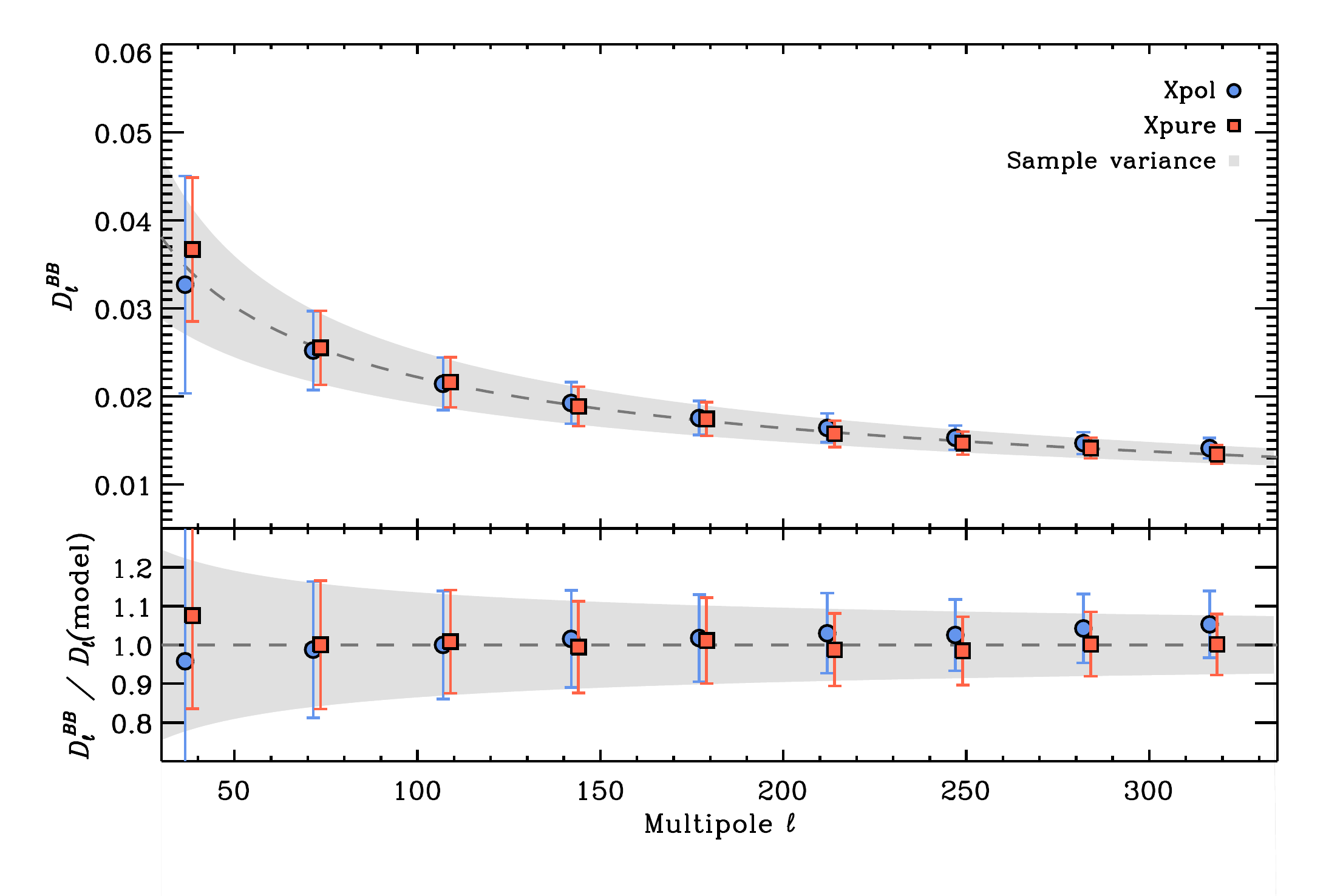}
\caption{
Upper parts of both panels: performance of the \Xpol\
(blue points, see Sect.~\ref{xpol}) and \Xpure\ (red squares,
see Sect.~\ref{xpure}) algorithms on the \dlbb\ power spectrum 
for Gaussian simulations of the dust polarization in the {\lwdpref}24
\window\ (top panel, defined in Sect.~\ref{intlat_masks})
and in \mb2\ (bottom panel, defined in Sect.~\ref{bicep2_mask}). The
signal input power spectrum is displayed as a dashed grey line and its
associated sample variance as a light grey shaded area. 
Lower parts of both panels: relative discrepancy
with respect to the input power spectrum.
\label{fig:simus}
}
\end{figure}

%%%%%%%%%%%%%%%  only one subsection %%%%%%%%%%%%%%%%%%%%%%%%%%%%%%%
%\subsection{Simulations}
%\label{simus}

Simulating the dust polarization for assessment of the
performance of polarization angular power spectra estimators is an
important
issue. One either relies on statistically isotropic, Gaussian simulations with
polarization power spectra similar to those of the dust or alternatively
uses non-Gaussian, anisotropic simulations. The problem in
the first case is that the hypotheses of Gaussianity and statistical isotropy
might apply only partially to a dust polarization map. The problem in
the second case is that one has to produce a non-Gaussian and
anisotropic simulation, which has a defined {\it input\/} angular
power spectrum, in order to be able to characterize what is retrieved
in the output. In this section, we will use statistically isotropic,
Gaussian simulations to
explore the performance of our algorithms for recovering the dust
polarization angular power spectra on a given masked sky. This choice is
motivated by the fact that cosmologists have tended to make this
assumption when analysing the CMB anisotropies from a sky contaminated by
non-Gaussian and anisotropic processes. 

We generated 1000 full sky $Q$ and $U$ Gaussian map simulations from
\clee\ and \clbb\ power spectra having
$C_\ell^{EE,BB}\propto\ell^{\,\alphaps}$,
the typical dust power spectrum shape that we measured in
Sect.~\ref{powerlawfit}. Additionally, we gave to these spectra the
$B$/$E$ amplitude hierarchy that we have measured in Sect.~\ref{EonB},
i.e., $C_\ell^{BB}=0.53C_\ell^{EE}$. For each of these simulated dust
polarized signal $Q$ and $U$ maps, we computed the \dlee\ and \dlbb\ angular
power spectra on the cut sky {\lwdpref}24 \window\ 
and the much smaller \mb2\ \patc.
Note that the structure in the simulations is driven by the assumed power
spectra and the random phases drawn for each mode and is not correlated with
the geometry of the mask.

The results are shown in Fig.~\ref{fig:simus}. For each region, we display the
\dlbb\ angular power spectra recovered by \Xpol\ and \Xpure\ from the
simulations, together with the input power spectrum and the sample variance
associated with each region.

\begin{figure*}[t!]
\centering
\includegraphics[height=0.6\textwidth]{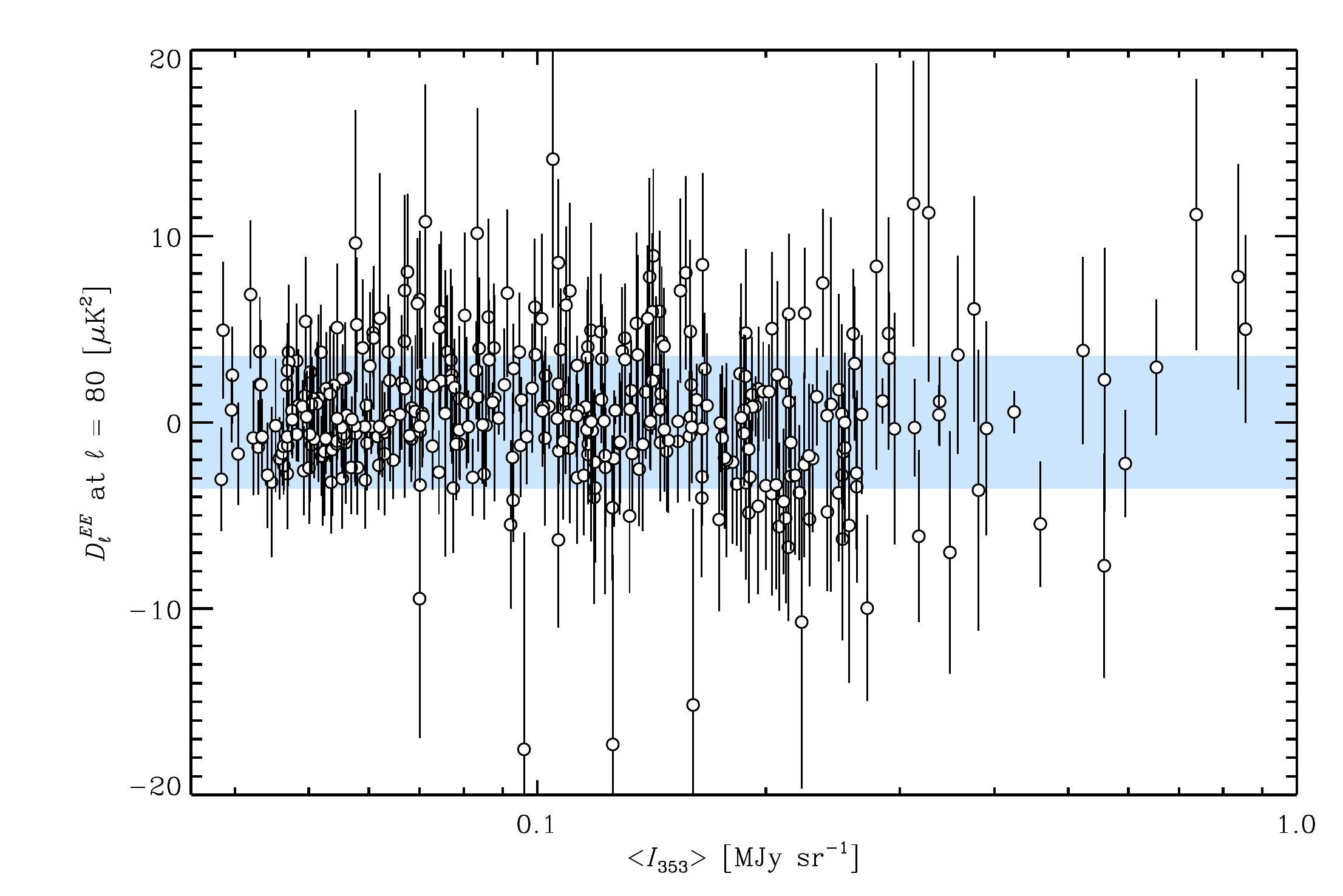}
\includegraphics[height=0.6\textwidth]{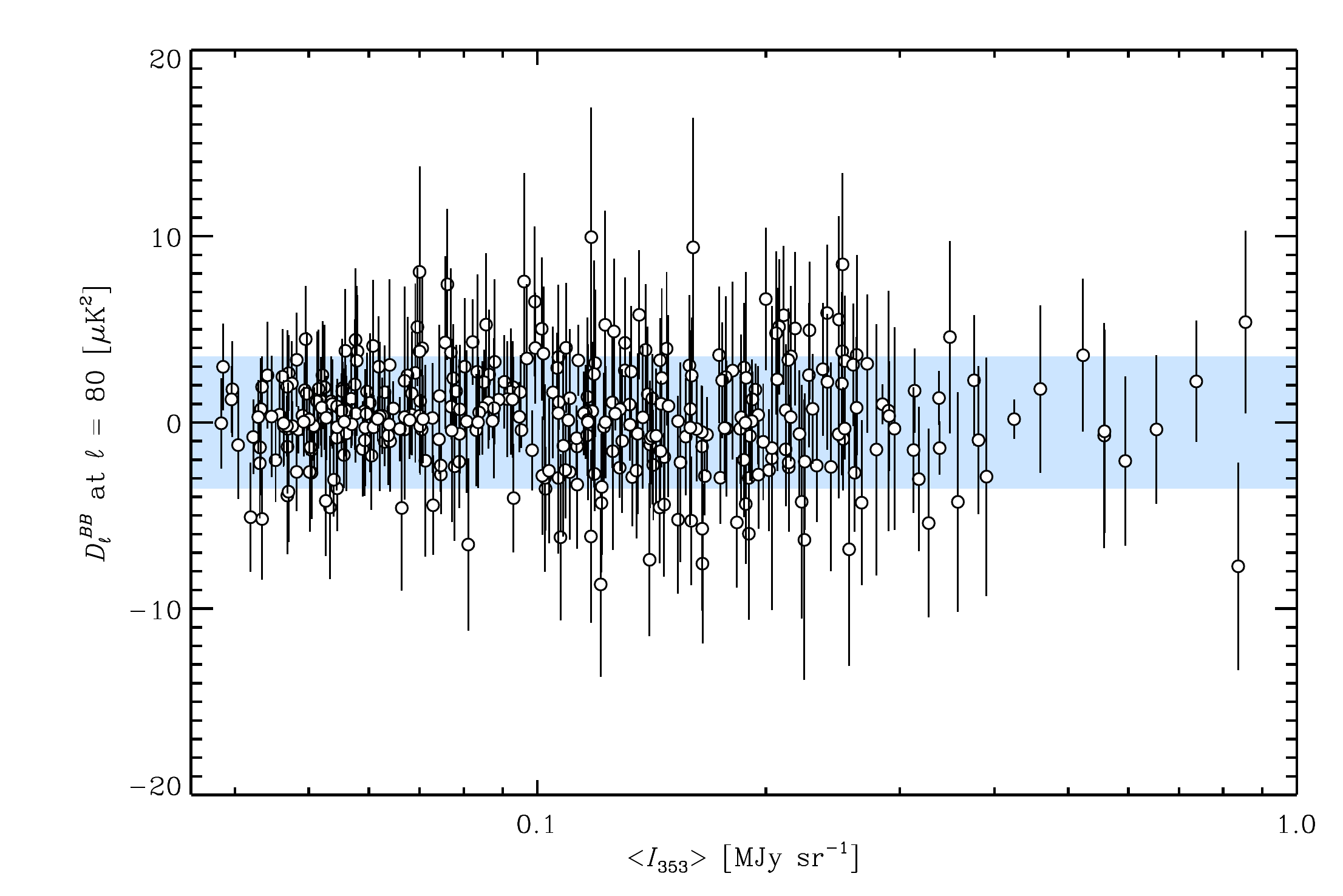}
\caption{
{\it Top}: same as Fig.~\ref{fig:highlat_bb}, but
instead of computing the \dlee\ (top panel) and \dlbb\ (bottom panel)
spectra from the 353\,GHz cross-correlation between DetSet1 and DetSet2,
we compute them from the cross-correlation between the HalfRing
half-differences for each detector set, i.e.,
$(353_{\rm DS1,HR1}-353_{\rm DS1,HR2})/2
 \times(353_{\rm DS2,HR1}-353_{\rm DS2,HR2})/2$.
The 3$\,\sigma$ expectations from \Planck\ inhomogeneous white-noise
Monte Carlo simulations are represented as light-blue shaded areas.
\label{fig:highlat_systematics}
}
\end{figure*}

\begin{figure*}[t!]
\centering
\includegraphics[height=0.6\textwidth]{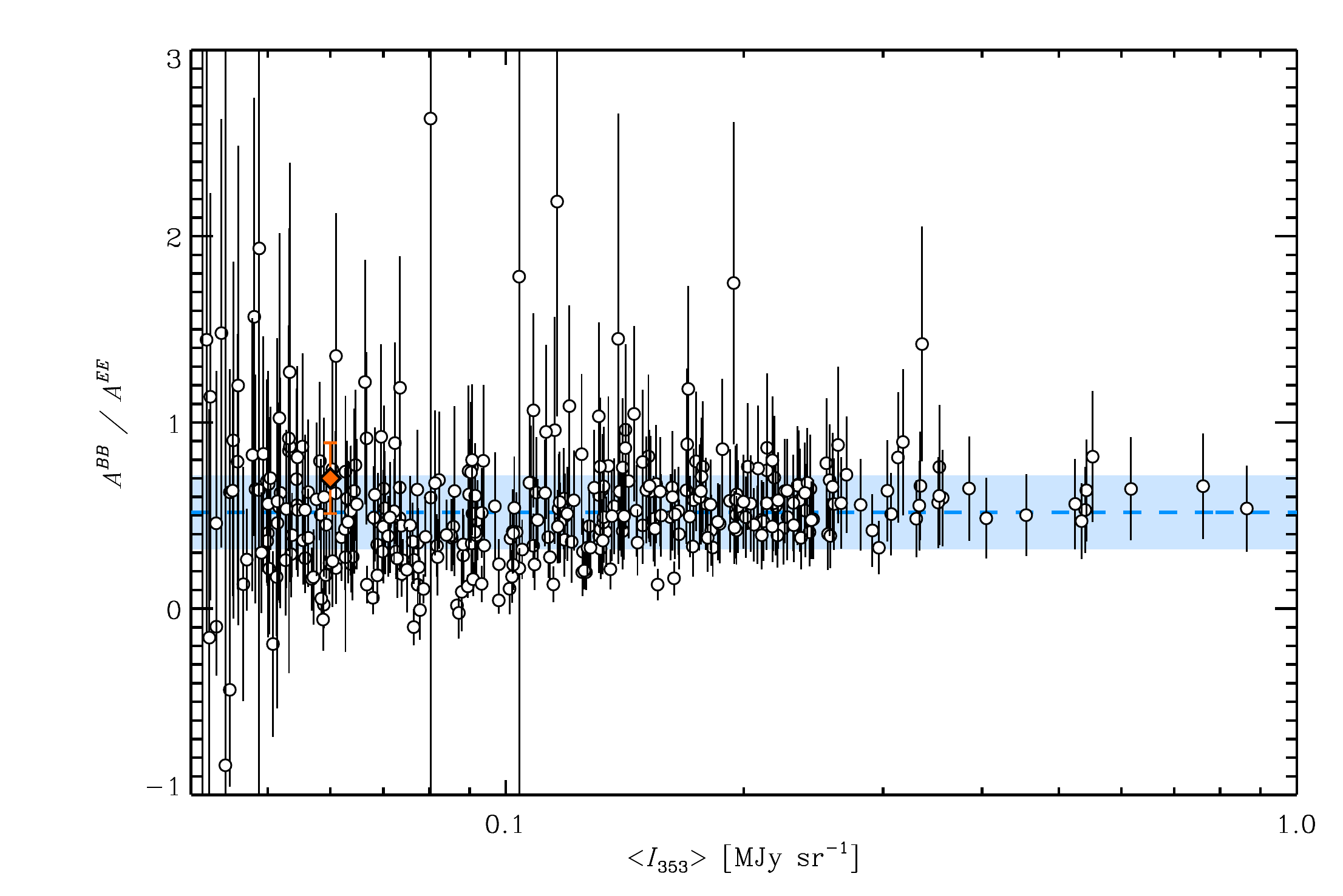}
\caption{Ratio of the dust $A^{BB}$ and $A^{EE}$ angular power spectra
amplitudes computed on the 352 \patcs\ at high Galactic latitudes (described
in Sect.~\ref{highlat_spectra}, circles), as a function of the mean dust
intensity \idust\ of the \patc. The mean value,
$\big\langle A^{BB}/A^{EE}\big\rangle=0.51\pm0.18$ here, is represented as a
dashed-blue line and the associated $\pm1\,\sigma$ dispersion as a light blue
shaded area. The $A^{BB}/A^{EE}$ ratio for the \mb2\ spectra presented in
Sect.~\ref{bicep2_ps353} is displayed as an orange diamond.
\label{fig:highlat_BonE}
}
\end{figure*}

We can see that for both methods there is no overall bias in the
recovered \dlbb\ angular power spectra, demonstrating that there is not
substantial leakage from $E$ to $B$. At multipoles above 40, the recovered
values agree with the input values and the maximum excursions (5\,\%)
seen in some bins are consistent with the expected accuracy for this
number of simulations. Moreover, \Xpol\ and \Xpure\ give very similar
results, in terms of the mean value and uncertainty.  
\hpeter{Simplified sentence.}
In the multipole bin $\ell\simeq40$ on \mb2\ the small offsets for both
\Xpol\ and \Xpure\ ($-4$\,\% and $8$\,\%, respectively) might be hinting
at a limitation of the estimation procedure in this multipole bin because of
the size of the \patc.
\hQ{ORIGINAL: On \mb2, at
$\ell\simeq40$, there is a negative offset for \Xpol\ and a positive offset
for \Xpure\ (around 10\,\%).  For both methods, this might be hinting at a
limitation in this multipole bin due to the size of the \sten.}

A related comparative assessment of the two methods on actual \Planck\ data
is presented in Appendix~\ref{bicep2_xpol-xpure}.

%%%%%%%%%%%%%%%%%%%%%%%
\section{Complementary dust spectra at 353\,GHz: \dlte, \dltb, and \dleb}
\label{te-tb-eb}

Using the same procedure as described in Sects.~\ref{part:computation}
and \ref{intlat_spectra} for the dust polarized angular power spectra for the
\lwdpref\ {\window}s at intermediate Galactic latitude, we have computed the
353\,GHz dust spectra involving temperature and polarization, \dlte\ and
\dltb, and the cross-spectrum for polarization, \dleb.  This completes the
entire set of polarization-related power spectra. 
Spectra such as \dltb\ and \dleb\ necessarily vanish for cosmic fields that
satisfy statistical isotropy and parity invariance.  
However, the emission from our Galaxy satisfies neither of these and so
the \dltb\ and \dleb\ spectra are generally expected to be non-zero.

\begin{figure}
\centering
\includegraphics[height=0.33\textwidth]{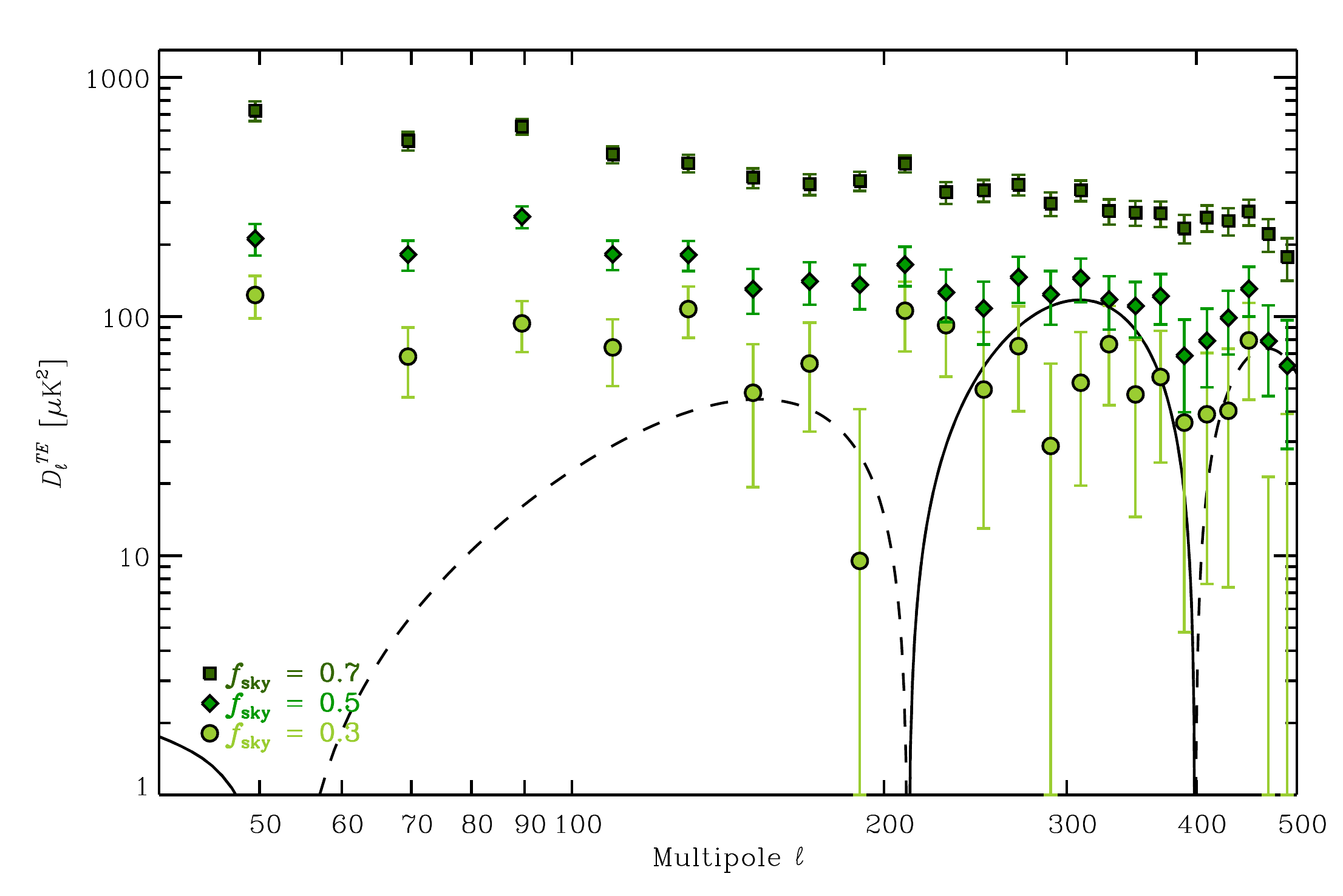}
\caption{\Planck\ HFI 353\,GHz \dlte\ power spectra (in $\mu$K$_{\rm CMB}^2$)
computed on three of the selected \lwdpref\ analysis regions that have
$\fsky=0.3$ (circles, lightest), $\fsky=0.5$ (diamonds, medium), and
$\fsky=0.7$ (squares, darkest).  The uncertainties plotted here are
$\pm 1\,\sigma$.  The \Planck\ 2013 best-fit $\Lambda$CDM \dlte\ expectation
\citep{planck2013-p11} is displayed as the black curve, solid where positive,
dashed where negative.
}
\label{fig:te_large}
\end{figure}

In Fig.~\ref{fig:te_large} we present the \dlte\ results for
$\fsky=\{0.3,0.5,0.7\}$ (note that the \Planck\ best-fit CMB \dlte\ power
spectrum was removed, as in Sect.~\ref{method_data_sets}).  
Again, the amplitudes of the spectra increase with increasing \fsky because
the polarized emission is brighter on average when more sky is retained. The
amplitudes of the spectra are about a factor of 3 higher than those reported
for \dlee\ in Fig.~\ref{fig:eebb_spectra}.  
For the largest region there is evidence of a power-law dependence on multipole
$\ell$ over the range $\ell=\lmin$ to $600$, with a slope compatible with what
is found for \dlee\ and \dlbb\ in Sect.~\ref{powerlawfit}
($\alpha_{TE} = -2.37, -2.43$, and $-2.47$ for \fsky = 0.3, 0.5, and 0.7,
respectively).  The fact that the slopes of the spectra are similar to those
found in Sect.~\ref{powerlawfit} for the dust polarization and in
\cite{planck2014-XXII} for dust intensity, indicates that the $TE$
cross-correlation is dominated by the dust correlations in this multipole
range, and not by the other components in the intensity maps (e.g., CIB
or point sources).
We also show the \dlte\ power spectrum computed from the \Planck\ 2013
best-fit $\Lambda$CDM model of the CMB temperature data \citep{planck2013-p11}.
Like in Fig.~\ref{fig:eebb_spectra} for \dlee, the CMB model is below the
353\,GHz angular power spectra at low
$\ell$ and about the same order of magnitude at $\ell>300$. 

\begin{figure}
\centering
\includegraphics[height=0.33\textwidth]{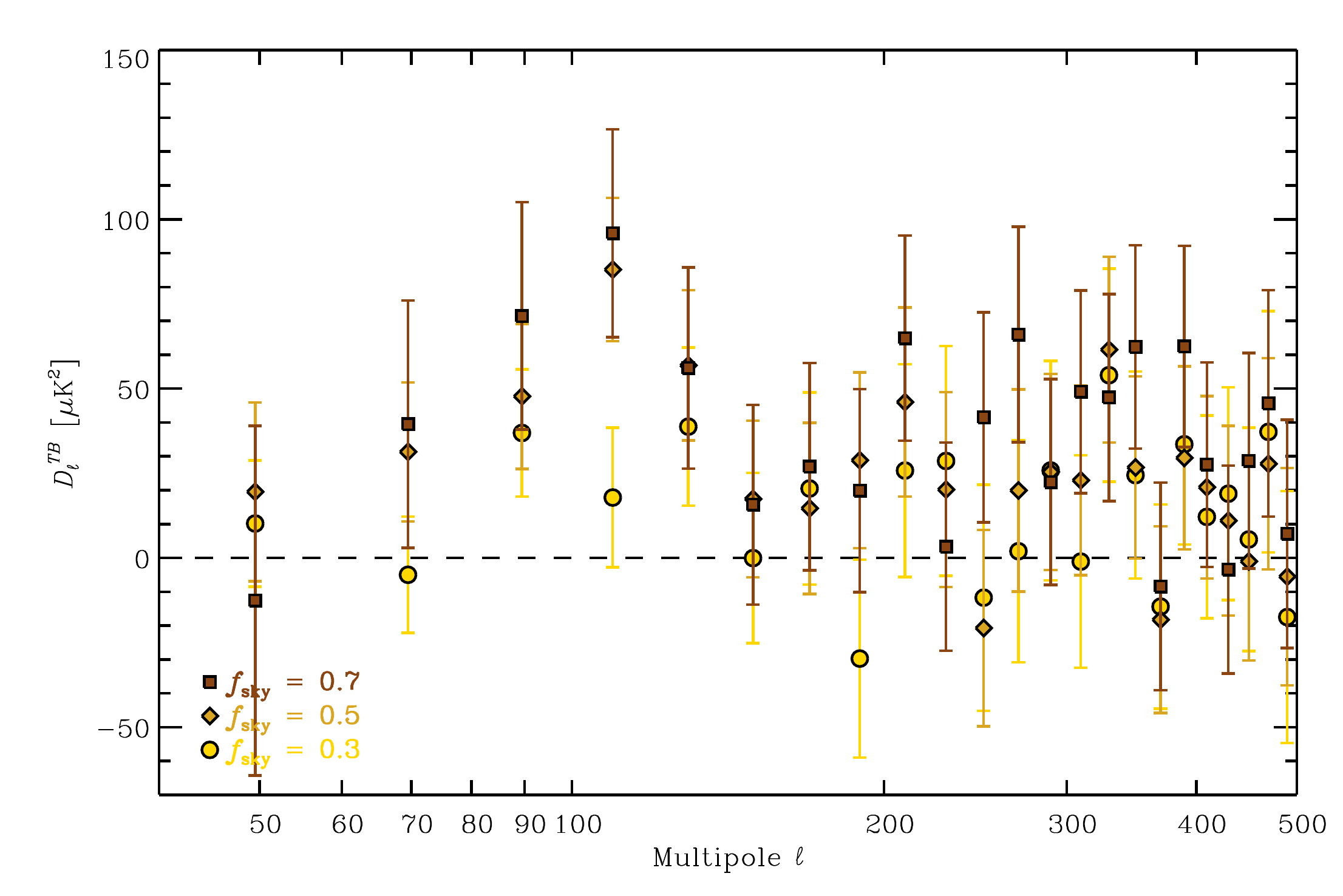}
\caption{\Planck\ HFI 353\,GHz \dltb\ power spectra (in $\mu$K$_{\rm CMB}^2$)
computed on three of the selected \lwdpref\ analysis regions that have
$\fsky=0.3$ (circles, lightest), $\fsky=0.5$ (diamonds, medium), and
$\fsky=0.7$ (squares, darkest).  The uncertainties plotted here are
$\pm 1\,\sigma$.}
\label{fig:tb_large}
\end{figure}

Similarly, the results for \dltb\ are shown in Fig.~\ref{fig:tb_large}, but
now on a linear scale.  For the two largest regions (\fsky = 0.5 and 0.7), we
detect a significant signal, increasing with the sky area, in the range
$\ell=60$--130.  In the same multipole range, the dust is only marginally
detected for the \fsky = 0.3 region.  At lower and higher $\ell$, the
individual binned
measurements of \dltb\ on all the regions are compatible with zero.

\begin{figure}
\centering
\includegraphics[height=0.33\textwidth]{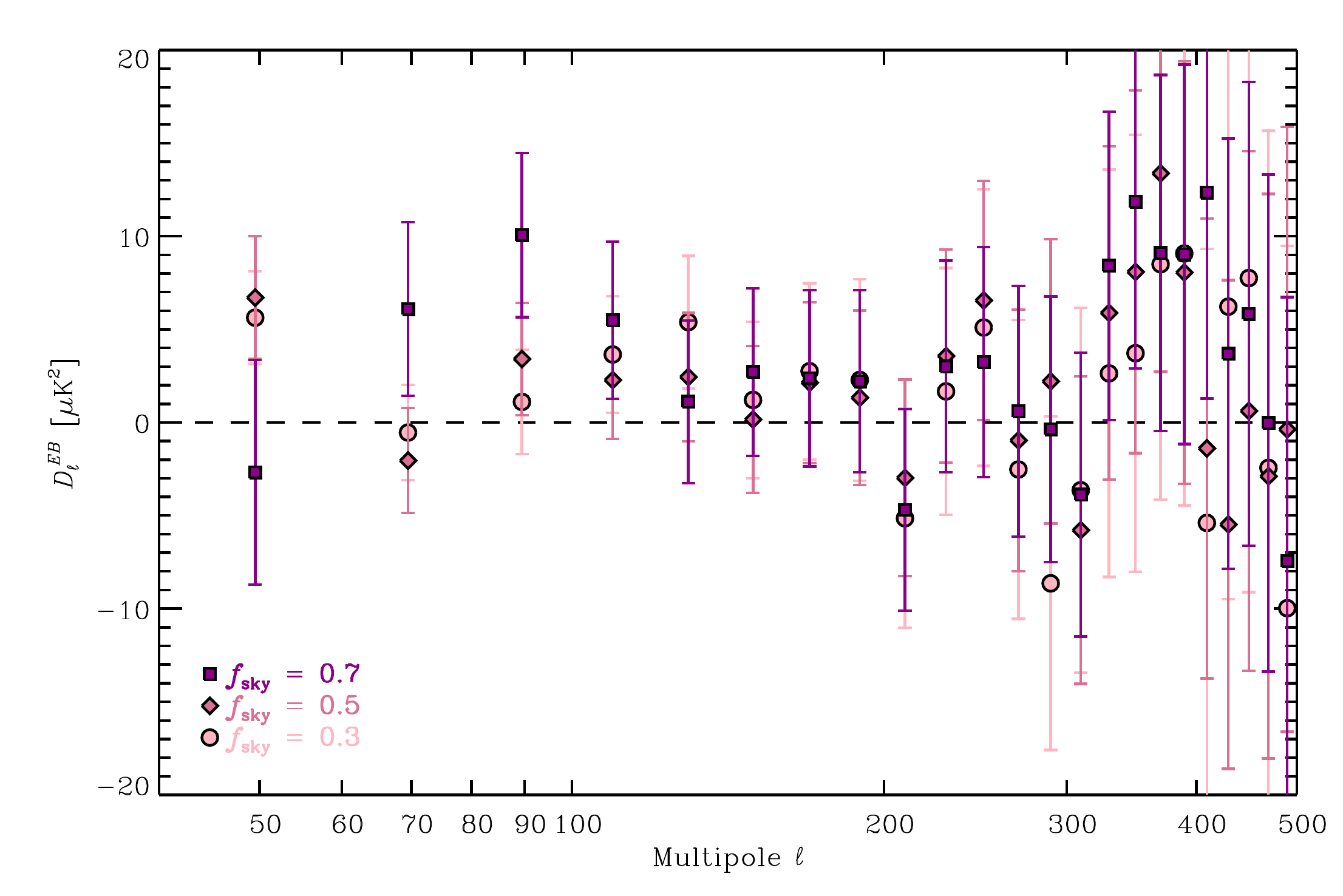}
\caption{\Planck\ HFI 353\,GHz \dleb\ power spectra (in $\mu$K$_{\rm CMB}^2$)
computed on three of the selected \lwdpref\ analysis regions that have
$\fsky=0.3$ (circles, lightest), $\fsky=0.5$ (diamonds, medium), and
$\fsky=0.7$ (squares, darkest).  The uncertainties plotted here are
$\pm 1\,\sigma$.}
\label{fig:eb_large}
\end{figure}

Finally we show the results for \dleb\ in Fig.~\ref{fig:eb_large}.  There is a
marginal detection of the \dleb\ spectrum in the range $\ell=60$--130 for the
\fsky = 0.7 mask.  For the other masks and at other multipoles, the \dleb\
spectra are compatible with zero.

%%%%%%%%%%%%%%%%%%%%%%%%%%%%%%%%%%%%%%%%%%%%%%%%%%%%%%%%%%%%%%%%%%
\section{Notes on the analysis of the high Galactic latitude \patcs}
\label{appendix_highlat}

\subsection{Assessment of \Planck\ polarization systematic uncertainties}
\label{appendix_highlat_spectra}

To assess the level of systematic effects in the results presented in
Sect.~\ref{highlat_spectra}, we use a null-test analysis (see also
Appendix~\ref{bicep2_syste}).
We repeated the statistical analysis for the 353\,GHz \dlee\ and \dlbb\
cross-spectra on the high-Galactic latitude \patcs. The steps of the analysis
are exactly the same as in Sect.~\ref{highlat_spectra_processing}, except that
instead of using the 353\,GHz $353_{\rm DS1}\times353_{\rm DS2}$ cross-power
spectra, we use the
$(353_{\rm DS1,HR1}-353_{\rm DS1,HR2})/2
 \times(353_{\rm DS2,HR1}-353_{\rm DS2,HR2})/2$ half-difference cross-spectra.
These cross-spectra are expected to be consistent with zero for uncorrelated
noise and subdominant systematics. 

The computed dust \dlee\ and \dlbb\ amplitudes $A^{EE,BB}$ (at $\ell=80$
computed on each of these \patcs) for the null-test spectra, in
$\mu$K$_{\rm CMB}^2$ at 353\,GHz, are displayed in
Fig.~\ref{fig:highlat_systematics} as a function of \idust.
We also report in this figure the expected noise level, computed from
Monte Carlo simulations of \Planck\ inhomogeneous white noise.

We can see that the fitted amplitudes for the null-test spectra on the 352
high Galactic latitude \patcs\ are scattered around zero, without any obvious
bias, for the entire range of \idust; this is the expectation for
half-difference cross-spectra. Taking into account the individual error bars,
there is no \patc\ that has a
null-test spectrum amplitude inconsistent with the $\pm3\,\sigma$
range of the instrumental noise dispersion. 

We thus conclude from this analysis that the results presented in
Sect.~\ref{highlat_spectra} are not significantly affected by instrumental
systematics.

%%%%%%%%%%%%%%%%%%%%%%%%%%%%%%%%%%%%%%%%%%%%%%%%%%%%%%%%%%%%%%%%%%
\subsection{Dust $B$-mode amplitudes compared to $E$-mode}
\label{appendix_highlat_BonE}

In Sect.~\ref{highlat_spectra}, we computed the 353\,GHz dust \dlee\ and
\dlbb\ amplitudes on the 352 high Galactic latitude \patcs\ defined in
Sect.~\ref{highlat_masks}. We present in Fig.~\ref{fig:highlat_BonE} the ratio
of the amplitudes of these spectra, $A^{BB}/A^{EE}$ as a function of the mean
dust intensity \idust\ of the \patc\ on which it was computed.

These ratios are scattered around the inverse noise variance weighted
$\big\langle A^{BB}/A^{EE}\big\rangle=0.52\pm0.19$ value. This value is
consistent with the one computed in Sect.~\ref{EonB} on the \lwdpref\
\window{s}. Nevertheless, it can be seen from Fig.~\ref{fig:highlat_BonE}
that there are \patcs\ with a significantly different
$A^{BB}/A^{EE}$ ratio, e.g., some smaller values for \idust\ in the range
0.08--0.1\,\MJysr.

We additionally report in Fig.~\ref{fig:highlat_BonE} the $A^{BB}/A^{EE}$
value determined for the \mb2\ spectra computed in Sect.~\ref{bicep2_ps353}.
The $A^{EE}$ and $A^{BB}$ amplitudes are computed in a similar way as in
Sect.~\ref{highlat_spectra}, fitting a power law in $\ell$ to the \dlee\ and
\dlbb\ spectra with a fixed \alphaexponent\ $\alpha_{EE,BB}=\alphaps$. Thus,
for \mb2, we find that under these assumptions $A^{BB}/A^{EE}=0.70\pm0.19$.
This value is higher than the average for such fields, but consistent with the
mean expected value.

\section{Systematic effects relating to the power spectrum estimate in the
\bicep\ \sten}
\label{methods_bicep}

%%%%%%%%%%%%%%%%%%%%%%%%%%%%%%%%%%%%%%%%%%%%%%
\subsection{Dust \dlbb\ spectrum measurements in the \bicep\ \sten\ using
\Xpol\ and \Xpure}
\label{bicep2_xpol-xpure}

\begin{figure}
\centering
\includegraphics[height=0.33\textwidth]{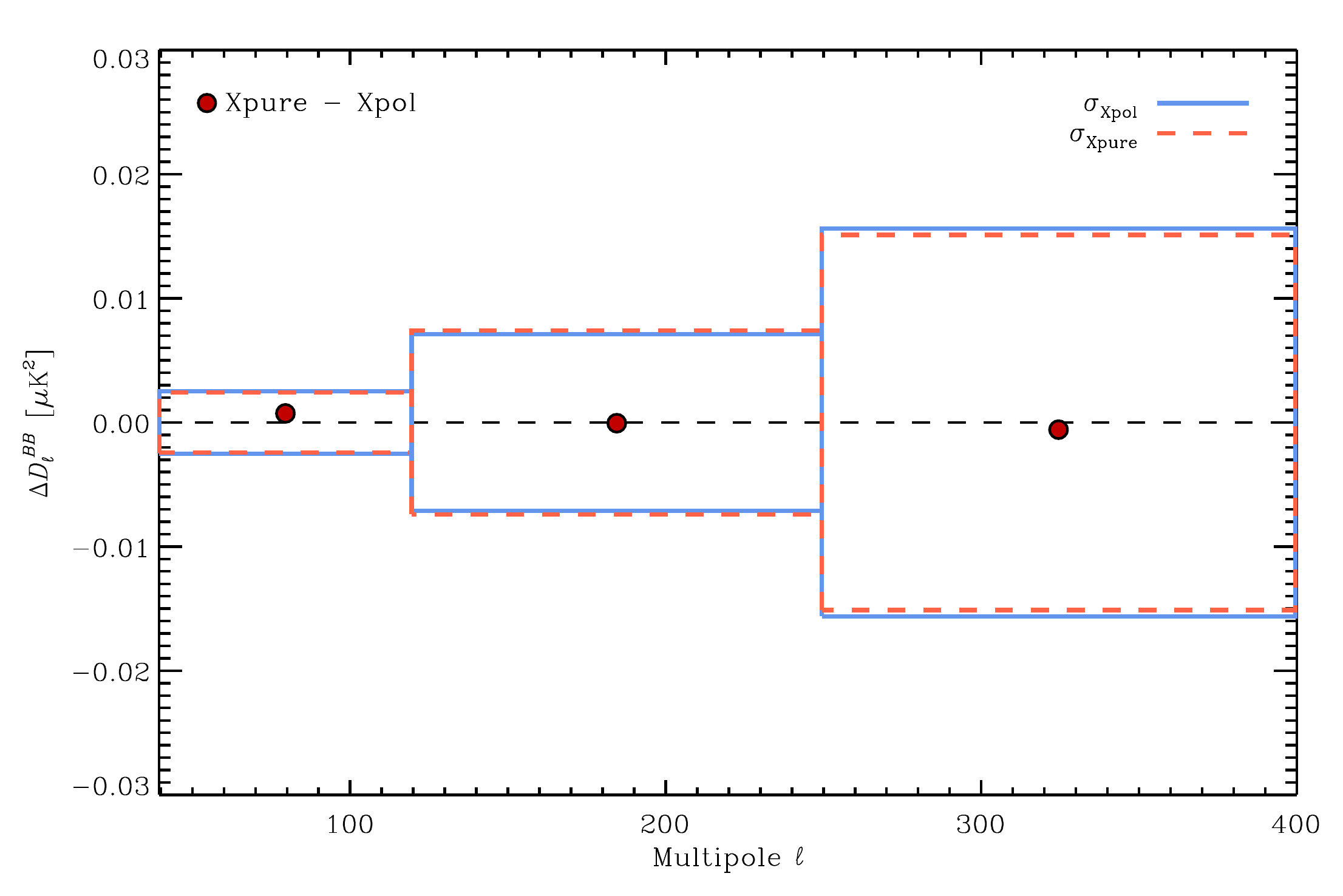}
\caption{
\Xpol\ minus \Xpure\ \Planck\ 353\,GHz \dlbb\ angular power spectrum
differences extrapolated to 150\,GHz, computed from the cross-DetSets on \mb2\
(red circles). The blue boxes represent the $\pm1\,\sigma$ errors computed
using \Xpol\ from the data on \mb2, while the dashed-orange boxes are the
\Xpure\ $\pm1\,\sigma$ errors coming from Monte Carlo simulations of
\Planck\ noise.
\label{fig:xpolxpure}
}
\end{figure}

We present in Fig.~\ref{fig:xpolxpure} the difference of the \dlbb\ power
spectra computed by \Xpol\ and \Xpure\ in
\mb2\ (${\cal D}_\ell^{BB,{\tt Xpure}}-{\cal D}_\ell^{BB,{\tt Xpol}}$).
The processing is the same as described in Sect.~\ref{bicep2_ps353}.

The \Xpol\ and \Xpure\ angular power spectra are consistent with each other in
all three $\ell$ bins. The maximum difference observed is in the $\ell=40$--120
bin, where the \Xpure\ spectrum is 10\,\% higher than the \Xpol\ spectrum
presented in Sect.~\ref{bicep2_ps353}. Moreover, the \Xpol\ recovered error
bars and the \Xpure\ error bars from Monte Carlo simulations of \Planck\
inhomogeneous noise, also presented in Fig.~\ref{fig:xpolxpure}, are
consistent. 

We conclude from this comparison that the results presented in
Sect.~\ref{bicep2_ps353} do not depend significantly on the method used for
estimating the dust \dlbb\ spectrum.

\begin{figure}
\centering
\includegraphics[height=0.33\textwidth]{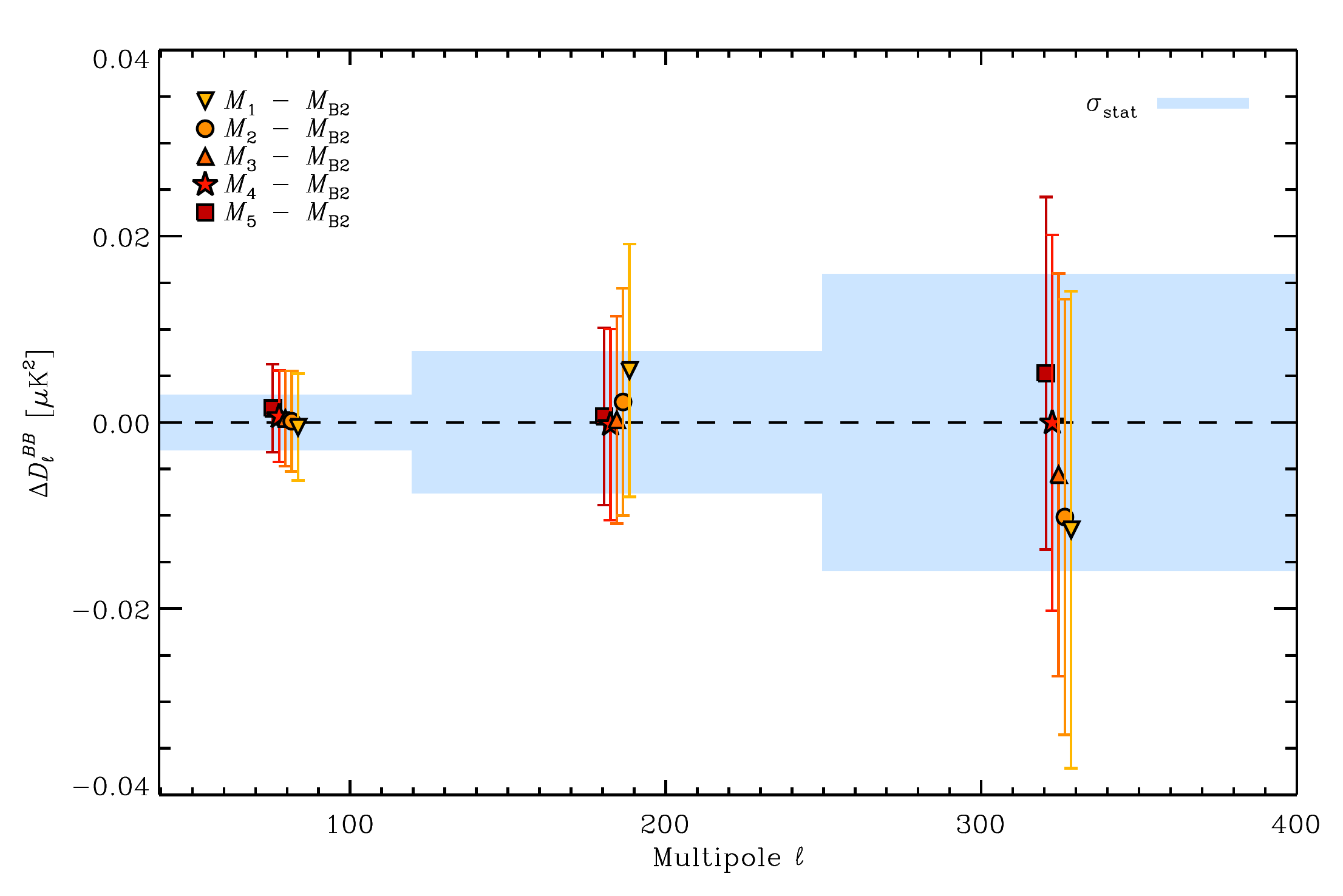}
\caption{
\Planck\ 353\,GHz \dlbb\ angular power spectrum differences extrapolated to
150\,GHz, determined from the spectra computed on the $M_{1-5}$ {\window}s
presented in Appendix~\ref{bicep2_mask_dependence} and the spectrum computed
on \mb2.  \dlbb\ difference values for the five {\window}s are displayed using
colours from yellow (for $M_1$) to dark red (for $M_5$). 
As in Fig.~\ref{fig:bicep2_bb}, the blue boxes represent the statistical
uncertainties from noise associated with the \mb2\ spectrum. They are centred
on zero, since we compute spectrum differences here.
}
\label{fig:test_masks}
\end{figure}

%%%%%%%%%%%%%%%%%%%%%%%%%%%%%%%%%%%%%%%%%%%%%%%%%%
\subsection{Impact of uncertainties in the specification of the \bicep\ \sten}
\label{bicep2_mask_dependence}

Our estimation of the dust \dlbb\ angular power spectrum in \mb2\ can
depend on the detailed definition of the \sten. To explore the dependence of
our results on the extent and specification of our approximation
of the \bicep\ \sten, we repeat
our computation of the dust \dlbb\ power spectrum on the 353\,GHz data
extrapolated to 150\,GHz, as in Sect.~\ref{bicep2_ps353}, for slight variations
in the definition of this \sten.

For this purpose, we start from the mask $M$ defined in
Sect.~\ref{bicep2_mask}, which consists of zeros and ones. We increase its size
(with a smoothing and a cut), in order to have an outline that extends exactly
$1^\circ$ further in every direction. This defines $M_5^{\rm b}$
(``b'' for binary, since it contains zeros and ones like $M$).
We then decrease the size of $M_5^{\rm b}$ (with smoothing and cut) by exactly
$1^\circ$ in every direction to obtain $M_4^{\rm b}$.  After that
$M_4^{\rm b}$ is shrunk in the same way to obtain $M_3^{\rm b}$ and we proceed
iteratively down to $M_1^{\rm b}$. Then to each $M^{\rm b}_i$ mask, we apply
exactly the same procedure for apodizing that we applied to $M$ to produce
$M_{\rm B2}$ in Sect.~\ref{bicep2_mask}, to yield the final $M_i$ \sten.
The effective area ($f_{\rm sky}^{\rm eff}4\pi$) is 805\,deg$^2$ for $M_5$,
677\,deg$^2$ for $M_4$, 564\,deg$^2$ for $M_3$, 460\,deg$^2$ for $M_2$,
and 365\,deg$^2$ for $M_1$. Note that due to the forward and backward
smoothing, \mb2\ defined in Sect.~\ref{bicep2_mask} is similar, but not
exactly equal to $M_4$.

We compute the 353\,GHz \dlbb\ power spectrum on all these five \stens\ using
\Xpol\ and extrapolate the results to 150\,GHz with the frequency dependence
from \cite{planck2014-XXII}, as in Sect.~\ref{bicep2_ps353}. We show
in Fig.~\ref{fig:test_masks} the difference between these spectra and the
\dlbb\ spectrum computed on \mb2\ presented in Sect.~\ref{bicep2_ps353}.
We also display the mean value and the error budget for \mb2,
centred on zero, since we are looking at differences between power spectra.

The recovered scatter is very low in the bin with $\ell=40$--120
(the standard deviation of the mean value among the \stens\ is
$0.04\times10^{-2}\,\mu$K$_{\rm CMB}^2$), much lower than the statistical
uncertainty in this particular bin. In the two other bins, the scatter is
more important, but still consistent with the statistical
uncertainty.

We conclude from this analysis that the main result presented in
Sect.~\ref{bicep2_ps353} does not depend on our precise definition of \mb2.

\begin{figure}
\centering
\includegraphics[height=0.33\textwidth]{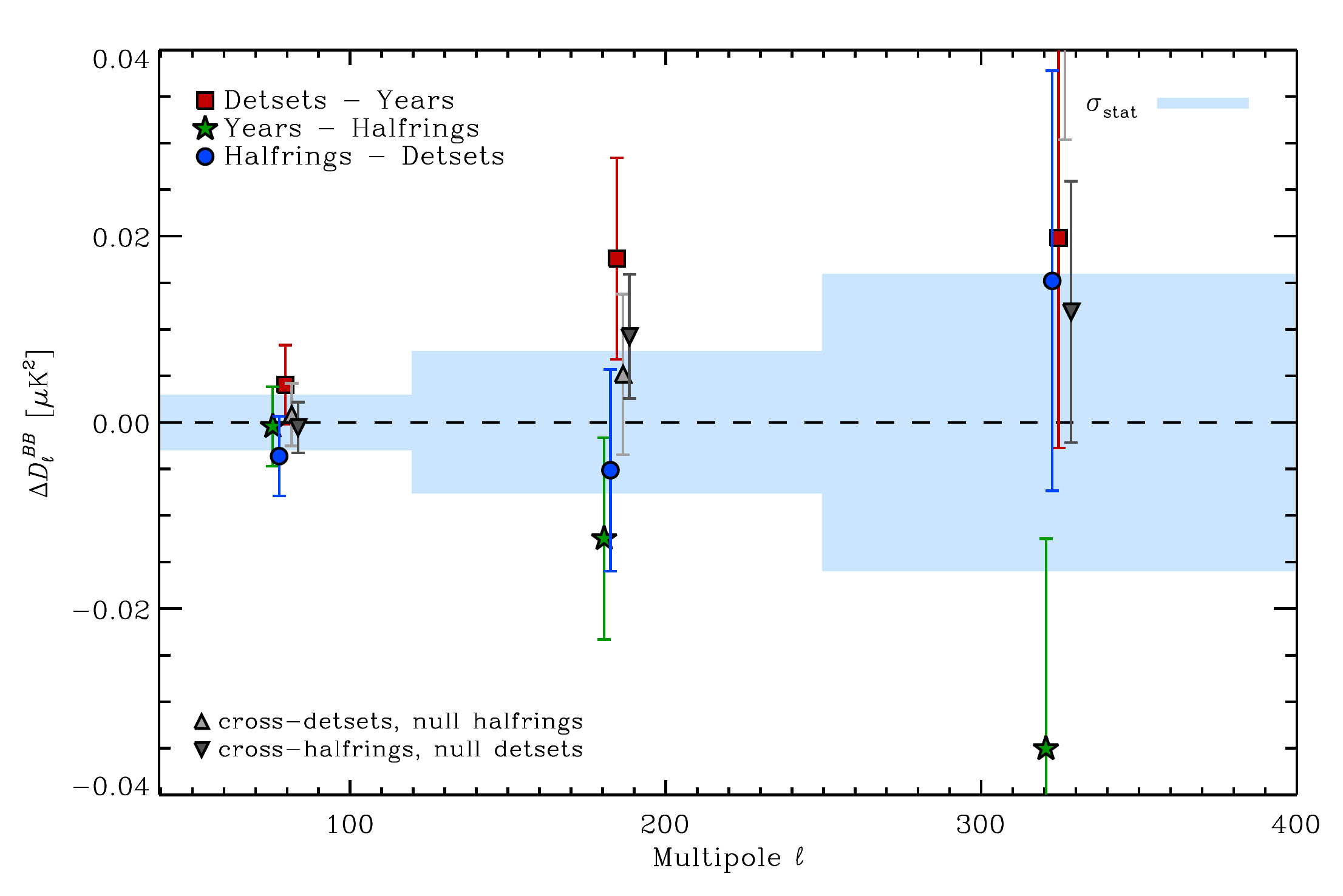}
\caption{
\Planck\ 353\,GHz \dlbb\ angular power spectrum differences extrapolated to
150\,GHz, computed from the following cross-spectra differences: DetSets
minus Years (red squares); Years minus HalfRings (green stars);
and HalfRings minus DetSets (blue circles). We also show the null test
performed using the cross-DetSets spectra of the HalfRing half-differences
(light grey triangle) and the cross-HalfRings of the DetSet half-differences
(dark grey reversed triangle) at 353\,GHz. As in Fig.~\ref{fig:bicep2_bb},
the blue boxes represent the statistical uncertainties from noise; these are
centred on zero, since we do not expect any signal for these null tests.
}
\label{fig:test_syste}

\end{figure}

%%%%%%%%%%%%%%%%%%%%%%%%%%%%%%%%%%%%%%%%%%%%%%%%%%%%%%%%%
\subsection{Assessment of the \Planck\ HFI systematics in the \bicep\ \sten}
\label{bicep2_syste}

To assess the potential contribution of systematic effects to the dust
353\,GHz \dlbb\ spectrum computed on \mb2, we repeat the analysis
presented in Sect.~\ref{bicep2_ps353} on null tests coming from subtracting
different \Planck\ 353\,GHz data subsets.

The cross-DetSets, cross-HalfRings, and cross-Years spectra are computed
for the 353\,GHz data extrapolated to 150\,GHz. The differences of these
cross-spectra (DetSets minus Years, Years minus HalfRings and HalfRings minus
DetSets) are presented in Fig.~\ref{fig:test_syste}. In the first $\ell$ bin
($\ell=40$--120), the three differences of the dust \dlbb\ power spectra are
consistent with zero, given the statistical uncertainty. They
exhibit a dispersion consistent with what is expected from statistical noise.
In the other bins, the dispersion is larger, but still consistent with the
noise.   We conclude that the systematic errors are subdominant.

In addition to these data cuts, we create null-test \dlbb\ spectra on
\mb2\ from the four data subsets at 353\,GHz involving detector sets and ring
halves.  We compute the cross-DetSet spectra between their two HalfRing
half-differences $(353_{\rm DS1,HR1}-353_{\rm DS1,HR2})/2
 \times(353_{\rm DS2,HR1}-353_{\rm DS2,HR2})/2$ and the cross-HalfRings
between the two DetSet half-differences
$(353_{\rm DS1,HR1}-353_{\rm DS2,HR1})/2
 \times(353_{\rm DS1,HR2}-353_{\rm DS2,HR2})/2$.
These null-test spectra are also displayed in Fig.~\ref{fig:test_syste}.
We can see that these spectra are consistent with zero in all the $\ell$ bins,
given the statistical and systematic uncertainties.

We report in Table~\ref{table_systes} the $\chi^2$ values for the null
hypothesis, as well as the $\pte$ for these null tests (plus additional ones,
not presented in Fig.~\ref{bicep2_syste}).
All these tests are consistent with the null hypothesis.

\begin{table}[htbp]
\begingroup
\newdimen\tblskip \tblskip=5pt
\caption{
Null hypothesis values of $\chi^2$ and $\pte$ for the null tests
presented in Sect.~\ref{bicep2_syste}.
}
\label{table_systes}
%\vskip -3mm
\footnotesize
\setbox\tablebox=\vbox{
 \newdimen\digitwidth
 \setbox0=\hbox{\rm 0}
 \digitwidth=\wd0
 \catcode`*=\active
 \def*{\kern\digitwidth}
 \newdimen\signwidth
 \setbox0=\hbox{+}
 \signwidth=\wd0
 \catcode`!=\active
 \def!{\kern\signwidth}
 \newdimen\smallwidth
 \setbox0=\hbox{\hskip 0.12em}
 \smallwidth=\wd0
 \catcode`?=\active
 \def?{\kern\smallwidth}
 \halign{\tabskip=0pt\hbox to 2.4in{#\leaderfil}\tabskip=1em&
 \hfil#\hfil&
 \hfil#\hfil\tabskip=0pt\cr
\noalign{\doubleline}
\omit\hfil 353\,GHz null test\hfil& $\chi^2$ ($N_{\rm dof}=3$)& PTE\cr
\noalign{\vskip 4pt\hrule\vskip 6pt}
${\rm ?DS1}\times{\rm ?DS2}- {\rm YR1}\times {\rm YR2}$& 3.76&  0.29\cr
${\rm YR1}\times{\rm YR2}  - {\rm HR1}\times {\rm HR2}$& 3.79&  0.28\cr
${\rm HR1}\times{\rm HR2}  - {\rm ?DS1}\times{\rm ?DS2}$& 1.46& 0.69\cr
\noalign{\vskip 4pt\hrule\vskip 6pt}
${\rm (?DS1-?DS2)}_{\rm HR1}\times{\rm (?DS1-?DS2)}_{\rm HR2}/4$&  2.67& 0.44\cr
${\rm (YR1 - YR2)}_{\rm ?DS1}\times{\rm(YR1 - YR2)}_{\rm ?DS2}/4$& 3.53& 0.32\cr
${\rm (?DS1-?DS2)}_{\rm YR1}\times{\rm (?DS1-?DS2)}_{\rm YR2}/4$&  2.04& 0.56\cr
${\rm (HR1 - HR2)}_{\rm YR1}\times{\rm (HR1 - HR2)}_{\rm YR2}/4$&  2.67& 0.45\cr
\noalign{\vskip 4pt\hrule\vskip 6pt}
}}
\endPlancktable
\endgroup
\end{table}

\subsection{Synchrotron contribution}
\label{sec:bicep2_synchrotron}

An estimate can be made of the level of a synchrotron contribution in \mb2\
by a simple extrapolation from the lowest frequency
\Planck\ LFI channels using a reasonable range of spectral indices.
Starting with the \Planck\ LFI 28.4\,GHz $Q$ and $U$ maps masked to
retain \mb2, a pure pseudo-$C_\ell$ power
spectrum estimator was used to calculate the $B$-mode band-powers
${\cal D}^{BB}_\ell$. The noise bias and significance was estimated
from 100 FFP8 noise simulations\footnote{Noise simulations for \Planck\ are
described in the Explanatory Supplement and in a forthcoming paper.},
which show the 28.4\,GHz power spectrum to be completely within the expected 
$\pm2\,\sigma$ noise envelope.  We therefore use the $3\,\sigma$ upper limit on
the scatter of the simulations
as the maximum allowable synchrotron signal at 28.4\,GHz.

A spectral index of $-3.12\pm0.04$ for synchrotron emission at high Galactic
latitudes was found by \citet{Fuskeland} using $TT$ plots of {\it
WMAP\/} 9\,yr data.  Adopting a spectral index of $-3.1$ for a direct
extrapolation to 150\,GHz, we find $3\,\sigma$ upper limits on the possible
contribution of synchrotron $B$-mode power at $\ell=45$ and $\ell=74$ 
to be 2.1\,\%, expressed relative to the primordial spectrum with $r=0.2$.
This estimate is consistent with the values
from \citet{Fuskeland}.  A more detailed and complete analysis of diffuse
polarized synchrotron emission will appear in a forthcoming \Planck\ paper.

%%%%%%%%%%%%%%%%%%
%%%%%%%%%%%%%%%%%% Separate these appendices specific to BICEP
%%%%%%%%%%%%%%%%%% presented in the order that they are referred to in the text
%%%%%%%%%%%%%%%%%%

\section{Dust polarization decorrelation}
\label{decorrelation}

\begin{figure}
\centering
\includegraphics[height=0.33\textwidth]{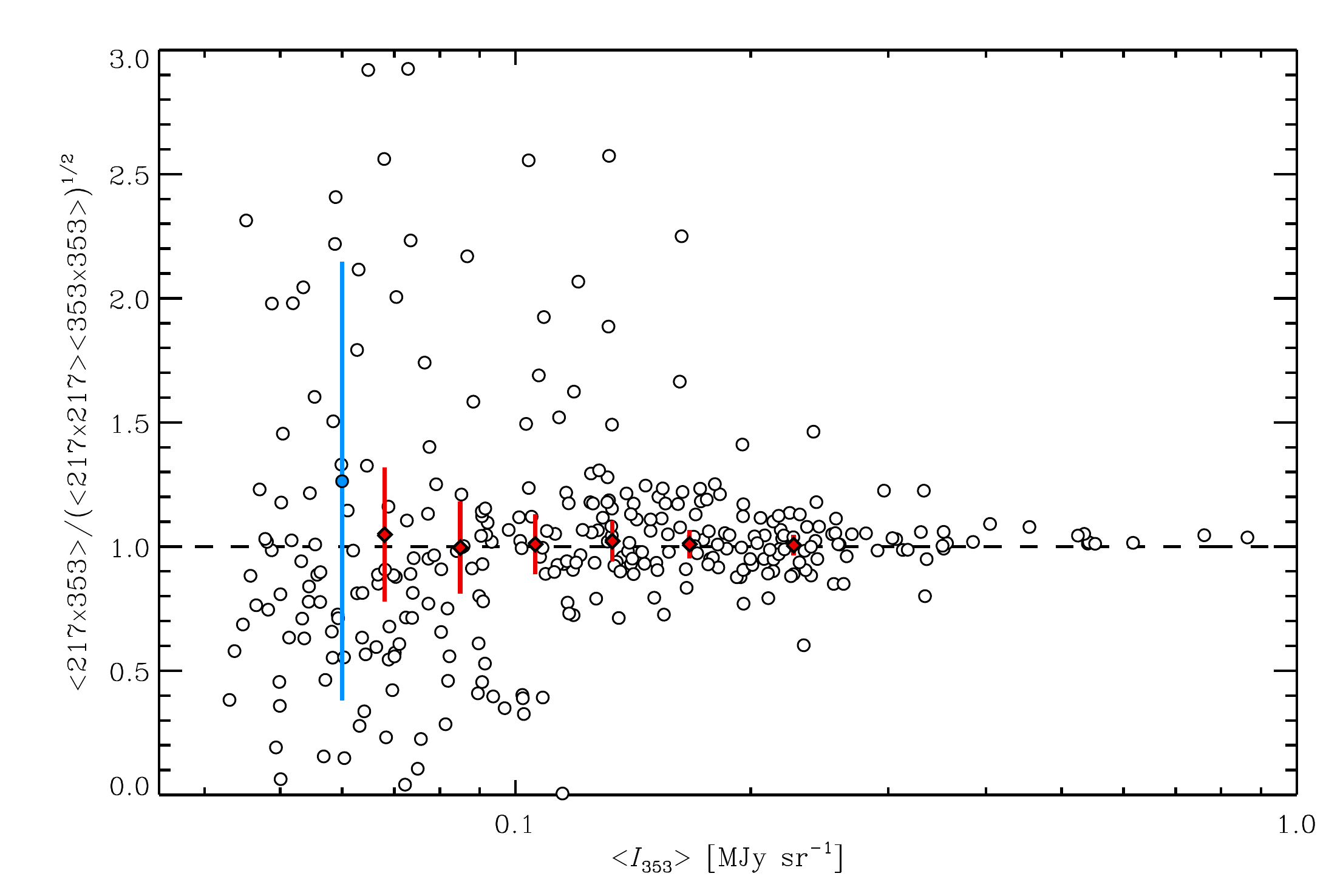}
\caption{Ratio of the $217\times353$ \dlbb\ cross-spectrum amplitude with
respect to the geometrical mean between the 217\,GHz and 353\,GHz \dlbb\
auto-spectrum amplitudes, as a function of \idust\ for the region on which the
spectra are computed.  These ratios are presented for the 352 patches of
Sect.~\ref{highlat_spectra} (black circles, with error bars not presented for
the sake of clarity), for the six LR regions of Sect.~\ref{intlat_spectra}
(red diamonds) and for the \mb2\ region of Sect.~\ref{sect:bicep2}
(blue diamond).}
\label{fig:freq_ratio}

\end{figure}

Extrapolation of the polarized dust signal observed at 353\,GHz to lower
frequencies (relevant to the CMB) hinges on the assumption that the pattern
of dust polarization on the sky is independent of frequency, with only the
amplitude changing, scaling as the SED.  However, because the dust emission is
the sum of many contributions from grains of different composition, size,
shape, and alignment, which will change along the line of sight, there can in
principle be some decorrelation of the signal from frequency to frequency.  We
have seen in Sect.~\ref{SED} that the dust polarization power spectra
amplitudes are compatible with the typical dust SED, indicating that if there
is such a decorrelation it has to be relatively weak.  Nevertheless, in this
appendix, we will assess this possible effect quantitatively.

If there is a decorrelation of the dust polarization with frequency, one
expects the cross-spectrum between two frequencies to be smaller than the
geometric mean of the two auto-spectra.
Hence, we compute the ratio between the cross- and auto-spectra for
217 and 353\,GHz, where we have the sensitivity to quantify the effect.
We compute \dlbb\ at these two frequencies and for the cross-spectrum and fit
a power-law amplitude $A^{BB}$ in each case, with a slope fixed to
$\alpha_{BB}=-2.42$, as in Sect.~\ref{powerlawfit}. 
 
The $d^{BB}_{217,353}\equiv
 A^{BB}_{217\times353}/(A^{BB}_{217\times217}\, A^{BB}_{353\times353})^{1/2}$
ratios are displayed in Fig.~\ref{fig:freq_ratio} as a function of \idust\,
for the 352 patches used in Sect.~\ref{highlat_spectra}.  Although
there is a large dispersion at low \idust\, one can see that the ratios are
centred on unity, indicating that on average there is no decorrelation;
among the 353 patches, we find a mean decorrelation ratio of
$d^{BB}_{217,353}=1.01\pm0.07$.

In Fig.~\ref{fig:freq_ratio} we also display this ratio for the six LR regions
of Sect.~\ref{intlat_spectra} and the \mb2\ region of Sect.~\ref{sect:bicep2},
finding no evidence for a significant decorrelation; for the LR regions, we
specifically find $d^{BB}_{217,353}=1.01\pm0.03$.

\end{document}